\documentclass[reprint,
superscriptaddress,
nobibnotes,
amsmath,
amssymb,
aps,
floatfix,
11pt
]{revtex4-2}

\pdfoutput=1

\usepackage[varg]{txfonts}
\usepackage[utf8]{inputenc}
\usepackage[english]{babel}
\usepackage[T1]{fontenc}
\usepackage{amsmath}
\usepackage{hyperref}
\usepackage{graphicx}
\usepackage{dcolumn}
\usepackage{bm}
\usepackage{dsfont}
\usepackage{mathrsfs}
\usepackage{enumitem}
\usepackage[dvipsnames]{xcolor}

\usepackage[acronym]{glossaries}
\glsdisablehyper
\makeglossaries

\newcommand*{\f}{\frac}
\newcommand*{\mc}{\mathcal}
\newcommand*{\dg}{\dagger}

\newcommand*{\mbb}{\mathds}
\newcommand*{\mscr}{\mathscr}
\newcommand*{\ex}{\mathrm{e}}
\newcommand*{\dimE}{d_\mathsf{E}}
\newcommand*{\dimS}{d_\mathsf{S}}
\newcommand*{\env}{\mathsf{E}}
\newcommand*{\syst}{\mathsf{S}}
\newcommand*{\ctrl}{\mathsf{ctrl}}

\DeclareMathOperator{\tr}{tr}
\DeclareMathOperator*{\Motimes}{\text{\raisebox{0.25ex}{\scalebox{0.6}{$\bigotimes$}}}}
\DeclareMathOperator*{\Moplus}{\text{\raisebox{0.25ex}{\scalebox{0.6}{$\bigoplus$}}}}
\DeclareMathOperator*{\Mcirc}{\text{\raisebox{0.25ex}{\scalebox{0.7}{$\bigcirc$}}}}
\DeclareMathOperator*{\llangle}{\langle\!\langle\!}
\DeclareMathOperator*{\rrangle}{\!\rangle\!\rangle}

\newacronym{rb}{RB}{Randomized Benchmarking}
\newacronym{spam}{SPAM}{State Preparation and Measurement}
\newacronym{asf}{ASF}{Average Sequence Fidelity}
\newacronym{dd}{DD}{Dynamical Decoupling}
\newacronym{udd}{UDD}{Universal Dynamical Decoupling}
\newacronym{cp}{CP}{Completely Positive}
\newacronym{tp}{TP}{Trace Preserving}
\newacronym{cptp}{CPTP}{Completely Positive Trace Preserving}
\newacronym{ptm}{PTM}{Pauli Transfer Matrix}
\newacronym{povm}{POVM}{Positive Operator Valued Measurement}
\newacronym{rc}{RC}{Randomized Compiling}

\setlist[itemize]{leftmargin=*, noitemsep}
\setlist[enumerate]{leftmargin=*, itemsep=0.1em}

\definecolor{C1}{HTML}{758BFD}
\definecolor{C2}{HTML}{EF2D56}
\definecolor{C3}{HTML}{16F4D0}
\definecolor{C4}{HTML}{731DD8}
\definecolor{iqmblue}{RGB}{117,157,235}
\definecolor{iqmgreen}{RGB}{95, 221, 151}

\hypersetup{colorlinks=true, linkcolor=iqmblue, citecolor=iqmblue, urlcolor=iqmblue}

\usepackage{tikz}
\usepackage{lipsum}

\newcounter{result}
\newenvironment{result}[1][]{\refstepcounter{result}\par\medskip\noindent%
   \underline{\textbf{Result~\theresult. #1}} \itshape}{\medskip}

\begin{document}

\title{Operational Markovianization in Randomized Benchmarking}

\author{P. Figueroa-Romero}
\email{pedro.romero@meetiqm.com}
\affiliation{IQM Germany GmbH, Georg-Brauchle-Ring 23-25, 80992 Munich, Germany}

\author{M. Papi\v{c}}
\affiliation{IQM Germany GmbH, Georg-Brauchle-Ring 23-25, 80992 Munich, Germany}
\affiliation{Department of Physics and Arnold Sommerfeld Center for Theoretical Physics, Ludwig-Maximilians-Universität München, Theresienstrasse 37, 80333 Munich, Germany}

\author{A. Auer}
\affiliation{IQM Germany GmbH, Georg-Brauchle-Ring 23-25, 80992 Munich, Germany}

\author{M.-H. Hsieh}
\affiliation{Hon Hai Quantum Computing Research Center, Taipei, Taiwan}

\author{K. Modi}
\affiliation{School of Physics and Astronomy, Monash University, Clayton, VIC 3800, Australia}
\affiliation{Centre for Quantum Technology, Transport for New South Wales, Sydney, NSW 2000, Australia}

\author{I. de Vega}
\affiliation{IQM Germany GmbH, Georg-Brauchle-Ring 23-25, 80992 Munich, Germany}
\affiliation{Department of Physics and Arnold Sommerfeld Center for Theoretical Physics, Ludwig-Maximilians-Universität München, Theresienstrasse 37, 80333 Munich, Germany}

\begin{abstract}
A crucial task to obtain optimal and reliable quantum devices is to quantify their overall performance. The average fidelity of quantum gates is a particular figure of merit that can be estimated efficiently by \gls{rb}. However, the concept of gate-fidelity itself relies on the crucial assumption that noise behaves in a predictable, time-local, or so-called Markovian manner, whose breakdown can naturally become the leading source of errors as quantum devices scale in size and depth. We analytically show that error suppression techniques such as \gls{dd} and Pauli-twirling can operationally \emph{Markovianize} \gls{rb}: \emph{i}) fast \gls{dd} reduces non-Markovian \gls{rb} to an exponential decay plus longer-time corrections, while on the other hand, \emph{ii}) Pauli-twirling generally does not affect the average, but \emph{iii}) it always suppresses the variance of such \gls{rb} outputs. We demonstrate these effects numerically with a qubit noise model. Our results show that simple and efficient error suppression methods can simultaneously tame non-Markovian noise and allow for standard and reliable gate quality estimation, a fundamentally important task in the path toward fully functional quantum devices.
\end{abstract}
\glsresetall

\maketitle

\section{Introduction}
The characterization of noise in quantum information processors will remain a necessary and unavoidable task to build fault-tolerant devices. Some of the most common sets of techniques to achieve this can be said to fall within an interval comprising tomographic techniques on one end, and \gls{rb} techniques on the other: while the first can provide a detailed description of noise with an exponential measurement and sampling overhead in system size, the latter can estimate coarse average error rates efficiently~\cite{kashefi_2020, QCVV_2021}.

Due to the simplicity of \gls{rb}-based protocols, they have become ubiquitous both for small systems~\cite{Emerson_2005, PhysRevA.85.042311, Knill_2008, PhysRevLett.109.240504, Wallman_2015, PhysRevA.97.032306}, as well as a steppingstone for scalable techniques for more ambitious learning of quantum noise~\cite{proctor_2021, cycle_2019, helsen_seq_2021, flammia_aces, harper2023learning}. However, the manageable analytical behavior of the \gls{rb} data --namely, an exponential decay capturing average gate-fidelities, with \gls{spam} errors isolated as multiplicative and offset constants-- is guaranteed only with highly-simplified and unrealistic assumptions about the noise. More realistic regimes have actively been under investigation~\cite{PRXQuantum.3.030320, helsen_general_2022} and can still benefit from \gls{rb}'s simplicity.

\begin{figure*}[ht!]
    \centering
    \includegraphics[width=\textwidth]{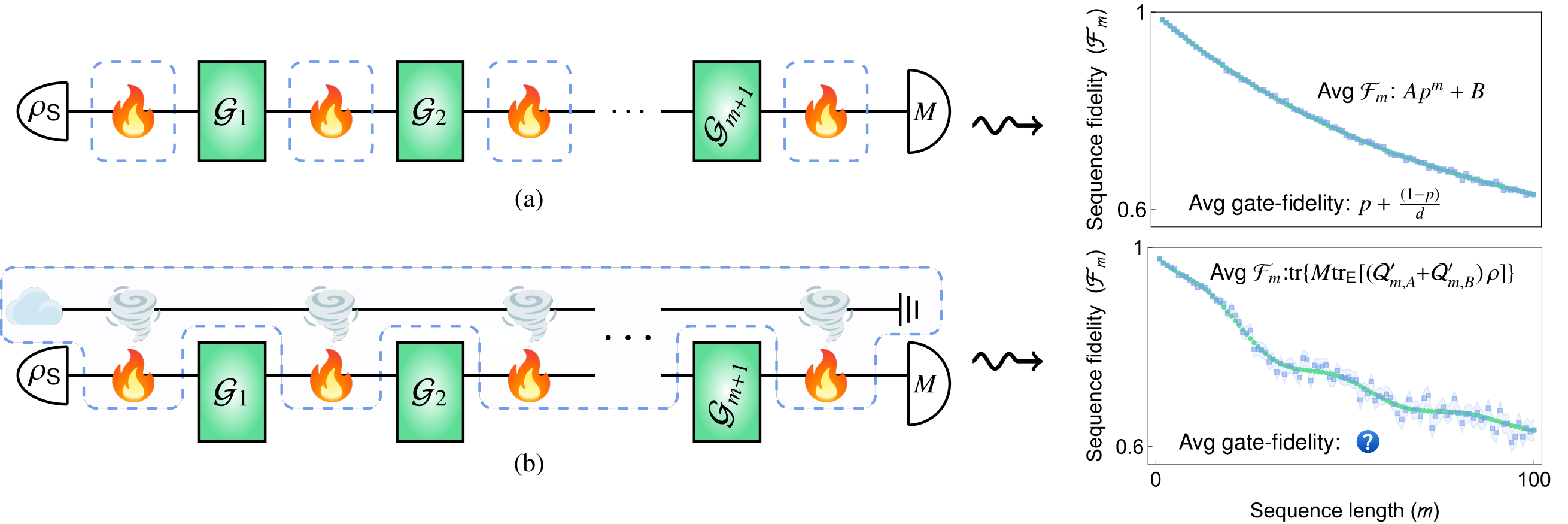}
    \caption{\textbf{Randomized Benchmarking (RB) under Markovian and non-Markovian noise}: Cartoons of sample circuits of the \gls{rb} protocol for a given initial state $\rho_\syst$, sequence of randomly sampled quantum gates $\{\mc{G}_i\}$ with ``undo'' (compiled inverse of the sequence) gate $\mc{G}_{m+1}$, and final measurement $M$, subject to (a) Markovian noise, where errors are uncorrelated with each other and can be associated to each individual gate, and (b) non-Markovian noise, where an external (quantum) system --i.e., an environment-- can serve as a memory, correlating errors in time and propagating their information to the final measurement. Example outputs of \gls{rb} associated with (a) and (b) are shown on the top and bottom right, correspondingly; detail in notation and meaning will be explained throughout the manuscript, here we only point out the distinction that non-Markovian \gls{rb} decays are generally non-trivially non-exponential, with the standard notion of an individual average gate-fidelity being ill-defined.
    }
    \label{fig: cartoon RB Markov vs nonMarkov}
\end{figure*}

Arguably, however, one of the most difficult simplifications to relax in any characterization technique is the one assuming that noise can be associated with each individual quantum gate separately. In reality, noise is e.g., introduced by the nature of the quantum device itself or by inherent limits in the control that the experimenter has, and errors occurring at a given time can propagate and affect future errors. These kinds of correlations and their effects are generally known as non-Markovianity~\cite{rivas_2014, RevModPhys.88.021002, RevModPhys.89.015001} and can be fully described in the quantum realm through multi-time generalizations of quantum channels~\cite{PhysRevA.97.012127, PhysRevLett.120.040405, PRXQuantum.2.030201}. A signature aspect of non-Markovian noise in \gls{rb} is the non-trivial non-exponential decay of the data rendered by it~\cite{limits_RB_2014, noise_corr_RB_2016, PRXQuantum.2.040351, figueroaromero2022general, ceasura2022nonexponential}, which makes the information about the noise and its correlations remarkably hard to analyze. Moreover, the key feature of \gls{rb} of robustness against \gls{spam} errors ceases to hold if any of these are correlated. This is depicted schematically in Figure~\ref{fig: cartoon RB Markov vs nonMarkov}.

While characterizing and better understanding non-Markovian noise is an active area of research~\cite{PhysRevApplied.13.034045, White2020demonstration, PhysRevA.104.022432, white2021nonmarkovian, PhysRevA.105.022605, Guo_2022, White2022}, various control and error suppression techniques, such as \gls{dd}~\cite{ezzell_2022} and Pauli-twirling, via \gls{rc}~\cite{winick2022RC}, have been shown to remove non-Markovian noise effects to a statistically significant extent. This can further be viewed with a resource-theoretic lens~\cite{berk2021extracting} as keeping local-system information in exchange of the consumption of temporal correlations. Generally, in the Markovian regime, both \gls{dd} and \gls{rc} are well-known to enhance the quality of quantum computations with a low resource overhead~\cite{DD_FT_Preskill, PhysRevLett.121.220502, PauliFR_2021, RC_superc_2021}, and combining \gls{rb} with quantum control has previously been done successfully, e.g., to optimize the quantum control itself~\cite{orbit_2014}, or to demonstrate enhanced average gate-fidelities~\cite{PhysRevA.92.062332}.

\underline{\textbf{Main Results (informal)}}: In this manuscript, we show that for a broad class of non-Markovian noise models, \gls{dd} can effectively and efficiently \emph{Markovianize} \gls{rb}, i.e., remove non-Markovian non-exponential deviations, allowing for a straightforward prediction of enhanced average gate-fidelities with low overhead. In other words, \gls{dd} converts non-Markovian correlations into more tractable quantum noise. We exemplify this effect numerically on a qubit with an $XY4$ sequence. Moreover, we analyze the effect of tailoring non-Markovian noise into Pauli noise within the accessible subsystem of interest in non-Markovian \gls{rb}, finding that, while Pauli-twirling does not Markovianize \gls{rb} in the same sense that \gls{dd} does, the uncertainty in the \gls{rb} outputs generally gets suppressed. We exemplify these effects with the same numerical model for a single qubit strongly interacting with another.

Our results show that coherent noise suppression and decoupling schemes in \gls{rb} can both efficiently Markovianize and allow the extraction of enhanced average gate fidelities that accurately capture --and do not overestimate-- all statistically-relevant error rates.

\section{RB and operational modeling of quantum noise}
\subsection{Setup and notation}
Detail on the notation we employ in the main text, as well as in the derivations for our results can be read in full in Appendix~\ref{appendix: notation}. Here we will consider a composite Hilbert space labeled $\syst\env$, comprised by a system of interest $\syst$, and an environment $\env$, which we assume to be inaccessible. $\env$ could simply be a subset of idle or ancillary qubits within a quantum device~\cite{ceasura2022nonexponential}. We will label their respective dimensions as $\dimS=2^{n_s}$, for $n_s$ qubits in $\syst$, and $\dimE$.

We present our results with the uniformly distributed $n_s$-qubit Clifford group as our gate-set and denote individual gates with the symbol $\mc{G}_i$, $i=1,2,\ldots$, although the general treatment with finite groups can be seen in the Appendices. We denote initial states as $\rho_\syst$ and measurements as $M$, both on system $\syst$, and which can be chosen arbitrarily, as long as the noiseless expectation $\tr(M\rho_\syst)$ is known. While the \gls{rb} protocol can be read in detail in Appendix~\ref{appendix: the RB protocol}, the three main input components in a \gls{rb} experiment are \emph{i}) the gate-set $\{\mc{G}_i\}_{i=1}^m$ for a given integer $m$, \emph{ii}) the initial state $\rho_\syst$, and \emph{iii}) the measurement element or observable $M$.

We will denote by $\mathbf{E}_\mc{G}$ the uniform average over the gate-set $\{\mc{G}_i\}$, and reserve the notation $\mc{F}_m$ for a so-called average-sequence fidelity, defined here as $\mc{F}_m:=\mathbf{E}_{\mc{G}}\tr[M^\prime\mc{G}_{m+1}^\prime\circ\mc{G}^\prime_m\circ\cdots\mc{G}^\prime_1(\rho_\syst^\prime)]$, where the primed terms denote the real (noisy) implementations of the corresponding initial state, measurement, and gates, and where $\mc{G}_{m+1}:=\mc{G}_m^{-1}\circ\cdots\mc{G}_1^{-1}$.

Finally, we refer to a quantum channel being a ``$\mathsf{X}$ channel'', or as a ``channel acting on $\mathsf{X}$'', when it maps inputs from space $\mathsf{X}$ to outputs in $\mathsf{X}$.

\subsection{RB: Markovian vs. non-Markovian}
Up to a few more parameters, like the number of gates sampled or the number of samples to generate, a full-fledged framework for \gls{rb} exists~\cite{helsen_general_2022} when the underlying noise is assumed to be effectively uncorrelated in time, i.e., Markovian. This allows us to estimate average gate fidelities, which despite being a rough figure of merit, are an essential component for the characterization of noise in a quantum device. In the Markovian, time-stationary, and gate-independent noise regime, the outputs of a \gls{rb} experiment over gate sequences of length $m$ are estimates that can be fitted to a function of the form
\begin{equation}
    \mc{F}_m = {A}\,p^m + B,
    \label{eq: Markov ASF (main)}
\end{equation}
so-called an average sequence fidelity, where here $p\lesssim1$ is a \emph{quality factor} capturing the noise solely associated to gates, and both $0\leq{A,B}\leq1$ are constants isolating the \gls{spam} errors. The average gate-fidelity~\textsuperscript{\footnote{ We point out that the average gate-fidelity of a quantum channel $\tilde{\mc{G}}$ with respect to a gate $\mc{G}$, is a measure of their ``average orthogonality'', rather than their distinguishability (which albeit related, are not quite the same thing): $\mathrm{F}_\mathsf{avg}:=\int{d}\psi\tr[\tilde{\mc{G}}(\psi)\mc{G}(\psi)]$, where $\psi$ are (uniformly distributed) pure states.}} of the noisy gates with respect to the ideal ones, $\mathrm{F}_\mathsf{avg}$, is then related to the quality factor $p$ simply as (see e.g.,~\cite{PhysRevA.85.042311})
\begin{equation}
    \mathrm{F}_\mathsf{avg} = p+\frac{(1-p)}{\dimS}.
\end{equation}

The Markovian assumption is effectively equivalent to an environment, $\env$, dissipating or \emph{forgetting} any information of its interaction with system $\syst$ at any given time, thereby just introducing noise locally at such time and reducing the purity of the noisy outputs, effectively as if $\env$ was not there at all. But several factors in the advancement and scaling up of quantum devices, including being able to probe smaller timescales and having larger systems strongly interacting with each other, make the Markov approximation implausible. Non-Markovianity, i.e., the presence of (non-negligible) temporal correlations, implies that anything other than $\syst$, in particular other qubits, can play the role of an $\env$ \emph{with memory} propagating undesired noise correlations in time~\cite{crosstalk_nM_White}, and that noisy processes cannot be thought of as individual quantum channels associated independently to the ideal quantum gates. 

A formal definition of non-Markovianity can be seen in Appendix~\ref{appendix: non-Markovianity def}; in particular, we work within the process tensor framework~\cite{PhysRevA.97.012127}, which generalizes the notion of stochastic processes to quantum theories~\cite{Milz2020kolmogorov} and contains both the classical notion and the various criteria for quantum non-Markovianity known hitherto~\cite{PhysRevLett.120.040405, PRXQuantum.2.030201}.

It has been shown in~\cite{figueroaromero2022general} that the generalization of Eq.~\eqref{eq: Markov ASF (main)} to the non-Markovian gate-independent~\textsuperscript{\footnote{ Non-Markovianity \emph{can} be said to introduce a type of gate-dependence; nevertheless, here by explicit gate-independence we mean noise associated to $\mc{I}_\env\otimes\mc{G}$, where $\mc{I}_\env$ is an identity map on $\env$ and $\mc{G}$ is an ideal gate, is not explicitly dependent on $\mc{G}$.}} noise case, has the (generally non-exponential) form
\begin{equation}
    \mc{F}_m = \tr\left\{M\tr_\env\left[\left(\mc{Q}^\prime_{m,A}+\mc{Q}^\prime_{m,B}\right)\rho\right]\right\},
    \label{eq: nM asf (main)}
\end{equation}
where the $\mc{Q}^\prime_{m,\bullet}$ are generalizations of quality factors, so-called \emph{quality maps} of associated sequence length $m$, and here $\rho$ is a noisy $\syst\env$ initial state depending on the prepared $\rho_\syst$ (which can generally get correlated with $\env$). Quality maps are, generically, \gls{cp} maps which can be further understood as averaged multi-linear maps (so-called process-tensors~\cite{PRXQuantum.2.030201}) taking the set of $m$ ideal digital gates $\{\mc{G}_i\}_{i=1}^m$ as input and being uniformly averaged over all of these. More generally than Markovian quality factors, they encode both \gls{spam} and fidelity-like information of the noise across a whole \gls{rb} process, reducing to $A$, $B$ and $p$ of Eq.~\eqref{eq: Markov ASF (main)} in the Markovian limit. In general, Eq.~\eqref{eq: nM asf (main)} is difficult to work with even in the time-stationary~\textsuperscript{\footnote{ Here with noise being time-stationary, albeit non-Markovian, we mean that noise associated to $\mc{I}_\env\otimes\mc{G}_i$, where $\mc{I}_\env$ is an identity channel on $\env$ and $\mc{G}_i$ is the ideal gate in $\syst$ at timestep $i$, is independent of the timestep $i$.}} noise regime, as opposed to the Markovian case, although it has been previously studied in~\cite{PRXQuantum.2.040351, figueroaromero2022general}.

To derive both the decays in Eq.~\eqref{eq: Markov ASF (main)} and Eq.~\eqref{eq: nM asf (main)}, and an explicit mathematical expression of all the quantities involved, we need a mathematical model of both Markovian and non-Markovian noise.

\subsection{Modeling of quantum noise}
Markovian noise can be generically modeled by defining noisy gates, now possibly non-unitary quantum channels, as $\tilde{\mc{G}}:=\Lambda\circ\mc{G}$, where $\mc{G}$ is the ideal digital gate, and $\Lambda$ is any \gls{cp} trace non-increasing map on $\syst$. Trace non-increasing maps account for processes such as leakage of information i.e., information loss, so we generally refer to this wider class of quantum maps as \emph{quantum channels} or \emph{noise channels}. In general, noise channels could depend on which specific gate is applied, $\tilde{\mc{G}}_i=\Lambda_{\mc{G}}\circ\mc{G}_i$, or at which time-step $i$ in a gate-sequence such gate is applied, $\tilde{\mc{G}}_i=\Lambda_{i}\circ\mc{G}$ (or, of course, depend on both, $\tilde{\mc{G}}_i=\Lambda_{\mc{G}_i}\circ\mc{G}$). To generalize this to non-Markovian (temporally correlated) noise we have to take $\env$ into account. That is, we now model the whole $\syst\env$ noisy map $\tilde{\mc{G}}:=\Lambda\circ(\mc{I}_\env\otimes\mc{G})$, with $\Lambda$ now a \gls{cp} map on $\syst\env$. Sequential applications $\tilde{\mc{G}}_j\circ\cdots\circ\tilde{\mc{G}}_k$ can give rise to temporal correlations among the corresponding $\Lambda$ noise maps, as formalized by the process tensor framework~\cite{PhysRevA.97.012127, PRXQuantum.2.030201}.

\begin{figure}[t]
    \centering
    \includegraphics[width=0.5\textwidth]{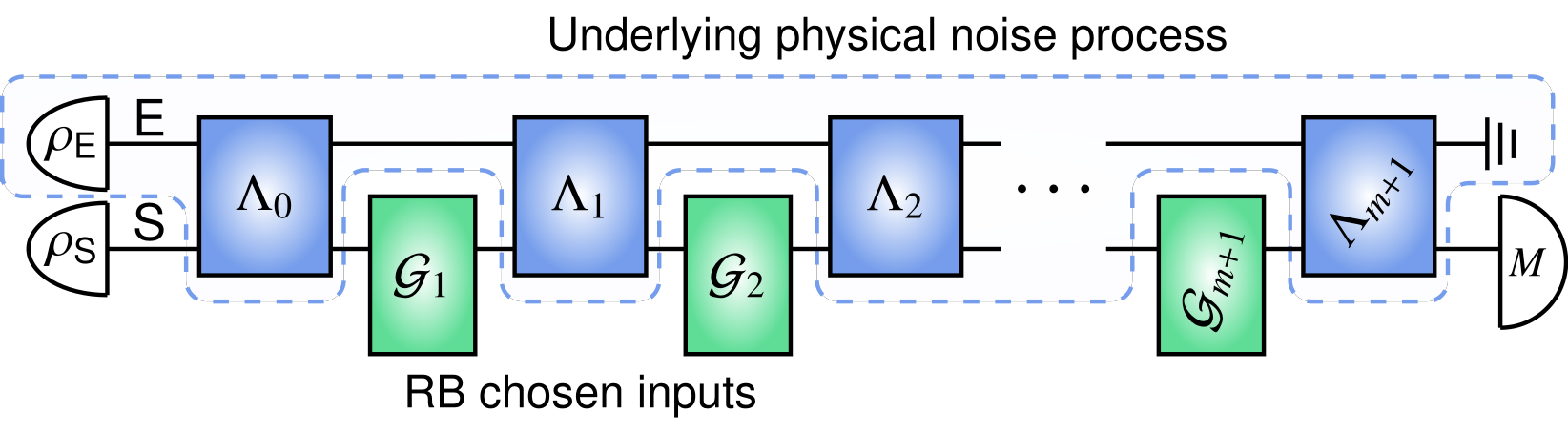}
    \caption{\textbf{Modeling \gls{rb} with non-Markovian noise}: Sample circuit of a \gls{rb} experiment on a quantum system $\syst$ with input state $\rho_\syst$, gate sequence $\mc{G}_1$,\ldots,$\mc{G}_m$, undo-gate $\mc{G}_{m+1}:=\mc{G}_1^{-1}\circ\cdots\circ\mc{G}_m^{-1}$, and measurement operator $M$; an underlying non-Markovian noise process can be represented by a fiducial initial state $\rho_\env$ on an environmental quantum system $\env$ together with a set of \gls{cp}, trace non-increasing maps $\Lambda_0$, \ldots, $\Lambda_{m+1}$ acting jointly on $\syst\env$.}
    \label{fig: nM RB}
\end{figure}

The average sequence fidelity in Eq.~\eqref{eq: nM asf (main)} corresponds simply to the explicit evaluation of the uniform average over Clifford gates, $\mathbf{E}_\mc{G}$, in
\begin{equation} \mc{F}_m:=\mathbf{E}_\mc{G}\!\left\{\tr\!\left[M\tr_\env\circ\Lambda_{m+1}\!\circ\!\mc{G}_{m+1}\!\circ\!\cdots\!\circ\!\Lambda_1\!\circ\!\mc{G}_1(\rho)\right]\right\},
\label{eq: asf noise model}
\end{equation}
where $\mc{G}_{m+1}:=\mc{G}_1^{-1}\circ\cdots\circ\mc{G}_m^{-1}$ and $\rho:=\Lambda_0(\rho_\env\otimes\rho_\syst)$ for given fiducial initial states of $\syst$ and $\env$, $\rho_\syst$ and $\rho_\env$, respectively. This is such that when $\Lambda_i=\mc{I}$, for all $i$, the average sequence fidelity equals $\tr\!\left(M\rho_\syst\right)$. Implicitly, together the gate-independent and time-stationary noise assumptions mean $\Lambda_{\mc{G}_i}=\Lambda_i=\Lambda$, for all gates $\mc{G}_i$ and time-steps $i$. A circuit representation of a \gls{rb} sequence sample with non-Markovian time-non-stationary noise can be seen in Fig.~\ref{fig: nM RB}; terms $\Lambda_0$ and $\Lambda_{m+1}$ are interpreted jointly as \gls{spam} noise terms.

As detailed in Appendix~\ref{appendix: RB Markov}, in the Markovian case, Eq.~\eqref{eq: asf noise model} leads to Eq.~\eqref{eq: Markov ASF (main)} with $p$ depending solely on the average gate-fidelity of $\Lambda$, as in Eq~\eqref{eq: Markov asf}, and $A$, $B$ depending solely on \gls{spam} errors, as in Eq.~\eqref{eq: Markov spam}. More generally, in~\cite{PRXQuantum.2.040351}, Eq.~\eqref{eq: asf noise model} was shown to be of the form of Eq.~\eqref{eq: nM asf (main)}, with the quality factors $\mc{Q}_{m,A}^\prime$ and $\mc{Q}_{m,B}^\prime$ given explicitly as in Eq.~\eqref{eq: appendix quality factors Clifford} of Appendix~\ref{appendix: RB nM to M}. This implies as well that, in the Markovian, time-stationary noise approximation regime, Eq.~\eqref{eq: nM asf (main)} reduces to the exponential decay of Eq.~\eqref{eq: Markov ASF (main)}.

We wish to be able to exploit the simplicity of the \gls{rb} protocol in spite of noise being non-Markovian. In the following, we show that precisely this reduction of the non-Markovian case to the Markovian one can be effectively achieved operationally by interleaving \gls{dd} sequences within the \gls{rb} protocol.

\section{Markovianization of RB with DD}
The way \gls{dd} works is by applying a sequence of pulses via a control Hamiltonian, $H_\ctrl$, in a way that effectively averages out undesired coupling terms in the total free-evolution Hamiltonian, $H$, which for finite-dimensional systems other than having a finite largest singular value, can be arbitrary~\cite{LidarCDD}. For infinite-dimensional $\env$, at least a frequency cut-off is required~\cite{LidarCDD}, but a broader classification for which Hamiltonians are amenable to \gls{dd} exists~\cite{Arenz_2017, DD_unbounded_2018}. Generally, \gls{dd} can effectively decouple, albeit partially, a wide class of $\syst\env$ Hamiltonians. Here we will consider \gls{udd}, which is universal in the sense that it averages out errors up to the first order in the Magnus expansion of the evolution~\cite{ezzell_2022}, independently of $H$. Concretely, we refer to a unitary group $\mbb{V}$ on $\syst$ as a universally decoupling group whenever $\sum_{v\in\mbb{V}}vXv^\dg=O_\env\otimes\mbb1_\syst$, for any $\syst\env$ operator $X$ and some $\env$ operator $O_\env$ and $\mbb1_\syst$ the identity on $\syst$.

In the setting of \gls{rb}, we can model free evolution at any given time-step as the underlying noise $\Lambda$ on the whole $\syst\env$, so \emph{Markovianizing} Eq.~\eqref{eq: nM asf (main)} can effectively be accomplished by applying \gls{udd} between \gls{rb} gates. We notice that doing so, implies that we will be benchmarking ``\gls{dd}-dressed'' Clifford gates, which has been shown to improve the benchmarked fidelities in the Markovian setting~\cite{PhysRevA.92.062332}. As detailed in Appendix~\ref{appendix: rbdd ideal}, we consider ideal pulses generated by $H_\ctrl(t)=\f{\pi}{2}\sum_k\delta(t\!-\!t_k)v_k$ at times $t_1<t_2<\ldots<t_\eta$, where $\delta$ is a Dirac delta and $\eta=|\mbb{V}|$ is the number of elements in the decoupling group, i.e., infinitely strong instantaneous pulses with decoupling operators $v_k$ at times $t_k$.

\begin{figure}[t]
    \centering
    \includegraphics[width=0.5\textwidth]{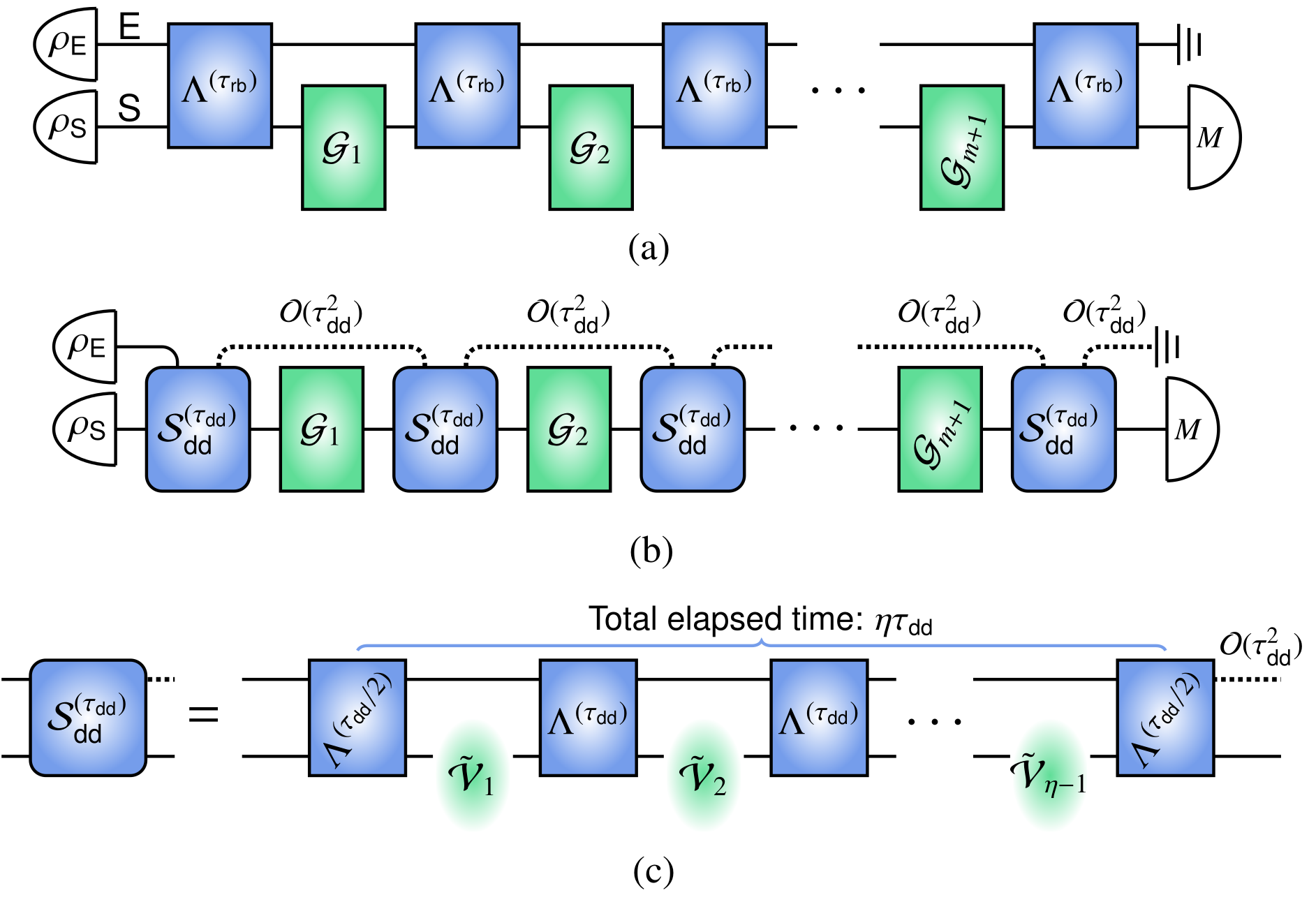}
    \caption{\textbf{Interleaved \gls{udd} on \gls{rb}}: Circuit depictions of (a) standard \gls{rb} with initial state $\rho_\syst$, measurement $M$, and time-intervals $\tau_\mathsf{rb}$ with noise $\Lambda^{(\tau_\mathsf{rb})}$ between application of gates $\mc{G}_i$, (b) the same \gls{rb}-sequence, now with a single interleaved \gls{udd} sequence $\mc{S}_\mathsf{dd}^{(\tau_\mathsf{dd})}$, leaving environment correlations only at $\mc{O}(\tau_\mathsf{dd}^2)$, and (c) the sequence $\mc{S}_\mathsf{dd}^{(\tau_\mathsf{dd})}$ with non-identity pulses $\tilde{\mc{V}}_i\in\mbb{V}$ applied in time-intervals of $\tau_\mathsf{dd}$, and total elapsed time $\eta\tau_\mathsf{dd}$, where $\eta=|\mbb{V}|$.}
    \label{Fig: DD RB}
\end{figure}

Let us label the noise maps by an elapsed-time $t$ between subsequent applications of any two gates as $\Lambda^{(t)}$. Denoting the channels associated to the ideal decoupling operators as $\mc{V}(\cdot):=v(\cdot)v^\dg$, these pulses can be applied at evenly-spaced time-intervals $\tau_\mathsf{dd}$ through $\mc{S}_\mathsf{dd}^{(\tau_\mathsf{dd})}:=\Mcirc_{\mc{V}\in\mbb{V}}\left(\mc{V}\circ\Lambda^{(\tau_\mathsf{dd})}\circ\mc{V}^\dg\right)$. Notice that we are taking the application of \gls{dd} pulses to be strictly evenly spaced. As we will interleave these among the random \gls{rb} gates, we will necessarily have free-evolution (noise) on the edge terms between the application of gates; taking either the first or last pulse as the identity, we can equivalently write
\begin{equation}
\mc{S}^{(\tau_\mathsf{dd})}_\mathsf{dd}=\Lambda^{(\tau_\mathsf{dd}/2)}\Mcirc_{\tilde{\mc{V}}}\left(\tilde{\mc{V}}\circ\Lambda^{(\tau_\mathsf{dd})}\circ\tilde{\mc{V}}^{\dg}\right)\circ\Lambda^{(\tau_\mathsf{dd}/2)},
\label{eq: dd sequence (main)}
\end{equation}
with $\tilde{\mc{V}}$ being non-identity elements of $\mbb{V}$.

For example, for a single-qubit, $\mbb{V}=\{\mbb1,X,Y,Z\}$, the Pauli group, so that $\mc{S}^{(\tau_\mathsf{dd})}_\mathsf{dd}=\mc{Z}\circ\Lambda^{(\tau_\mathsf{dd})}\circ\mc{Z}\circ\mc{Y}\circ\Lambda^{(\tau_\mathsf{dd})}\circ\mc{Y}\circ\mc{X}\circ\Lambda^{(\tau_\mathsf{dd})}\circ\mc{X}\circ\Lambda^{(\tau_\mathsf{dd})}$, where curly letters here are the maps associated to each Pauli operator, and which can be seen to be equivalent to the so-called $XY4$ and $XZ4$ sequences.

If we label by $\tau_\mathsf{rb}$ the original elapsed time between application of \gls{rb} Clifford gates, and if $\tau_\mathsf{rb}$ further turns out to be the \emph{minimum} time between application of two subsequent gates on the given device, then we would generally need to consider no less than $\tau_\mathsf{dd}=2\tau_\mathsf{rb}$, due to the edge terms in Eq.~\eqref{eq: dd sequence (main)}. Other considerations could come into play here in order to optimize $\tau_\mathsf{dd}$, e.g., the fact that Clifford gates are composite or in general whether \gls{dd} pulses, in particular, can be applied with $\tau_\mathsf{dd}<\tau_\mathsf{rb}$.

We can model the application of ideal \gls{udd} within a \gls{rb} protocol by interleaving a single sequence $\mc{S}^{(\tau_\mathsf{dd})}_\mathsf{dd}$ among each ideal gate of a \gls{rb} circuit, i.e., $\mc{S}^{(\tau_\mathsf{dd})}_\mathsf{dd}\circ\mc{G}_{m+1}\circ\cdots\circ\mc{S}^{(\tau_\mathsf{dd})}_\mathsf{dd}\circ\mc{G}_1\circ\mc{S}^{(\tau_\mathsf{dd})}_\mathsf{dd}$, as depicted in Fig.~\ref{Fig: DD RB}. Notice this will change the circuit depth, and in particular it will modify the total time between application of the \gls{rb} gates to $\eta\,\tau_\mathsf{dd}$; nevertheless, we obtain the following:

\begin{result}\label{result: markovianized rb}
Let $\Lambda^{(t)}=\mathrm{e}^{t\mc{L}}$, where $\mc{L}(\cdot):=-i[H,\cdot]+\mc{D}(\cdot)$ for a $\syst\env$ time-independent Hamiltonian $H$ and a dissipator $\mc{D}=\mc{D}_\syst+\mc{D}_\env$ with only local $\syst$ and/or $\env$ contributions, and let $\gamma$ be the diagonal matrix of relaxation rates of $\mc{D}_\syst:=\sum_k\gamma_k[L_k(\cdot)L_k^\dg-\f{1}{2}\{L_k^\dg{L}_k,\cdot\}]$, for some (unit-less) traceless and orthonormal operators $L_k$. Then for $\tau_\mathsf{dd}\ll1/\tr(\gamma)$, the average sequence fidelity of length $m$ for a \gls{rb} experiment under time-stationary noise with single interleaved \gls{udd} sequences of the form of Eq.~\eqref{eq: dd sequence (main)} satisfies,
\begin{equation}
    \mc{F}_m = A\,p_{\tau_\mathsf{dd}}^m + B + \mc{O}(\tau_\mathsf{dd}^2),
    \label{eq: dd asf analytical result}
\end{equation}
where $A,B$ are \gls{spam} constants for fixed $\tau_\mathsf{dd}$, and the quality factor is a $\mc{O}(\tau_\mathsf{dd})$ term
\begin{equation} p_{\tau_\mathsf{dd}} = 1 - \eta\,\tau_\mathsf{dd}\, \f{\tr(\gamma)}{\dimS-\f{1}{\dimS}},
\end{equation}
where $\eta=|\mbb{V}|$ is the number of elements of the decoupling group $\mbb{V}$.
\end{result}

The proof is shown in Appendix~\ref{appendix: rbdd ideal} and the value of all constants is shown explicitly in Eq.~(\ref{eq: appendix p_tau}-\ref{eq: appendix B}).

The main message in Result~\ref{result: markovianized rb} is that ideal-pulse \gls{udd} Markovianizes \gls{rb} for a broad class of non-Markovian noise models, in the sense of reducing the average sequence fidelity of Eq.~\eqref{eq: nM asf (main)} to that of Eq.~\eqref{eq: Markov ASF (main)}, to the first order in the \gls{dd} sequence interval time $\tau_\mathsf{dd}$. Notice the limiting case of $\tau_\mathsf{dd}\to0$ would be that when both the ideal \gls{dd} pulses \emph{and} the random gates in the \gls{rb} sequence are implemented infinitely fast, so that, of course, $\mc{F}_m=1$, as no added idling time is being considered.

The continuous dynamical model for the noise as a $\syst\env$ Lindblad evolution encompasses a broad class of noise models, where the $\syst\env$ dynamics are Markovian while the reduced ones on $\syst$ are non-Markovian. The only restriction is that the dissipator term, $\mc{D}$, cannot have global $\syst\env$ contributions to the first $\tau_\mathsf{dd}$ order for our result to hold, which would be the case if correlated $\syst\env$ information is dissipated. While in Section~\ref{Sec: numerical dd} we take as an example a two-qubit system, motivated by more realistic two-level defect noise models relevant in superconducting systems, our results hold for a wide class of noise models, in particular for the infinite-dimensional environment case whenever it can be modeled by a finite effective environment dimension~\cite{PhysRevLett.122.160401}. A limiting factor to our results, however, are models having parameter domains where the dynamics cannot be dynamically decoupled~\cite{nM_cannot_decouple_2021}, i.e., for which the given dynamical decoupling scheme does not converge to the average of the dynamical evolution.

In the absence of dissipation terms, $\mc{F}_m=1+\mc{O}(\tau_\mathsf{dd}^2)$, i.e., the average sequence fidelity is identically one, albeit only to the first $\tau_\mathsf{dd}$ order. Notice that the modeling of noise in Result~\ref{result: markovianized rb} explicitly takes into account the presence or absence of dissipation terms between applications of \gls{rb} gates and \gls{dd} pulses: e.g., Result~\ref{result: markovianized rb} will hold with a dilated Hamiltonian $H^\prime$ on a larger system $\syst\env\env^\prime$, only if the absence of global dissipation on $\syst\env$ is also implied.

The quality factor $p_{\tau_\mathsf{dd}}$ is derived in Appendix~\ref{appendix: rbdd ideal} and it is related to the average gate-fidelity of the channel $\mc{I}+\tau_\mathsf{dd}\,\mc{D}_\syst$ with respect to the identity $\mc{I}$; i.e., it quantifies the noise contribution from the dissipator $\mc{D}_\syst$, which is the sole generator of Markovian noise, once the Hamiltonian has been averaged out in $\syst$, and it is a trace-zero map, not \gls{cp} in general. Notice that this quality factor corresponds to the first order in $\tau_\mathsf{dd}$, Markovian part, of the \gls{dd}-dressed gates and not of the original bare gates. Notice too that $p_{\tau_\mathsf{dd}}$ is the leading term of the decay of $\mc{F}_m$ for small times $\tau_\mathsf{dd}$, where it takes a value close to one. The final expression in terms of the relaxation rates, $\gamma_k$, can then be seen to follow.

Higher-order, $\mc{O}\left(\tau_\mathsf{dd}^2\right)$ terms, albeit still also containing Markovian contributions, will, in general, be non-Markovian; so while non-exponential deviations may get suppressed and allow to fit an exponential, the resulting decay is purely Markovian only to first order. Thus we can interpret the efficacy of digital \gls{udd} sequences, in both Markovianization of the average sequence fidelity and improvement in overall average gate-fidelity, mainly in terms of how fast they can be applied, relative to the time between application of the \gls{rb} gates, $\tau_\mathsf{rb}$. Given that multi-qubit Cliffords are composite gates, it might be possible to implement \gls{dd} pulses in time-scales $\tau_\mathsf{dd}<\tau_\mathsf{rb}$. As we see in Section~\ref{Sec: numerical dd}, in practice this might be the main limiting factor, together with considerations such as non-ideal, finite-width pulses, as well as imperfect applications of these.

Finally, realistic \gls{dd} cannot be implemented as ideal instantaneous pulses, and these themselves can introduce control errors~\cite{qi2022efficacy}. Furthermore, it is known that finite-width \gls{dd} does not achieve perfect decoupling even to first time-order~\cite{ezzell_2022}. However, as long as pulses are sufficiently narrow, the noise they introduce is sufficiently small~\cite{qi2022efficacy} and local, \gls{udd} will Markovianize \gls{rb} to an extent close to the one predicted by Result~\ref{result: markovianized rb}. One can furthermore employ more elaborated \gls{dd} techniques such as Concatenated \gls{dd}~\cite{LidarCDD}, or devised optimized decoupling sequences, although we do not pursue that here.

\section{RB under Pauli-twirled noise}\label{sec: rb and rc}
While the Markovianizing effect of \gls{dd} in \gls{rb} is somehow expected, a prominent error-suppression technique that has recently been shown to suppress non-Markovian noise in a statistically significant way~\cite{winick2022RC, Hashim_2023} is \glsreset{rc}\gls{rc}~\cite{RC_2016, RC_superc_2021}. \gls{rc} can be understood as the operational way of tailoring arbitrary Markovian noise quantum channels into Pauli channels, which mathematically corresponds to a mapping known as Pauli-twirling. As opposed to \gls{dd}, \gls{rc} is not a quantum control technique, but instead it relies on compiling a set of logically-equivalent circuits (i.e., with no increase in depth and effectively implementing the same quantum operations) where noisy gates are dressed with uniformly sampled random single-qubit Pauli gates; averaging over all such circuits, approximately and efficiently implements a Pauli-twirl.

Similar to \gls{dd}, where we worked with the ideal pulse limit, here we focus on the case where \gls{rc} fully and perfectly tailors $\Lambda$ into $\Lambda^\mathsf{P}$, i.e., Pauli-twirling. While for a single-qubit this can be done exactly, in general, the tailoring of noise into Pauli noise by \gls{rc} can be quantified to occur to a large percentage with a small number of samples~\cite{RC_2016}, generally scaling as an inverse fraction of the number of samples~\cite{Hashim_2023}. The only caveats to keep in mind is that this sampling overhead of \gls{rc} would compound with that of \gls{rb}~\footnote{In the case of of \gls{rc} and \gls{rb}, the number of samples compound as their product, and both techniques have at most a linear sampling complexity in system size; this contrasts with the case of \gls{rc} and GST employed in~\cite{Hashim_2023}, where they reported requiring 40 hours to collect data for 100 \gls{rc} randomizations to characterize a 2-qubit gate in a superconducting system.} and that the Pauli gates could themselves be significantly noisy.

\begin{figure}[t]
    \centering
    \includegraphics[width=0.5\textwidth]{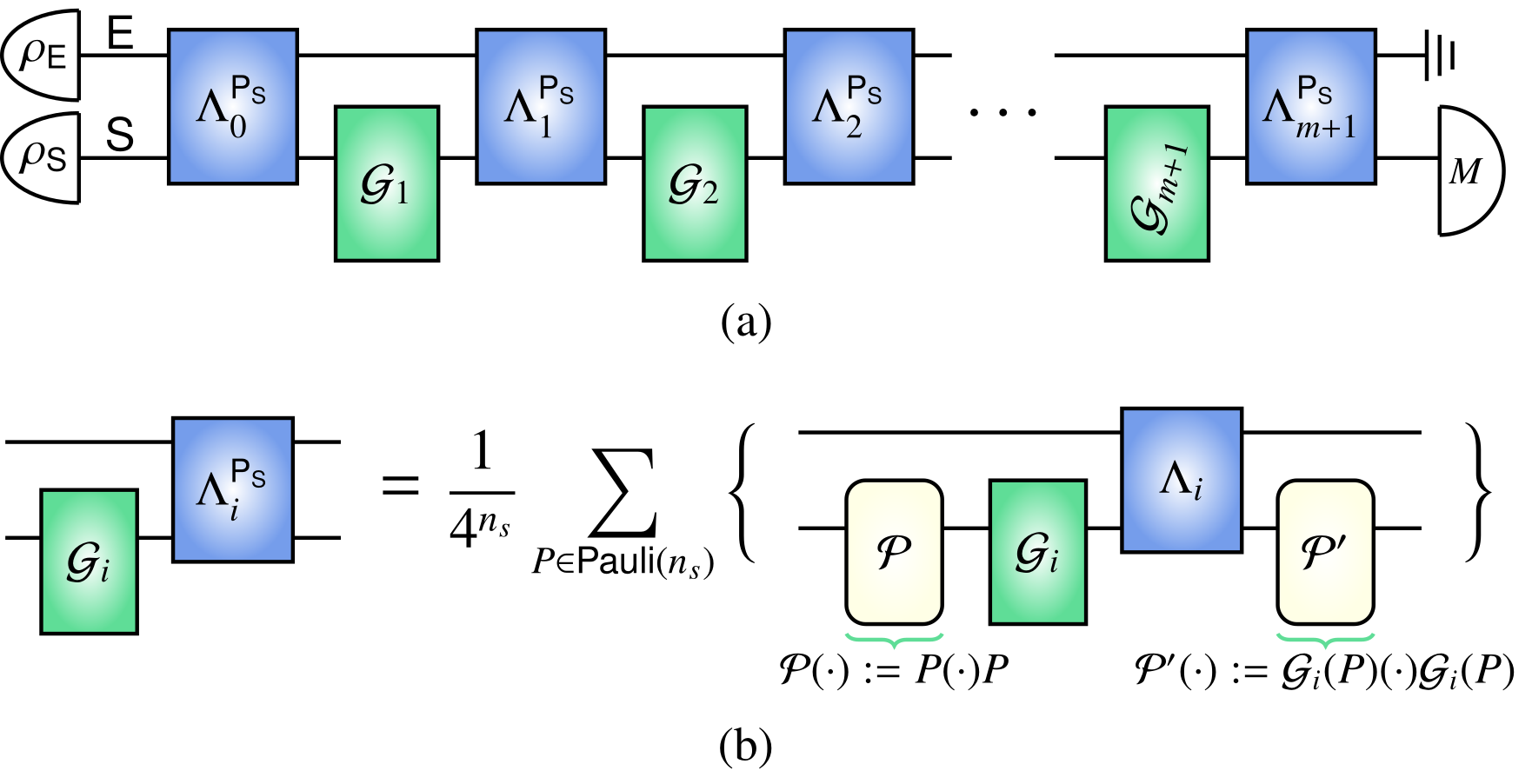}
    \caption{\textbf{RB under non-Markovian $\syst$-Pauli-twirled noise}: In (a), a sample circuit for a standard \gls{rb} experiment where noise $\Lambda_i^{\mathsf{P}_\syst}$ is time-non-stationary, non-Markovian, and has been Pauli-twirled on subsystem $\syst$, while in (b) the operational definition for $\Lambda_i^{\mathsf{P}_\syst}$ is shown explicitly as an average over ideal $n_s$-qubit Pauli terms $P$; this is known as a $\mc{G}_i$-twisted twirl~\cite{flammia_aces} on $\syst$, and \gls{rc} accurately approximates it efficiently by randomly sampling single-qubit Pauli terms and compiling the outputs~\cite{RC_2016}.}
    \label{fig: PT RB and RC}
\end{figure}

Given any $n_s$-qubit quantum channel $\Phi$, a Pauli-twirl maps its $\chi$-matrix representation, $\Phi(\cdot):=\sum_{ik}\chi_{ik}P_i\,(\cdot)P_k$, to $\Phi^{\mathsf{P}}(\cdot)=\sum_i\chi_{ii}P_i\,(\cdot)P_i$, where here $P$ are $n_s$-qubit Pauli operators. The channel $\Phi^{\mathsf{P}}$ is generally a non-unitary (stochastic or incoherent) quantum channel known as a Pauli channel, and the $\alpha_i:=\chi_{ii}$ define a probability distribution called Pauli error rates. It can be seen that average gate-fidelity (and thus \gls{rb}) estimates precisely a $\chi_{00}=\alpha_0$ term, corresponding to the identity term, $P_0:=\mbb1$ (probability of no error happening). Crucially, Pauli-twirling suppresses all off-diagonal terms of the $\chi$ matrix, which leaves the average gate-fidelity unchanged, but nevertheless removes terms associated with coherent errors, which can give worst-case error rates orders of magnitude higher than stochastic errors~\cite{Sanders_2015}.

For the non-Markovian case, a Pauli-twirl solely on system $\syst$ on any given $\syst\env$ quantum channel $\Lambda$, in the $\chi$-matrix representation, takes the form
\begin{equation}
    \Lambda^{\mathsf{P}_\syst}(\cdot) := \sum\chi_{\mu\nu,ii}(R_\mu\otimes{P}_i)(\cdot)(R_\nu^{\dg}\otimes{P}_i),
\end{equation}
for some basis operators $R$ on $\env$, and coefficients $\chi_{\mu\nu,ii}$ of the corresponding $\syst\env$ Hermitian $\chi$-matrix. If we were to trace out $\env$ on this channel, then, of course, the reduced channel is a Pauli channel, $\tr_\env\Lambda^{{\mathsf{P}_\syst}}\left(\rho_\env\otimes\cdot\right)=\sum\alpha_i(\rho_\env^\prime)P_i(\cdot)P_i$, with $\alpha_i$ being Pauli error-rates dependent on the $\env$-reduced output state $\rho_\env^\prime$. This tracing, however, generally occurs at the end of a computation or any multi-time process.

Operationally, however, the mapping $\Lambda\mapsto\Lambda^{\mathsf{P}_\syst}$ can only be implemented by modifying the noisy gates $\tilde{\mc{G}}:=\Lambda\circ(\mc{I}_\env\otimes\mc{G})$, as we don't have direct access to $\Lambda$. This can be done through a so-called $\mc{G}$-twisted twirl~\cite{flammia_aces} on system $\syst$, defined as
\begin{equation}
    \tilde{\mc{G}}(\cdot)\mapsto4^{-n_s}\sum_{P}\mc{G}(P)\,\tilde{\mc{G}}\!\left(P\,(\cdot)\,P\right)\mc{G}(P),
\end{equation}
with sum over $P$ all $n_s$-qubit ideal Pauli operators. This is depicted schematically in Fig.~\ref{fig: PT RB and RC}. \gls{rc} thus approximates $\syst$-Pauli-twirls by randomly sampling $\mc{G}_i$-twisted twirls on all time steps and recompiling the Pauli gates, which can be done with a number of samples much smaller than $4^{n_s}$ for larger $n_s$~\cite{RC_superc_2021}. As mentioned above, here we will correspondingly focus on the limit of perfect $\syst$-Pauli-twirling.

We now point out the following:

\begin{result}\label{result: rc and rb}
    For any \gls{rb} sequence fidelity given as $\mathrm{f}_m[\{\Lambda_i\}_{i=0}^{m+1}] := \tr[{M}\tr_\env\Mcirc_{i=1}^{m+1}(\Lambda_i\circ\mc{G}_i)\,\rho]$ of length $m$, where here $\rho := \Lambda_0(\rho_\env\otimes\rho_\syst)$, the corresponding average sequence fidelity with $\syst$-Pauli-twirled noise, $\mathbf{E}_\mc{G}\mathrm{f}_m[\{\Lambda_i^{\mathsf{P}_\syst}\}_{i=0}^{m+1}]$, remains in general non-exponential.
    
    Furthermore, when averaging is over uniformly distributed $n_s$-qubit Clifford gates,
    \begin{equation}
    \mathbf{E}_\mc{G}\mathrm{f}_m\left[\{\Lambda_i\}_{i=0}^{m+1}\right] \!=\! \mathbf{E}_\mc{G}\mathrm{f}_m\left[\{\Lambda_0,\Lambda_{m+1}\}\!\cup\!\{\Lambda_i^{\mathsf{P}_\syst}\}_{i=1}^{m}\right],
    \label{eq: pauli asf (main)}
    \end{equation}
    where here $\cup$ denotes the union of sets. On the other hand,
    \begin{equation}
\mathbf{V}_\mc{G}\mathrm{f}_m\left[\{\Lambda_i^{\mathsf{P}_\syst}\}_{i=0}^{m+1}\right] \leq \mathbf{V}_\mc{G}\mathrm{f}_m\left[\{\Lambda_i\}_{i=0}^{m+1}\right],
\label{eq: variance pauli (main)}
    \end{equation}
    where $\mathbf{V}_\mc{G}$ is variance over arbitrary gates $\mc{G}$.
\end{result}

The proof can be seen in Appendix~\ref{appendix: rc and rb}.

The first statement means that Pauli-twirling does not Markovianize the average sequence fidelity as \gls{dd} does, i.e., it does not turn non-Markovian non-exponential \gls{rb} decays into exponential ones. In other words, in general the average sequence fidelity remains of the form of Eq.~\eqref{eq: nM asf (main)} under Pauli-twirling, as shown in Appendix~\ref{appendix: asf Pauli-twirled}. In particular, this statement also holds for any gate set other than the multi-qubit Clifford group, at least as long as it forms a finite group. On the other hand, the second statement in Eq.~\eqref{eq: pauli asf (main)} holds only for the Clifford group but it implies that \gls{rc}, or any Pauli-twirling technique, would at most have an effect (not necessarily Markovianizing or increasing the average sequence fidelity) on \gls{spam} contributions when it comes to non-Markovian \gls{rb} data.

The reason behind this result, and for the apparent contradiction with those of~\cite{winick2022RC, Hashim_2023}, is that average gate-fidelity as a figure of merit only takes into account the probability of no error happening on $\syst$ and not all the other error terms that Pauli-twirling eliminates. That is, average gate-fidelity by definition is proportional only to the zero\textsuperscript{th} (identity) element of the $\chi$-matrix; explicitly, see Eq.~(\ref{eq: nM asf time-dep form (appendix)},\ref{eq: nM p time-dep form (appendix)}) in Appendix~\ref{appendix: asf Pauli-twirled}. On the other hand, as pointed out in~\cite{winick2022RC}, non-linear metrics in the noise (whose complexity to estimate typically scale exponentially in system size) can decrease significantly under Pauli-twirling. This being said it is still possible that there is a class of noise profiles where it is sufficient to $\syst$-Pauli-twirl the \gls{spam} noise in order to Markovianize the average sequence fidelity.

Despite not Markovianizing the average sequence fidelity, Eq.~\eqref{eq: variance pauli (main)} establishes that Pauli-twirling will, in general, reduce the variance of sequence fidelities~\textsuperscript{\footnote{ The statement in Eq.~\eqref{eq: variance pauli (main)} holds more generally for $\mathrm{f}_m$ being noisy expectation values.}}. The importance of this is twofold, if non-exponential deviations are present, it allows for confidence as to whether these are due to non-Markovianity or e.g., sample uncertainty, and if the decay is exponential, it allows for an accurate estimation of meaningful error rates. While mathematically this result can be seen to follow simply because Pauli-twirling removes additive terms to the variance, it can be argued that, physically, the reason is that it reduces the coherence of noise, as precisely the expectation of the squared sequence fidelity is proportional to the so-called average unitarity. The unitarity, $u$, is a figure of merit quantifying loss of purity due to noise~\cite{Wallman_2015}, and satisfies $u=1$ if noise is coherent, i.e., due to a unitary map, and $u\leq1$ otherwise. In the Markov case, $u\geq{p^2}$~\cite{PhysRevA.99.012315}, which is saturated for depolarizing noise, i.e., Pauli channels with all $\chi_{ii}$ terms, for $i\neq0$, being equal. In the non-Markov case, it is \emph{expected} as well that $\syst$-Pauli-twirling would only decrease or leave the total unitarity unchanged~\textsuperscript{\footnote{ In the $\syst\env$ case, it is unclear whether $\syst$-Pauli-twirling \emph{indeed will always} either only decrease or leave the total unitarity unchanged, due precisely to $\env$; here the role of the average sequence fidelity would also come into play to always decrease the variance.}}. This, in a sense, was already pointed out in the original unitarity benchmarking proposal of~\cite{Wallman_2015}, although as also mentioned there, it is less straightforward to give a concrete bound.

Finally, similar to the case of \gls{dd}, here we have considered an exact implementation of Pauli-twirling, i.e., a perfect application of \gls{rc}. While this is unrealistic, in general, \gls{rc} incurs in only a small Pauli sampling overhead to closely approximate an exact Pauli-twirl~\cite{RC_superc_2021}. We now give two concrete numerical examples of our results for both \gls{dd} and Pauli-twirling within non-Markovian \gls{rb}.

\begin{figure*}[t!]
    \centering    \includegraphics[width=1\textwidth]{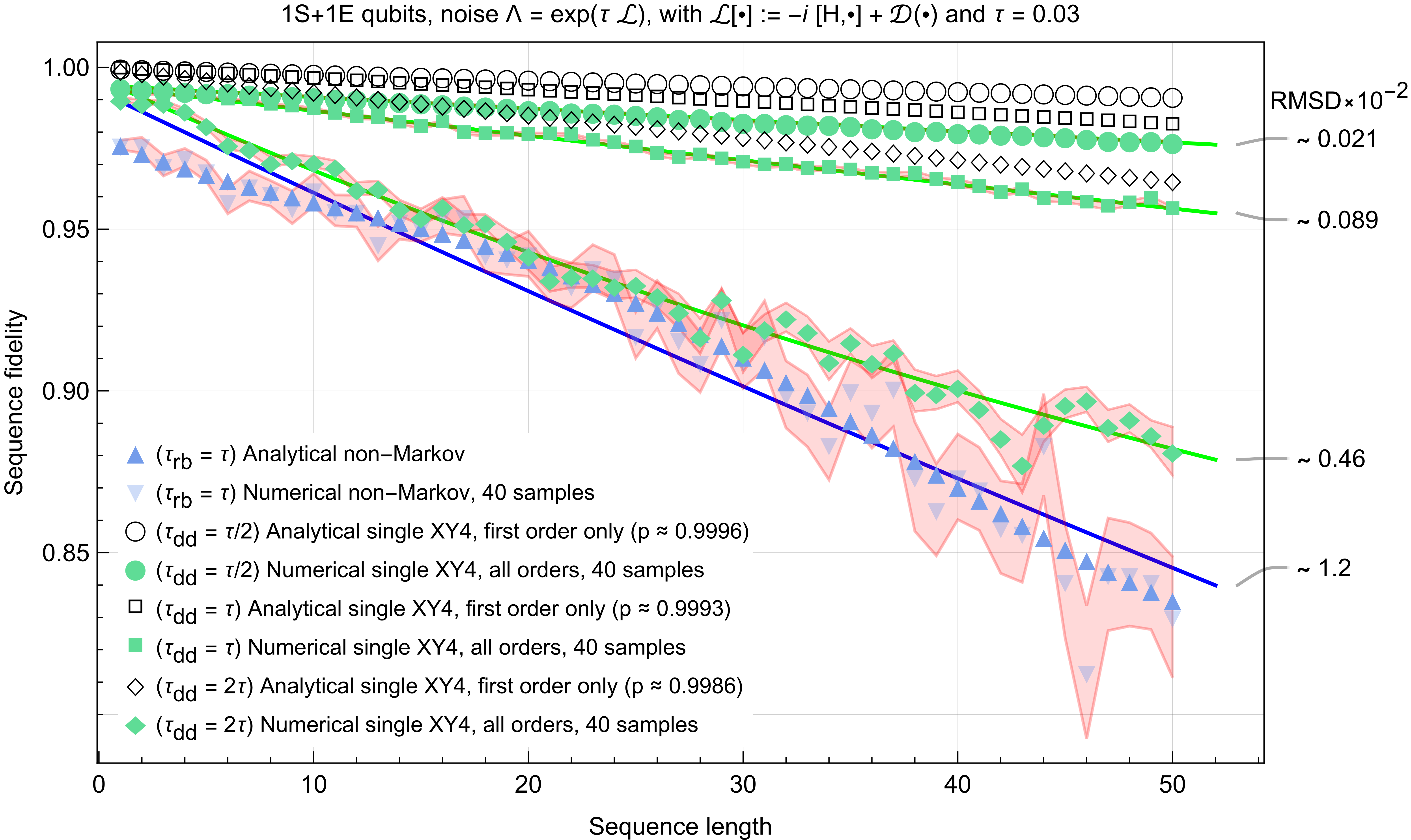}
    \caption{\textbf{Interleaved DD in non-Markovian \gls{rb}}: Average sequence fidelities ($\mc{F}_m$) for \gls{rb} experiments on increasing sequence lengths ($m$) with the full, generally non-Markovian noise model $\Lambda^{(\tau_\mathsf{rb})}$ (blue triangles), and interleaved ideal $XY4$ sequences (green closed markers and black open markers) with time-intervals between pulses of $\tau_\mathsf{dd}=\tau/2$, $\tau$ and $2\tau$, for a fixed $\tau=\tau_\mathsf{rb}=0.03$. All numerical averages were taken over 40 samples, with red bands denoting uncertainty of mean; non-Markovian analytical decay (blue upper triangle) computed according to~\cite{PRXQuantum.2.040351} and $\mc{O}(\tau_\mathsf{dd})$ analytical decays (black open markers) up to first time-order obtained according to Eq.~\eqref{eq: dd asf analytical result}. On the right, the Root-Mean-Squared Deviations (RMSD) $\times10^{-2}$ --defined as the square root of the mean of the squares of the deviations-- of the numerical averages, with respect to fitted exponentials of the form of Eq.~\eqref{eq: Markov ASF (main)}, are shown. The 1$\syst$+1$\env$ qubit model employed is given by $\Lambda^{(t)}=\ex^{t\mc{L}}$ where $\mc{L}(\cdot):=-i[H,\cdot]+\mc{D}_\syst(\cdot)$ for $H=J XX + h_x (XI + IX) + h_y (YI + IY)$ and local $\syst$ dissipator $\mc{D}_\syst(\cdot):=\sum_k\gamma_k[L_k(\cdot)L_k^\dg-\f{1}{2}\{L_k^\dg{L}_k,\cdot\}]$; chosen parameters $\rho_\syst=M=|0\rangle\!\langle0|$, $J=1.7$, $h_x=1.47$, $h_y=-1.05$ and Lindblad terms $L_0=X$, $\gamma_0=0.002$, $L_1=Z$, $\gamma_1=0.007$; Lindblad evolution approximated up to $\mc{O}(t^{10})$.}
    \label{fig: rb dd numerical}
\end{figure*}

\section{Numerical Examples}
\subsection{XY4 to Markovianize a qubit}\label{Sec: numerical dd}
As proof-of-principle, we consider both one $\syst$ qubit and one $\env$ qubit with time-stationary noise $\Lambda^{(t)}=\ex^{t\mc{L}}$ acting jointly on both qubits, where $\mc{L}(\cdot):=-i[H,\cdot]+\mc{D}_\syst(\cdot)$ for a Hamiltonian $H=J XX + h_x (XI + IX) + h_y (YI + IY)$ with some constants $J,h_x,h_y$ and local $\syst$-dissipator $\mc{D}_\syst(\cdot):=\sum_k\gamma_k[L_k(\cdot)L_k^\dg-\f{1}{2}\{L_k^\dg{L}_k,\cdot\}]$ with $L_0=X$ and $L_1=Z$. While this is an arbitrary dynamical model, similar qubit-to-qubit noise mechanisms are conceivable to arise e.g., with two-level system defects in superconducting qubits~\cite{PRXQuantum.3.040332}, albeit with distinct spin interactions, dissipators, and varying strengths.

We compare interleaved \gls{dd} pulses given by the $XY4$ sequence, briefly described below Eq.~\eqref{eq: dd sequence (main)}  and depicted in Fig.~\ref{Fig: DD RB}, for \gls{dd} sequences of $\tau_\mathsf{dd}=\tau/2$, $\tau$, and $2\tau$ time-intervals, where we fixed $\tau:=\tau_\mathsf{rb}$ to be the evolution time of the noise between applications of random Clifford gates in the original non-Markovian \gls{rb} experiment. Outputs for particular constants in the noise model $\Lambda^{(t)}$ can be seen in Fig.~\ref{fig: rb dd numerical}, where the non-Markovian analytical average sequence fidelity was computed according to~\cite{PRXQuantum.2.040351}, and the analytical decays to first $\tau_\mathsf{dd}$ orders were computed according to Eq.~\eqref{eq: dd asf analytical result} where $\eta=4$.

The main message is that \gls{dd} effectively removes non-Markovian non-exponential deviations for a time scale at which \gls{dd} sequences are applied $\tau_\mathsf{dd}\ll1/\tr(\gamma)$, and how well it does so --whether it outputs decays close to the purely exponential $\mc{O}(\tau_\mathsf{dd})$ decay--, depends mainly on whether the $\tau_\mathsf{dd}$ is small relative to that of the noise, $\tau_\mathsf{rb}$, in the original non-Markovian \gls{rb} experiment.

In particular, interleaving \gls{dd} pulses in time-intervals of $\tau_\mathsf{dd}=2\tau_\mathsf{rb}$ does reduce non-exponential deviations, but still, non-Markovian noise dominates the decay. In general, this can be assessed by inspecting the separation of the numerical, all-$\tau_\mathsf{dd}$-order average decays with respect to the corresponding only first-$\tau_\mathsf{dd}$-order analytical decays; further discussion and an analysis in the time-scales for $\tau_\mathsf{dd}$ relative to the chosen $\tau_\mathsf{rb}$ can be seen in Appendix~\ref{appendix: rbdd decoupling times}, as captured by Fig.~\ref{fig: tau_dd}, where it is clear that this separation is suppressed and decays asymptotically become purely Markovian for $\tau_\mathsf{dd}<\tau_\mathsf{rb}$.

Taking into account that pulses will have a finite width and might themselves be noisy due to implementation errors, the ability to perform good pulses in time scales shorter than that of the \gls{rb} gates, would lead to a useful reduction of non-exponential deviations and enhancement of average sequence fidelities, as shown by the plots for time-intervals $\tau_\mathsf{dd}=\tau_\mathsf{rb}$ and $\tau_\mathsf{dd}=\tau_\mathsf{rb}/2$. Comparison with the corresponding analytical $\tau_\mathsf{dd}$ decays, up to the first time-order, can help to estimate the contribution of non-Markovian and higher-order Markovian terms in the model.

\subsection{Subsystem Pauli-twirled noise}
We now consider the same 1$\syst$+1$\env$ qubit model but where instead of interleaving $XY4$ pulses, we consider noise that has been Pauli-twirled on qubit $\syst$. We now fix a value for $\tau$ and simply denote $\Lambda$ as the noise channel for all time steps. We distinguish between \gls{spam} noise terms $\Lambda_0=\Lambda_{m+1}$ and the \emph{bulk} noise terms $\Lambda_1=\Lambda_2=\ldots=\Lambda_m$ (i.e., all non-\gls{spam} noise). Since we deal with a single-qubit, and \gls{rc} does not change the logical structure of quantum circuits (e.g., does not increase their depth), we take the limit of perfect Pauli-twirled noise on $\syst$.

\begin{figure}[t!]
    \centering
    \includegraphics[width=0.52\textwidth]{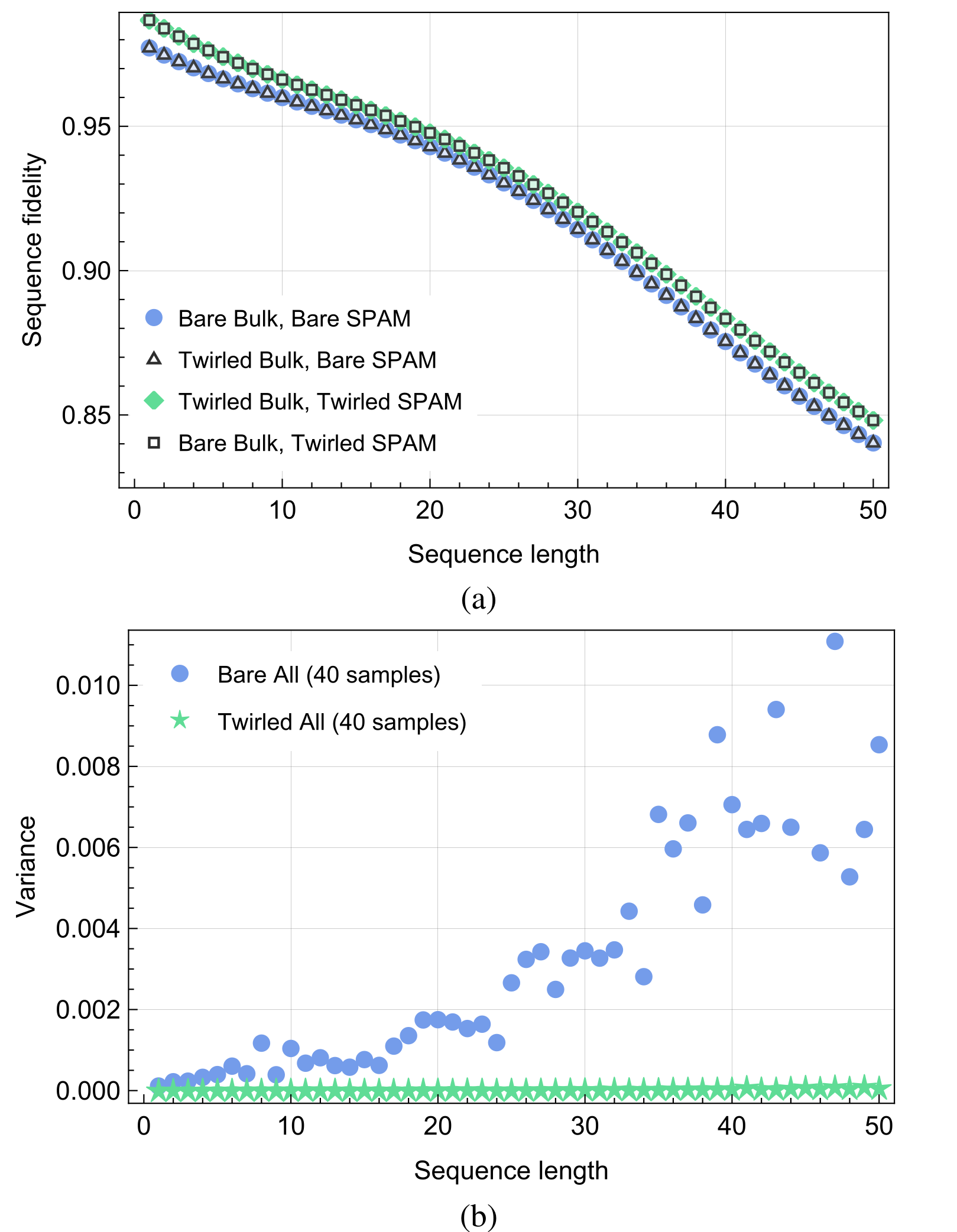}
    \caption{\textbf{The effect of Pauli-twirling on non-Markovian \gls{rb}}: For the same model and parameters (fixed $\tau_\mathsf{rb}=0.03$ and $\Lambda^{(\tau_\mathsf{rb})}$ truncated at $\mc{O}(\tau_\mathsf{rb}^{10})$) as Fig.~\ref{fig: rb dd numerical}, in (a), a comparison of the analytical average sequence fidelity decay (computed according to~\cite{PRXQuantum.2.040351}) for combinations of bare noise ($\Lambda$) and $\syst$-Pauli-twirled noise ($\Lambda^{\mathsf{P}_\syst}$) on bulk terms $\Lambda_1=\Lambda_2=\ldots=\Lambda_m$ and \gls{spam} terms $\Lambda_0=\Lambda_{m+1}$, showing that decays coincide up to \gls{spam} terms; while in (b) the numerical variance over 40 samples of \gls{rb} circuits, either all under bare noise or $\syst$-Pauli-twirled noise, showing that the variance generally gets suppressed for the latter.}
    \label{fig: rb rc numerical}
\end{figure}

In Fig.~\ref{fig: rb rc numerical}, we numerically demonstrate Result~\ref{result: rc and rb} by comparing the analytical average sequence fidelities (computed according to~\cite{PRXQuantum.2.040351}) with different choices of either bare noise or $\syst$-Pauli-twirled noise for the bulk terms and the \gls{spam} ones; we also show the behavior of the variance of the sequence fidelity with respect to sequence length for a set of 40 numerical samples, all either with bare noise or $\syst$-Pauli-twirled noise.

For the behavior of the average sequence fidelity with respect to \gls{spam}, we should point out that an overall increase in fidelity due to $\syst$-Pauli-twirled noise is incidental in this example; other than this, it is clear that decays remain non-exponential and that whenever \gls{spam} terms ($\Lambda_0$ and $\Lambda_{m+1}$) coincide, it is irrelevant for the decay whether all other terms were $\syst$-Pauli-twirled or not. While we do not argue about the feasibility of operationally twirling \gls{spam} noise, it is conceivable that this could be achieved at least partially on either the state preparation term or the measurement one (which in practice are indistinguishable).

For the case of the variance, we employed numerical averages on 40 \gls{rb} samples. The $\syst$-Pauli-twirled variance gets clearly suppressed, and is lower than that of bare noise, for all sequence lengths. It stands out that the variance for bare noise itself changes considerably with increasing sequence length. While the variance for twirled noise is not entirely vanishing, it is considerably suppressed, given its relation to the unitarity, as mentioned above in Section~\ref{sec: rb and rc}, so deriving such guarantees mathematically would make a strong point for the employment of a Pauli-twirling technique, such as \gls{rc}, in estimating average gate fidelities.
\section{Conclusions and Discussion}
We have shown that noise suppression techniques, such as \gls{dd} and Pauli-twirling, can be effective tools for dealing with non-Markovian noise sources in \gls{rb}. In particular, \emph{i}) \gls{udd} applied with fast and narrow pulses reduces a wide class of non-Markovian non-exponential \gls{rb} decays to an exponential decay plus perturbative corrections in time, \emph{ii}) Pauli-twirling does not Markovianize \gls{rb} in the same sense that \gls{dd} does, and in fact leaves the sequence fidelity decay invariant up to \gls{spam} noise, however, \emph{iii}) Pauli-twirling is ensured to decrease, or at worst leave unchanged, the variance of \gls{rb} sequence fidelities.

Our results imply that standard noise suppression techniques can be valuable tools in taming, benchmarking, and optimizing non-Markovian noise. In particular, while we have dealt with the \emph{standard} (or original) \gls{rb} protocol as a benchmarking framework, this approach is amenable to be adapted to any \gls{rb}-based technique considering e.g., scalability or specific purpose metrics~\cite{helsen_general_2022}.

In the case of \gls{dd}, while Markovianization is always achieved at short times, the full effective \gls{rb} sequence fidelities can still be dominated by non-Markovian terms, so the main limitation for \gls{dd} is the time-scale at which pulses can be applied, which for an enhancement of fidelities would need to be shorter than those between application of the \gls{rb} gates. Our main Result~\ref{result: markovianized rb} regarding \gls{dd} is stated for a broad class of continuous noise models with mild restrictions on the global Hamiltonian and dissipators. More generally, however, for any noise dynamics, the timescales at which decoupling can be efficiently achieved are connected to such dynamics~\cite{Addis_2015, DArrigo2019}, but not necessarily to whether they display non-Markovianity~\cite{nM_cannot_decouple_2021}. Furthermore, while they might be generally non-Markovian, some dynamics can hide non-Markovian effects~\cite{hidden_nM_PhysRevA.103.012203, elusive_nM_PhysRevA.104.L050404}, in particular for our case in \gls{rb}, and not display significant deviations. Finally, there is the fact that realistic \gls{dd} pulses have a finite width and themselves can introduce errors; for our results, however, it is sufficient for them to be narrow and contain only local, time, and gate-independent small errors.

For Pauli-twirling in our main Result~\ref{result: rc and rb}, it is somewhat surprising that $\syst$-Pauli-twirling does not Markovianize \gls{rb} data, since Pauli-twirling also effectively decouples by averaging out all non-Pauli error terms~\cite{winick2022RC}. Concretely, \gls{rb} with the Clifford group, while it is clear that in the Markovian case it leaves average gate-fidelities unchanged, as opposed to \gls{dd}, it was expected for it to have a decoupling effect that would impact \emph{directly} the average of the non-Markovian \gls{rb} outputs. Nevertheless, Pauli-twirling can be extremely valuable in suppressing the uncertainty in the average sequence fidelity, which can ensure an accurate diagnosis of non-exponential deviations (as opposed to statistical fluctuations) and reliable estimation of error rates. Furthermore, as argued before, this reduction in variance due to Pauli-twirling is related to the amount of coherence of the noise, which in turn is intimately connected and relevant to average gate-fidelity and fault tolerant-relevant metrics such as diamond norm~\cite{Sanders_2015}.

Noise suppression techniques, such as \gls{dd} and Pauli-twirling, are a vital ingredient allowing for the possibility to take quantum computing beyond a noise-intermediate regime. Our results highlight the importance of incorporating such basic noise suppression techniques to deal with one of the most complicated sources of errors, namely non-Markovianity, not only for deployment but for basic error diagnostics and benchmarking. As a perspective, we expect these ideas to be useful, easily adapted, and enhanced, to other scalable and holistic benchmarking and characterization techniques, e.g.,~\cite{proctor_2021, cycle_2019, helsen_seq_2021, flammia_aces, harper2023learning}, most of which are the intellectual progeny of \gls{rb}.

\begin{acknowledgments}
We thank Jay Nath for valuable discussions, and PFR thanks Robin Blume-Kohout for helpful comments during the APS March Meeting 2023. KM acknowledges support from the Australian Research Council Discovery Projects DP210100597 and DP220101793. PFR, MP, AA, and IdV acknowledge support from the German Federal Ministry of Education and Research (BMBF) under Q-Exa (grant No. 13N16062) and QSolid (grant No. 13N16161).
\end{acknowledgments}

\newpage

\printglossary

\onecolumngrid
\appendix
\section{Preliminaries}\label{appendix: preliminaries}
\subsection{Notation}\label{appendix: notation}
Throughout the manuscript, we refer to an environment $\env$ and a system $\syst$: these are quantum systems with Hilbert spaces $\mscr{H}_\env$ and $\mscr{H}_\syst$, respectively, where we take $\dim(\mscr{H}_\syst):=\dimS<\infty$. When $\env$ is assumed to be finite, we denote $\dim(\mscr{H}_\env):=\dimE$. When we refer to the composite $\syst\env$, we mean the space $\mscr{H}_\env\otimes\mscr{H}_\syst$. When referring to generic finite-dimensional Hilbert spaces $\mscr{H}$, we normally use $d=\dim(\mscr{H})$.

We mainly use either the capital Greek letters $\Phi,\,\Lambda$, or a curly font, e.g., $\mc{Q},\,\mc{P}$, to denote quantum channels. We use the term quantum channel to mean \gls{cp} map, and the term quantum gate to mean a digital unitary map. One particular exception is the symbol $\mc{F}_m$ which will be used specifically in the context of \gls{rb}. Finally, we use $\mc{I}$ for an identity channel and $\mbb1$ for an identity operator.

We will employ a vectorized representation of density matrices (which we often refer to simply as quantum states). Specifically, we employ
\begin{align}
    |\cdot\rrangle &:= \mathrm{vec}(\cdot),\quad\text{where}\quad\mathrm{vec}(|i\rangle\!\langle{j}|)=|ij\rangle\quad\text{for any}\quad\,|i\rangle,|j\rangle\in\mscr{H},
\end{align}
i.e., $|\cdot\rrangle$ is a vectorization of matrices sequentially stacking rows as columns, with the corresponding dual being $\llangle\cdot|:=|\cdot\rrangle^\dg$ where $\dg$ is a conjugate transpose. Notice, thus, that we can now express interior products as $\llangle{A}|B\rrangle=\tr[A^\dg{B}]$. Similarly then, for any quantum channel between bounded operator spaces, $\Phi:\mscr{B}(\mscr{H})\to\mscr{B}(\mscr{H})$, with Kraus operators $\phi_\mu$ (where $\mu$ can run on up to $d^2$ terms), we define
\begin{align}
    |\Phi(\cdot)\rrangle &:= \hat{\Phi}|\cdot\rrangle,\qquad\text{where}\qquad\hat{\Phi} = \sum_\mu\phi_\mu\otimes\phi_\mu^*,
    \end{align}
i.e., $\hat{\cdot}$ turns quantum channels into matrices. Here we will always assume \emph{at least} a trace non-increasing property $\sum\phi_\mu^\dg\phi_\mu\leq\mbb1$ (meaning the eigenvalues of such sum of Kraus operators are positive and upper-bounded by unity) on quantum channels and otherwise mention when a full \gls{tp} property, for which $\sum\phi_\mu^\dg\phi_\mu=\mbb1$, is being considered.

\subsection{Non-Markovianity}\label{appendix: non-Markovianity def}
Here we briefly define formally what we mean by the term \emph{non-Markovianity}, as summarized in~\cite{figueroaromero2022general}. A comprehensive review and discussion on the topic can be seen in~\cite{PRXQuantum.2.030201}.

Within classical probability theory, non-Markovianity can be described by a stochastic process, $\{X_t\}$, where generally probabilities are conditionally dependent on the past, i.e.
\begin{equation}
    \mbb{P}(x_k|x_{k-1},\ldots,x_0) = \mbb{P}(x_k|x_{k-1},\ldots,x_{k-\ell}),
    \label{eq: classical nM definition}
\end{equation}
for any integers $0\leq{\ell}\leq{k}$ and sequences of event outcomes $x_i$, with $\mbb{P}(\cdot|\cdot)$ denoting a conditional probability. In particular, when $\ell=1$, the process is called Markovian and when $\ell=0$ it is called random; otherwise, the process is non-Markovian with Markov order $\ell$. The fact that the complexity in describing non-Markovian processes increases exponentially in increasing Markov-order can be seen from joint probabilities requiring up to $\ell$-point correlations within the respective conditional probabilities.

Quantum mechanically, the process tensor (a.k.a. quantum comb~\cite{PhysRevLett.101.060401, PhysRevA.80.022339}, causal box~\cite{Portmann_causal}, correlation kernel~\cite{nurdin2021heisenberg}, process matrix~\cite{Costa_2016}, channel with memory~\cite{PhysRevA.72.062323}, or strategy~\cite{Gutoski_games}, to mention some) framework~\cite{PhysRevA.97.012127, PhysRevLett.120.040405, Milz_2017, Milz2020kolmogorov, Taranto_2020, PRXQuantum.2.030201, PhysRevLett.123.040401}, takes into account the invasive nature of observation to unambiguously provide a generalization of the condition in Eq.~\eqref{eq: classical nM definition}, as shown in~\cite{PhysRevLett.120.040405}. In this case, the medium for information to be sent across timesteps is an environment $\env$, part of a bipartite system $\syst\env$, with $\syst$ being the system of interest. Then, for an initial state $\rho$ of $\syst\env$, and upon measuring a {\gls{povm}} $\mc{J}_k:=\{M_{x_n}^{(k)}\}_{x_n}$ on system $\syst$, we may describe the probability of observing a sequence of quantum events $x_k,\ldots,x_0$ by
\begin{equation}
    \mbb{P}(x_k,\ldots,x_0|\mc{J}_k,\ldots,\mc{J}_0) := \tr\left[M_{x_k}\,\rho^{(k)}\right],
    \label{eq: joint prob quantum}
\end{equation}
where $\rho^{(k)}:=\tr_\env\left[\Mcirc_{i=1}^{k-1}\left(\mc{U}_i\circ\mc{A}_{x_i}\right)\rho\right]$ is the state of system $\syst$ at the $k$\textsuperscript{th} timestep, with $\mc{U}_i$ being dynamical maps on $\syst\env$ describing the evolution of the full system-environment between any two timestep, and $\mc{A}_{x_i}$ being \gls{cp} maps acting on system $\syst$ alone at timestep $i$: precisely, each $\mc{J}_i:=\{\mc{A}^{(i)}_{x_n}\}_{x_n}$ is called an instrument, where $\mc{A}^{(i)}_{x_n}$ is an experimental intervention represented by a \gls{cp} map with state outcome $x_n$, and such that $\sum_{x_n}\mc{A}^{(i)}_{x_n}=\mc{A}^{(i)}$ is a \gls{cptp} map.

We drop the super-indices in Eq.~\eqref{eq: joint prob quantum} for clarity, which we may write more succinctly  as the inner product
\begin{align}
    \mbb{P}(x_k,\ldots,x_0|\mc{J}_k,\ldots,\mc{J}_0) = \tr\left(\Upsilon_k\,\Theta_k^\mathrm{T}\right),
    \label{eq: probabilities PT contraction}
\end{align}
where $\mathrm{T}$ denotes a transpose, and $\Upsilon_k$ and $\Theta_k$ are tensors containing all dynamics $\{\mc{U}_i\}$ and all interventions $\{\mc{A}_i\}$; in the Choi-Jamio\l{}kowski representation, these take the form
\begin{align}
    \Upsilon_k &:= \tr_\env\left\{\left[\Mcirc_{i=1}^k(\mc{U}_i\otimes\mc{I}_{\mathsf{aux}}\circ\mscr{S}_i)\right]\rho\otimes\psi^{\otimes{k}}\right\},
    \label{eq: Choi state pt}\\
    \Theta_k &:= M_{x_k} \otimes \left[\Motimes_{i=1}^{k-1}\left(\mbb1_{\mathsf{A}_i}\otimes\mc{A}_{x_i}\right)\right]\psi^{\otimes{k}},
\end{align}
where $\mathsf{aux}:=\mathsf{A}_1\mathsf{B}_1\ldots\mathsf{A}_k\mathsf{B}_k$, with $\mathsf{A}_i$, $\mathsf{B}_i$ being $\dimS$-dimensional auxiliary spaces, $\mscr{S}_i$ being a swap map between $\syst$ and $\mathsf{A}_i$, and $\psi=\sum|ii\rangle\!\langle{jj}|$ being an unnormalized maximally entangled state.

The process tensor framework thus allows to neatly separate the underlying dynamical source for any given quantum process, including all temporal correlations therein, from all experimentally controllable operations. This description is entirely general as a quantum stochastic process framework~\cite{Milz2020kolmogorov, PRXQuantum.2.030201}, and similarly the instruments used to describe interventions are entirely general and can be temporally correlated themselves. Similar to the case of quantum states, the choice of employing a Choi state representation in Eq.~\eqref{eq: Choi state pt} allows us to readily deduce properties of the process. In particular, temporal properties get codified as spatial properties within the Choi state, so that a Markovian process takes an uncorrelated form,
\begin{equation}
    \Upsilon^{(\mathrm{M})} := \Motimes_i\mscr{Y}_{i:i-1}\otimes\rho_\syst,    
\end{equation}
with $\mscr{Y}_{i:i-1}$ being individual Choi states of dynamics connecting the $(i-1)$\textsuperscript{th} and $i$\textsuperscript{th} steps. This implies that we may quantify the non-Markovianity of a process by simply quantifying its distinguishability from the closest Markovian one, i.e., $\mc{N}:=\min_{\Upsilon^{(\mathrm{M})}}d(\Upsilon,\Upsilon^{(\mathrm{M})})$, for any operationally meaningful distance measure $d(\cdot,\cdot)$. Here we won't focus on the multi-time processes themselves, either $\Upsilon$ or $\Upsilon^{(\mathrm{M})}$, but rather on the consequences of each kind within \gls{rb}.

\subsection{The Randomized Benchmarking protocol and the standard Markovian analysis}\label{appendix: the RB protocol}\glsreset{asf}\glsreset{rb} 
A standard \gls{rb} protocol on a system of interest, $\syst$, of dimension $\dimS$ proceeds as follows:
\begin{enumerate}
    \item Prepare an initial state $\rho$ on $\syst$.
    \item Sample a sequence of $m$ distinct quantum gates, $\mc{G}_1,\mc{G}_2,\ldots,\mc{G}_m$, uniformly at random from a given gate set $\mbb{G}$ forming a finite group, and compile the inverse of the sequence in $\mc{G}_{m+1} =\mc{G}_1^{-1}\circ\cdots\circ\mc{G}_m^{-1}$, where here $\circ$ denotes composition of maps (i.e., $\mc{A}\circ\mc{B}$ reading $\mc{A}$ \emph{after} $\mc{B}$). We refer to $\mc{G}_{m+1}$ as an undo-gate.
    \item Apply the composition $\mc{S}_m:=\mc{G}_{m+1}\circ\mc{G}_m\circ\cdots\circ\mc{G}_1$ on $\rho$.
    \item Estimate the probability $f_m := \tr\left[M\mc{S}_m\left(\rho\right)\right]$ via a \gls{povm} element $M$. Generally, a real physical implementation of these steps renders $f_m\neq\tr(M\rho)$.
    \item Repeat $n$ times the steps 1 to 4 for the same sequence length $m$, same initial state $\rho$, same \gls{povm}~element $M$, and different sets of gates chosen uniformly at random, $\left\{\mc{G}_i^{(1)}\right\}_{i=1}^m,\,\left\{\mc{G}_i^{(2)}\right\}_{i=1}^m,\ldots,\left\{\mc{G}_i^{(n)}\right\}_{i=1}^m$ from $\mbb{G}$, to obtain the corresponding probabilities $f_m^{(1)},f_m^{(2)},\ldots{f}_m^{(n)}$. Estimate the average probability $\mc{F}_m=1/n\sum_{i=1}^nf_m^{(i)}$. We refer to $\mc{F}_m$ as an \gls{asf}.
    \item Examine the behavior of the estimated \gls{asf}, $\mc{F}_m$, over different sequence lengths $m$, with respect to the ideal expected $\tr(M\rho)$.
\end{enumerate}
The protocol up to this point is independent of any assumptions for the underlying sources of noise causing the outputs to deviate from the ideal expected one, as well as the choices for the gate-set (which is only restricted to be a finite group, hence contain inverses), the dimension of the system $\syst$ and the initial state and final measurement.

For several practical reasons, the system $\syst$ usually consists of one or two-qubits, i.e., $\dimS=2^n$ for either $n=1$ or $n=2$, the initial state is taken to be the ground state in the computational basis $\rho=|0\rangle\!\langle0|$ and $M$ the corresponding projector $M=|0\rangle\!\langle0|$, so that the \gls{asf} is rather simply a survival probability of the ground state.

\subsection{Modeling for Markovian RB}\label{appendix: RB Markov}
To make the analysis tractable and connect the \gls{rb} protocol with the estimation of so-called average gate fidelities, in a manner robust to \gls{spam} errors, some assumptions are commonly made for the ideal gate-set and the underlying noise affecting the physical protocol: the ideal gate-set $\mbb{G}$ is usually taken to be the multi-qubit Clifford group~\textsuperscript{\footnote{ The Clifford group on $n$-qubits is defined as the set of unitaries normalizing the $n$-qubit Pauli group, $\mbb{P}_n^*$ modulo the identity, i.e., $\mbb{C}_n:=\{G:\mbb{U}(2^n)|P\in\pm\mbb{P}_n^*\,\Rightarrow\,GPG^\dg\in\pm\mbb{P}_n^*\}$. The crucial property for \gls{rb} is that the $n$-qubit Clifford group forms a so-called unitary 2-design, while the main limitation to scalability comes from the fact that $n$-qubit Clifford gates are composite gates, with both the number of elements and elementary components scaling non-favorably in $n$.}} and noise assumed to be at least \emph{Markovian}, i.e., uncorrelated in time, and thus associated to each gate independently. Moreover, noise is also commonly assumed to be independent of both \emph{when} a gate is applied (time-stationary) and \emph{which} gate is applied (gate-independent). These can be made mathematically concrete by modeling individual noisy gates as $\tilde{\mc{G}}:=\Lambda\circ\mc{G}$, or equivalently, defining noise quantum channels by $\Lambda:=\tilde{\mc{G}}\circ\mc{G}^{-1}$, where $\tilde{\mc{G}}$ and $\Lambda$ are in general \gls{cp} maps sending inputs in $\syst$ to outputs in $\syst$. Notice that, physically, here we are assuming that the ideal digital gates $\mc{G}$ are precisely instantaneous noiseless quantum gates, while the maps $\Lambda$ can model \emph{the effects} of any possible Markovian quantum noise process. Furthermore, we could have equivalently defined $\tilde{\mc{G}}=\mc{G}\circ\Lambda$ and we would be modeling the same phenomena. Another, non-equivalent choice, could be to model noisy gates with dynamical generators (Hamiltonians) so that errors occur \emph{during} a physical (e.g., a simple pulse) but error-free gate; this, albeit physically plausible, has limitations: a deeper discussion on Markovian errors and their modeling can be seen in~\cite{taxonomy}.

\begin{figure}[t!]
    \includegraphics[width=0.8\textwidth]{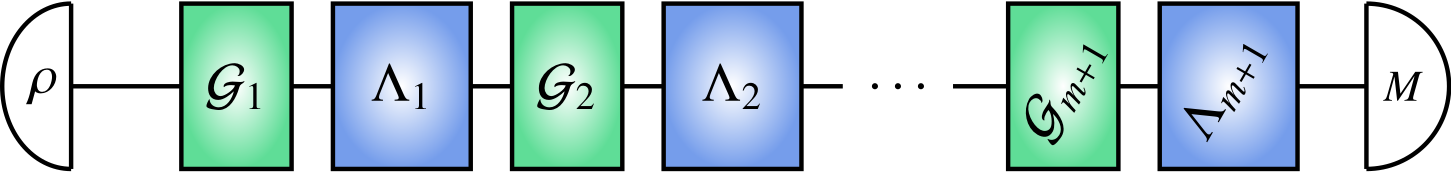}
    \caption{\textbf{Modeling Markovian noise in a quantum circuit}: An initial state $\rho$, is acted on with an ideal gate $\mc{G}_1$, which itself can be affected by a quantum channel $\Lambda_1$; subsequently, an ideal gate $\mc{G}_2$ is applied and can be affected by another quantum channel $\Lambda_2$, and so on, until a measurement with an element $M$ is performed. Modeling of noisy gates is described by the composition $\Lambda_i\circ\mc{G}_i$, or equivalently, $\mc{G}_i\circ\Lambda_i$, for any \gls{cp} and trace non-increasing map $\Lambda_i$. When $\Lambda_i=\Lambda_j=\Lambda$ for all $i\neq{j}$ time-step labels, we say that noise in the sequence is time-stationary.}
\label{appendix fig: Markov PT}
\end{figure}

With these assumptions of Markovianity, time-independence and gate-independence, we can model noisy initial states as $\tilde\rho:=\Lambda_0(\rho)$, \gls{rb} sequences as $\tilde{\mc{S}}_m:=\tilde{\mc{G}}_{m+1}\circ\tilde{\mc{G}}_m\circ\cdots\circ\tilde{\mc{G}}_1$, and noisy probabilities as $f_m:=\llangle{M}|\hat{\tilde{\mc{S}}}_m|\tilde{\rho}\rrangle$. Notice that measurement noise is indistinguishable from that of the initial state. Then, this allows to show that the analytical \gls{asf}, i.e., the analytical average of the sequences $f_m$ over uniformly distributed Clifford gates (or indeed over any so-called unitary 2-design~\textsuperscript{\footnote{A unitary $t$-design is an ensemble of unitary matrices that reproduce up to the $t$\textsuperscript{th} statistical moment of the unitary group with the uniform, so-called Haar measure; see e.g.~\cite{PhysRevA.85.042311}\label{footnote: unitary designs}}}) will take an exponential form in the sequence length $m$,
\begin{equation}
    \mc{F}_m = A p^m + B,\quad\text{where}\quad\,p=\f{\dimS\mathfrak{F}_\Lambda-1}{\dimS-1},\quad\text{with}\quad\mathfrak{F}_\Lambda:=\f{\dimS+\tr\left(\hat{\Lambda}\right)}{\dimS(\dimS+1)},
    \label{eq: Markov asf}
\end{equation}
and $0\leq{A, B}\leq1$ being constants capturing \gls{spam} errors,
\begin{equation}
    A = \llangle{M}|\hat{\Lambda}^\prime|\rho_\syst-\mbb1/\dimS\rrangle = \tr[M\Lambda^\prime(\rho_\syst-\mbb1/\dimS)],\qquad{B} = \llangle{M}|\hat{\Lambda}^\prime|\mbb1/\dimS\rrangle = \tr[M\Lambda^\prime(\mbb1/\dimS)],
    \label{eq: Markov spam}
\end{equation}
where $\Lambda^\prime=\Lambda\circ\Lambda_0$. In the case $\rho_\syst=M=|0\rangle\!\langle0|$, then $A=\langle0|\Lambda^\prime(|0\rangle\!\langle0|)|0\rangle-B$, and $B=\langle0|\Lambda^\prime(\mbb1)|0\rangle/\dimS$, so that $B$ can be seen as quantifying non-unitality of $\Lambda^\prime$ and $A$ as a corresponding fidelity quantifier minus such unitality, so that for \emph{low} \gls{spam} noise (unital and close to identity), both $A,B\approx1/\dimS$.

Detail of this result can be seen in standard \gls{rb} literature, such as~\cite{PhysRevA.85.042311}. The quantity ${p}\leq1$ is sometimes labeled the \emph{noise strength} of $\Lambda$, while $\mathfrak{F}_\Lambda\leq1$ is the so-called \emph{average gate-fidelity} of $\Lambda$ with respect to the identity $\mc{I}$. Notice that $\mathfrak{F}_\Lambda$ can equivalently be read as the average gate-fidelity of any noisy gate, $\tilde{\mc{G}}_i$, with respect to the respective ideal gate, $\mc{G}_i$: generically, the average gate-fidelity is defined as $\mathfrak{F}_{\Lambda}:=\mathbf{E}_\psi\left[\langle\psi|\Lambda(|\psi\rangle\!\langle\psi|)|\psi\rangle\right]$, where here $\mathbf{E}_\psi$ means uniform average over all pure states $|\psi\rangle$ on $\syst$, which can be shown to be equivalent to the definition in Eq.~\eqref{eq: Markov asf}. The noise strength is rigorously lower-bounded by $1/\left(1-\dimS^2\right)$ and the average gate-fidelity by $1/(\dimS+1)$, but this would be of course not sensible in practice and $p$ is at least assumed to be positive.

Many generalizations and different versions of \gls{rb} estimating various figures of merit exist, and in particular many of the assumptions before can be relaxed~\cite{helsen_general_2022}. Here we will focus on the \emph{non-Markovian} case, i.e., where the noise channels $\Lambda$ can be correlated with each other in time, and where furthermore, the character of these correlations is allowed to be quantum mechanical~\cite{time_entanglement}.

\section{Randomized Benchmarking for non-Markovian noise and the reduction to the Markovian case}\label{appendix: RB nM to M}
We now move on to considering a system-environment $\syst\env$ composite, where we assume gates can only be applied directly in system $\syst$, and similarly state-preparation and measurement can solely be done on $\syst$. Following up on the definitions made in Appendix~\ref{appendix: the RB protocol} for the Markovian case, we can generalize this to (temporally-correlated) non-Markovian noise, by letting the noise channels act on the whole $\syst\env$ composite. That is, we now take $\Lambda$ to be a quantum channel from $\syst\env$ quantum states to $\syst\env$ quantum states, which can either be a unitary channel or a \gls{cp} map, depending on whether $\syst\env$ is a closed system or there is dissipation to a larger environment, respectively. The environment $\env$ can thus be now understood as a quantum memory allowing for quantum information to propagate between different instances of $\Lambda$. We can further relax the time-stationary noise constraint, and associate a distinct $\Lambda_i$ (acting on $\syst\env$) to each ideal gate $\mc{I}_\env\otimes\mc{G}_i$ where here explicitly $\mc{G}_i$ acts on $\syst$ alone. Notice that, importantly, we cannot associate noise to each individual gate $\mc{G}_i$ separately, but we have to take into account the environment, and as we don't have direct access to it, we would need to track how it changes all intermediate noisy gates throughout a given sequence. Many in-depth reviews about non-Markovianity exist, and the topic is vast enough to be dealt with here, but the main notions and formalism we employ can be consulted in~\cite{PRXQuantum.2.030201}. Notice that the \gls{rb} protocol, of course, remains the same regardless of the underlying type of noise, it is just our modeling that changes and gives a more general analytical behavior of the \gls{asf}.

\begin{figure}[t!]
    \centering
    \includegraphics[width=0.8\textwidth]{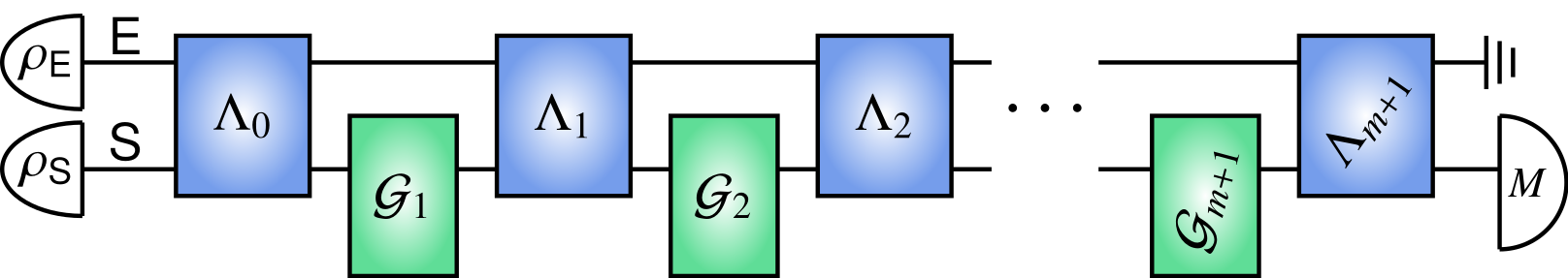}
    \caption{\textbf{Modeling non-Markovian noise in a quantum circuit}: Similar to the case of Fig.~\ref{appendix fig: Markov PT}, a sequence of ideal gates $\mc{G}_1,\mc{G}_2,\ldots,\mc{G}_{m+1}$ is applied on a system $\syst$ with initial state $\rho_\syst$ and final measurement operator $M$. Now, however, noise is modeled as a \gls{cp} $\Lambda_i$ associated to maps $\mc{I}_\env\otimes\mc{G}_i$, where $\mc{I}_\env$ is an identity channel on a quantum environment (i.e. an external, in principle inaccessible Hilbert space) $\env$, initialized in some fiducial state $\rho_\env$. Temporal correlations, i.e., non-Markovianity, between noise is mediated by $\env$, and \gls{spam} terms, $\Lambda_0$ and $\Lambda_{m+1}$ can correlate either preparation or measurement on $\syst$ and $\env$.}
    \label{fig: non-Markovian pt}
\end{figure}

Hence we now model the global initial noisy state as $\tilde{\rho}:=\Lambda_0(\rho_\env\otimes\rho_\syst)$ for some ideal prepared state $\rho_\syst$, and some fiducial state of the environment $\rho_\env$. Similarly, now individual noisy gates at time-step $i$ are taken to $\Lambda_i\circ\mc{G}_i$, where the identity map on $\env$ is implied (henceforth, we do this when clear by context). The final measurement remains as an operator $M$ acting solely on $\syst$. An \gls{asf} modeling a \gls{rb} experiment under non-Markovian~\textsuperscript{\footnote{ The only assumption we don't explicitly relax is gate-dependence, i.e., the noise maps $\Lambda_i$ at any given time-step $i$ do not depend explicitly on which gate $\mc{G}_i$ was applied.}} noise would then take the form $\mc{F}_m=\mathbf{E}_\mbb{G}\llangle{M}|\hat{\tr}_\env\hat{\Lambda}_{m+1}\hat{\mc{G}}_{m+1}\hat{\Lambda}_{m}\hat{\mc{G}}_{m}\cdots\hat{\Lambda}_{1}\hat{\mc{G}}_{1}|\tilde{\rho}\rrangle$, where the sequence of gates and noise maps can be depicted as in Fig.~\ref{fig: non-Markovian pt}; the average over the gate-set, $\mathbf{E}_\mbb{G}$, can then be computed analytically at least either when $\mbb{G}$ is the multi-qubit Clifford group (generally any unitary 2-design)~\cite{PRXQuantum.2.040351} or when it is a finite group~\cite{figueroaromero2022general} (extensions beyond finite groups or groups in general, have also been proposed, at least within the Markovian case~\cite{kong2021framework, PRXQuantum.3.030320}).

Explicitly, when $\mbb{G}$ is a finite group admitting a multiplicity-free representation~\textsuperscript{\footnote{ A representation of a finite group $\mbb{G}$ can be defined as a map from $\mbb{G}$ to a vector space of unitary linear operators, e.g., that of $d$-dimensional complex matrices $\phi:\mbb{G}\to\mbb{U}(\mbb{C}_{d\times{d}})$. We call a representation reducible if it can be expressed as a direct sum of irreducible representations (irreps), and we say this representation is multiplicity-free if it contains no more than a single copy of such irreps: $\phi=\Moplus_\mu\phi_\mu^{\otimes{n_\mu}}$ with $n_\mu=1$ for all labels $\mu$. This is a technical constraint that can potentially be relaxed~\cite{PRXQuantum.2.010351, helsen_general_2022} with some added but potentially non-crucial complications.\label{fn: multiplicity-free}}}, the analytical \gls{asf} for a \gls{rb} experiment with $m$ gates drawn uniformly from $\mbb{G}$, under gate-independent non-Markovian noise, takes the form (\cite{figueroaromero2022general})
\begin{equation}
    \mc{F}_m = \sum_{\pi\in{R}_{\mbb{G}}}\! \llangle{M}|\hat{\tr}_\env\hat{\Lambda}_{m+1}\left(\hat{\mc{Q}}_{m,\pi}\otimes\hat{\mc{P}}_\pi\right)|\tilde{\rho}\rrangle,
    \label{eq: non-Markov asf}
\end{equation}
where $R_\mbb{G}$ is a set of labels for the spaces associated to each irreducible representation of $\mbb{G}$, the $\hat{\mc{P}}_\pi$ are projector operators onto these~\textsuperscript{\footnote{ We use the notation $\hat{\mc{P}}_\pi$ in order to not excessively conflate terms in Eq.~\eqref{eq: non-Markov asf}, however, all $\hat{\mc{P}}_\pi$ here are literally projection operators, not stemming \emph{by definition} from a quantum channel. Of course one can still also associate a channel $\mc{P}_\pi$ to them, which however, does not necessarily has the action of a projection superoperator.}}, and $\mc{Q}_{m,\pi}$ is a so-called length-$m$ quality map associated to the sequence of noise maps on the $\pi$\textsuperscript{th} irreducible subspace. We define explicitly what quality maps are below, in Eq.~\eqref{eq: quality factor}, but their qualitative significance is essentially a generalization of the noise strength $p$, in Eq.~\eqref{eq: Markov asf}, from the standard Clifford case: indeed for the Markovian noise case, quality \emph{maps} are really just quality \emph{scalar factors}, and with the Clifford group there are just two irreducible subspaces, with one quality factor corresponding to 1 and the other to $p$ (these concepts, including the definition of a quality factor, are explained e.g., in~\cite{helsen2019character}).

Noise maps $\mc{Q}_{m,\pi}$ can be written in superoperator form as
\begin{align}
    \hat{\mc{Q}}_{m,\pi} = \sum_{\{\epsilon_j,\epsilon_j^\prime\}_{j=0}^{m}}^{\dimE} \f{\prod_{i=m-1}^{0}\tr\left(\langle\epsilon_{i+1}\epsilon_{i+1}^\prime|\hat{\Lambda}_{i+1}|\epsilon_i\epsilon_i^\prime\rangle\hat{\mc{P}}_\pi\right)}{\tr\left(\hat{\mc{P}}_\pi\right)^m}|\epsilon_m\epsilon_m^\prime\rangle\!\langle\epsilon_0\epsilon_0^\prime|,
    \label{eq: quality factor}
\end{align}
where here all $\{|\epsilon_i\rangle,|\epsilon_i^\prime\}_{i=0}^m$ are basis vectors of $\env$. The derivation of this expression for quality maps can be seen in detail in~\cite{figueroaromero2022general}, specifically in Appendices B.1 and B.3.

For the particular case of the Clifford group cited in the main text, also a full detail of derivation can be seen in Appendix B.2 of~\cite{figueroaromero2022general}: in this case, there are two irreducible subspaces with $\hat{\mc{P}}_1=\Psi$ and $\hat{\mc{P}}_2=\mbb1-\Psi$ where $\Psi=\sum_{i,j=1}^{\dimS}|ii\rangle\!\langle{jj}|/\dimS$ and $\mbb1$ is a $\dimS^2$ identity, so there are two corresponding quality factors, $\mc{Q}_{m,1}$ and $\mc{Q}_{m,2}$, which are explicitly defined through
\begin{align}
    \mc{Q}_{m,1} \otimes\mc{P}_1 &:= \left[(\$_\Lambda-\Theta_\Lambda)^{\circ{m}}\otimes\mc{I}_\syst\right]\left(\rho-\rho_\env\otimes\f{\mbb1}{\dimS}\right), \quad \text{and} \quad
    \mc{Q}_{m,2} \otimes\mc{P}_2 := \Theta_\Lambda^{\circ{m}}(\rho_\env)\otimes\f{\mbb1}{\dimS},
\end{align}
where
\begin{align}
    \Theta_\Lambda(\cdot) = \tr_\syst\left[\Lambda\left(\cdot\otimes\f{\mbb1}{\dimS}\right)\right], \quad \text{and}\quad
    \$_\Lambda(\cdot) := \sum_\mu\tr_\syst(\lambda_\mu)(\cdot)\tr_\syst(\lambda_\mu^\dg),
    \label{eq: theta map}
\end{align}
with $\lambda_\mu$ being the Kraus operators of $\Lambda$, i.e. $\hat{\Lambda}=\sum_\mu\lambda_\mu\otimes\lambda_\mu^*$. In the main text, we express this in a simplified way as
\begin{align}
    \mc{Q}^\prime_{m,A} = \Lambda_{m+1}\circ(\mc{Q}_{m,1}\otimes\mc{P}_1), \quad \text{and} \quad \mc{Q}^\prime_{m,B} = \Lambda_{m+1}\circ(\mc{Q}_{m,2}\otimes\mc{P}_2).
    \label{eq: appendix quality factors Clifford}
\end{align}

Here, however, we are not interested in studying general quality maps, but rather whether we can operationally reduce them to the Markovian quality factors, i.e., if we can \emph{Markovianize} the non-Markovian \gls{asf} of Eq.~\eqref{eq: non-Markov asf} through an experimental protocol embedded in \gls{rb}.

Thus, before considering the question of how to do this operationally, let us show that indeed, if the noise across all steps is uncorrelated on $\syst\env$, so that no information leaks to $\env$ and then gets propagated in time, we recover the Markovian \gls{asf}. That is, let us consider noise of the form $\Lambda=\Theta\otimes\Phi$, where $\Theta$ is a channel acting solely from and to $\env$, while $\Phi$ is a channel between $\syst$ systems, then
\begin{align}
    \hat{\mc{Q}}_{m,\pi} &= \sum_{\{\epsilon_j,\epsilon_j^\prime\}_{j=0}^m}^{\dimE} \f{\prod_{i=m-1}^{0}\langle\epsilon_{i+1}\epsilon_{i+1}^\prime|\hat{\Theta}_{i+1}|\epsilon_i\epsilon_i^\prime\rangle\tr\left(\hat{\Phi}_{i+1}\hat{\mc{P}}_\pi\right)}{\tr\left(\hat{\mc{P}}_\pi\right)^m}|\epsilon_m\epsilon_m^\prime\rangle\!\langle\epsilon_0\epsilon_0^\prime| \nonumber\\
    &= \f{\tr\left(\hat{\Phi}_1\hat{\mc{P}}_\pi\right)\cdots\tr\left(\hat{\Phi}_m\hat{\mc{P}}_\pi\right)}{\tr\left(\hat{\mc{P}}_\pi\right)^m}\,\prod_{i=m-1}^{0}\hat{\Theta}_{i+1},
    \label{eq: quality map uncorr noise}
\end{align}
for all $m$, and since $\hat{\tr}\,\hat{\Theta}_{m+1}\hat{\Theta}_m\cdots\hat{\Theta}_1\hat{\Theta}_0|\rho_\env\rrangle=\tr\left[\Theta_{m+1}\circ\cdots\circ\Theta_0(\rho_\env)\right]=1$, it follows that
\begin{align}
    \mc{F}_m &= \sum_{\pi\in{R}_{\mbb{G}}}\prod_{i=1}^{m}\left(\f{\tr\hat{\Phi}_i\hat{\mc{P}}_\pi}{\tr\hat{\mc{P}}_\pi}\right)\llangle{M}|\hat{\Phi}_{m+1}\hat{\mc{P}}_\pi\hat{\Phi}_0|\rho_\syst\rrangle \nonumber\\
    &= \sum_{\pi\in{R}_{\mbb{G}}} f_{1,\pi}f_{2,\pi}\cdots f_{m,\pi}\llangle{M}^\prime|\hat{\mc{P}}_\pi|\rho_\syst^\prime\rrangle,
\end{align}
where we defined
\begin{equation}
    f_{i,\pi} := \f{\tr\left(\hat{\Phi}_i\hat{\mc{P}}_\pi\right)}{\tr\hat{\mc{P}}_\pi},
\end{equation}
which are precisely the Markovian \emph{quality factors}~\cite{helsen2019character} mentioned before, and where $M^\prime$ and $\rho_\syst^\prime$ are the \gls{spam}-affected measurement and initial state. This is precisely the Markovian generalization of Eq.~\eqref{eq: Markov asf} under time-non-stationary noise and with the gate-set being any finite group; indeed it can be shown that it reduces to it in the time-stationary noise case and multi-qubit Clifford group~\cite{figueroaromero2022general}.

We can now proceed to show that this \emph{Markovianization} effect can be achieved, approximately, by interleaving so-called ideal \gls{udd} sequences into a non-Markovian \gls{rb} experiment.


\section{Non-Markovian Randomized Benchmarking with Ideal Pulse Universal Dynamical Decoupling}\label{appendix: rbdd ideal}
\gls{dd} is in general a quantum control technique (i.e., one involving physical control via pulses applied on a control Hamiltonian) allowing to decouple a system of interest from undesired degrees of freedom in its evolution; as such, it can be seen as an essential quantum error suppression technique. A digital, so-called \emph{ideal} \gls{dd} scheme, assumes that the control pulses applied on system $\syst$ are infinitely strong and narrow; i.e., that they are achieved by a time-non-stationary Hamiltonian of the form $H_\ctrl^{(i)}(t) = \f{\pi}{2}\delta(t-t_0)v_i$, where $\delta$ is a Dirac delta, $t_0$ is the time at which the pulse is applied, and $v_i$ is a system $\syst$ operator element of a group $\mbb{V}$ such that
\begin{equation}
\f{1}{|\mbb{V}|}\sum_{i=1}^{|\mbb{V}|}\mc{V}_i(Z) = W\otimes\mbb1_\syst,
    \label{eq: decoupling def}
\end{equation}
for any operator $Z$ in $\syst\env$ and some operator $W$ on $\env$ alone, where we defined
\begin{equation}
    \mc{V}_i(\cdot):=(\mbb1_\env\otimes{v}_i)(\cdot)(\mbb1_\env\otimes{v}_i^\dg).
\end{equation}

We only consider here $\mbb{V}$ to be a unitary group of dimension $\dimS$, i.e., the operators $v_i$ are unitary $v_iv_i^\dg=v_i^\dg{v}_i=\mbb1_\syst$.

In the case of so-called \gls{udd}, the goal is to apply the pulses encompassing the decoupling group can be applied between free evolution of the whole $\syst\env$ (i.e., between the application of the random \gls{rb} gates) in order to make use of the decoupling property in Eq.~\eqref{eq: decoupling def}. This can further be done in a more elaborated way (e.g., periodically or in a concatenated fashion~\cite{LidarCDD}) over some given time interval. While many schemes of \gls{dd} are known (not necessarily universal)~\cite{ezzell_2022}, here we only employ \gls{udd}. Similarly, while the dimension of system $\syst$, can in principle be kept general (as in \gls{rb}), practical considerations, such as noise and the implementation itself, imply that one normally considers just a single-qubit (for which we can set $\mbb{V}=\{\mbb1,X,Y,Z\}$, the single-Pauli group) or quite small-dimensional systems.

We now consider noise on the full $\syst\env$ which itself might experience both global and local dissipation to another external environment; this model covers a broad class of dynamical processes that may be used to model quantum noise with both a coherent (unitary) and a stochastic (dissipative) part. A similar argument to the one we make below can be seen in~\cite{Nurdin_2017, berk2021extracting}. We can describe this noise model, on a given timescale, as an open system evolution on $\syst\env$ as
\begin{equation}
    \Lambda^{(t)}(\cdot)=\ex^{t\mc{L}}(\cdot),
\end{equation}
where
\begin{align}
    \mc{L}(\rho) &:= - i[H,\rho] + \mc{D}(\rho), \\\text{with}&\quad\mc{D}:=\mc{D}_{\env\syst} + \mc{D}_\env + \mc{D}_\syst,\quad\text{where}\quad\mc{D}_\mathsf{A}(\cdot):=\sum_{k=1}^{d_{\mathsf{A}}^2-1}\gamma^{(\mathsf{A})}_{k}\left(L^{(\mathsf{A})}_k\rho{L}^{(\mathsf{A})\dg}_k - \f{1}{2}\{L_k^{(\mathsf{A})\dg}{L}^{(\mathsf{A})}_k,\rho\}\right),
    \label{eq: Lind and Dissip}
\end{align}
for a given $\syst\env$ noise total Hamiltonian $H$ and so-called dissipation maps $\mc{D}_{\mathsf{A}}$, which we take in a form such that dissipation operators $L^{(\mathsf{A})}_k$ are traceless and orthonormal ($\tr[L^{(\mathsf{A})}_k]=0$ and $\tr~[L^{(\mathsf{A})}_iL^{(\mathsf{A})\dg}_j]=\delta_{ij}$), and with positive constants $\gamma^{(\mathsf{A})}_k$. For notation, here $[a,b]:=ab-ba$ and $\{a,b\}:=ab+ba$, and we denote by $\mc{D}_{\env\syst}$ the global $\syst\env$ dissipator term.

For small time intervals $\delta{t}:=\tau$, this implies that to the first order in $\tau$ we have
\begin{align}
    \Lambda^{(\tau)}(\rho) \approx \rho - i\tau\, [H,\rho] + \tau\, \mc{D}(\rho).
\end{align}

Now we consider an ideal \gls{udd} sequence from $\mbb{V}$ applied with the decoupling operators acting between small windows of free noise evolution $\tau$. For now, suppose that the operators $v_i\in\mbb{V}$ themselves are noiseless (see Remarks in Sec.~\ref{Sec: Appendix Remarks}). Define $\Lambda_{v_j}^{(\tau)}:=\mc{V}_j\circ\Lambda^{(\tau)}\circ\mc{V}^\dg_j$, then,
\begin{equation}
    \Lambda_{v_j}^{(\tau)}(\rho) \approx \rho - i \tau \left[\mc{V}_j(H),\,\rho\right] + \tau\,\mc{V}_j\circ\mc{D}\circ\mc{V}_j^\dg(\rho),
\end{equation}
so that we may compose all the ``$v$-dressed'' maps, $\Mcirc_{i=1}^{|\mbb{V}|}\Lambda_{v_i}^{(\tau)}(\rho)$, and approximate this as a single evolution under an average Lindbladian $\overline{\mc{L}}:=\sum_{i=1}^{|\mbb{V}|}\mc{V}_i\circ\mc{L}\,\circ\mc{V}_i^\dg$ (again, with the assumption of a small enough $\tau$), as
\begin{align}
\Mcirc_{i=1}^{|\mbb{V}|}\Lambda_{v_i}^{(\tau)} \approx \mathrm{e}^{\tau\overline{\mc{L}}},
\end{align}
whose action on $\rho$ is, in turn, to the first order in $\tau$,
\begin{equation}                
    \mathrm{e}^{\tau\overline{\mc{L}}}(\rho) \approx \rho - i \tau \left[\sum_j\mc{V}_j(H),\,\rho\right] + \tau \sum_j\mc{V}_j\circ\mc{D}\circ\mc{V}^\dg_j(\rho).
\label{eq: lindblad dd}
\end{equation}

The decoupling property in Eq.~\eqref{eq: decoupling def} implies that within this infinitesimally small time window $\tau$, the second term on the right-hand-side in Eq.~\eqref{eq: lindblad dd}, containing the Hamiltonian, will be uncorrelated. Regarding the remaining terms, this does not occur in general for the global $\syst\env$ dissipation term. This is problematic, analytically, because the quality maps, and hence the \gls{asf} too, will remain as general non-Markovian objects unless we assume there is no global $\syst\env$ dissipation contribution in the noise or it is otherwise negligible for small times.

Considering thus, only local dissipation, i.e., $\mc{D}\approx\mc{D}_\env+\mc{D}_\syst$, in this (infinitesimally) small time-interval $\tau$ limit, writing $\tilde{L}^{(\syst)}_{j}:=v_j{L}_k^{(\syst)}v^\dg_j$, we end up with an evolution, on an initially uncorrelated state, of the form
\begin{align}
    \mathrm{e}^{\tau\overline{\mc{L}}}(\rho) &\approx \rho + -i\tau|\mbb{V}|\,[B,\rho_\env]\otimes\rho_\syst + \tau\, \mc{D}_\env(\rho_\env)\otimes\rho_\syst + \tau\,\rho_\env\otimes\mc{D}_\syst^{(\mathsf{dd})}(\rho_\syst)
    \label{eq: dd lind noise decoupled},
\end{align}
where here $\rho=\rho_\env\otimes\rho_\syst$ and we defined $\mc{D}_\syst^{(\mathsf{dd})}$ by
\begin{align}
    \mc{D}_\syst^{(\mathsf{dd})}(\rho_\syst) := \sum_j\sum_{k}\gamma^{(\syst)}_{k} \bigg(\tilde{L}^{(\syst)}_{j,k}\rho_\syst\tilde{L}^{(\syst)\dg}_{j,k} - \f{1}{2}\left\{\tilde{L}^{(\syst)\dg}_{j,k}\tilde{L}^{(\syst)}_{j,k},\rho_\syst\right\}\bigg).
\end{align}

Now for the quality map in \gls{rb}, suppose we have $\syst\env$ noise of the form $\Lambda_k=\ex^{\tau_k\overline{\mc{L}}_k}$ at time-step $k$, for given small time-intervals $\tau_k$ and \gls{dd}-averaged Lindbladians $\overline{\mc{L}}_k$ such that $\Lambda_k$ acts as in Eq.~\eqref{eq: dd lind noise decoupled} with corresponding operators labeled by subscripts $k$. Let us now label the three summands in Eq.~\eqref{eq: dd lind noise decoupled} as
\begin{align}  
\Theta_k^{(0)}\otimes\Phi_k^{(0)}&:=\mc{I}
,\\
\Theta_k^{(1)}\otimes\Phi_k^{(1)}(\rho_\env\otimes\rho_\syst) &:= -i[B_k,\rho_\env]\otimes\tau_k|\mbb{V}|\rho_\syst + \mc{D}_{k,\env}(\rho_\env)\otimes\tau_k\rho_\syst,\\
\Theta_k^{(2)}\otimes\Phi_k^{(2)}(\rho_\env\otimes\rho_\syst) &:= \rho_\env\otimes\tau_k\,\mc{D}_{k,\syst}^{(\mathsf{dd})}(\rho_\syst),
\end{align}
so that $\Lambda_k\approx\sum_{j_k=0}^2\Theta_k^{(j_k)}\otimes\Phi_k^{(j_k)}$, and thus we obtain the quality map
\begin{align}
    \hat{\mc{Q}}_{m,\pi}
    &= \sum_{j_1,\ldots,j_m=0}^2\f{\displaystyle{\prod_{i=m-1}^{0}}\tr\left(\hat{\Phi}_{i+1}^{(j_{i+1})}\hat{\mc{P}}_\pi\right)\,\hat{\Theta}_{i+1}^{(j_{i+1})}}{\tr\left(\hat{\mc{P}}_\pi\right)^m} + \mc{O}(\tau_k^2),
\end{align}
where the first term can be of order up to $\tau_1\tau_2\cdots\tau_m$, with individual $\mc{O}(\tau_k^2)$ terms, for all $k=1,2,\ldots,m$, neglected. Upon entering this noise map in the \gls{asf} in Eq.~\eqref{eq: non-Markov asf}, we obtain the term 
\begin{align}
\hat{\tr}\,\hat{\Theta}_{m+1}^{(j_{m+1})}\cdots\hat{\Theta}_0^{(j_0)}|\rho_\env\rrangle=\tr\left[\Theta_{m+1}^{(j_{m+1})}\circ\cdots\circ\Theta_0^{(j_0)}(\rho_\env)\right] := \begin{cases}
    0,\quad\text{if any}\quad j_i=1\\
    1,\quad\text{otherwise}
\end{cases},
\end{align}
so there are no contributions from $\env$ at order $\tau_k^1$, and quality factors will only take the identity and the \gls{dd}-dressed dissipators $\mc{D}_{k,\syst}^{(\mathsf{dd})}$. Thus to $\mc{O}(\tau_k^1)$ we have a purely Markovian decay of the form
\begin{align}
    \mc{F}_m &= \sum_{\pi\in{R}_{\mbb{G}}}\sum_{j_i\in\left\{0,2\right\}}\left(\f{\tr\hat{\Phi}^{(j_1)}_1\hat{\mc{P}}_\pi}{\tr\hat{\mc{P}}_\pi}\right)\cdots\left(\f{\tr\hat{\Phi}^{(j_m)}_m\hat{\mc{P}}_\pi}{\tr\hat{\mc{P}}_\pi}\right)\llangle{M}|\hat{\Phi}^{(j_{m+1})}_{m+1}\hat{\mc{P}}_\pi\hat{\Phi}^{(j_0)}_0|\rho_\syst\rrangle + \mc{O}(\tau_k^2),
\end{align}
which can be recast as
\begin{align}
    \mc{F}_m &= \sum_{\pi\in{R}_{\mbb{G}}}\left(\f{\tr\hat{\Omega}_1\hat{\mc{P}}_\pi}{\tr\hat{\mc{P}}_\pi}\right)\cdots\left(\f{\tr\hat{\Omega}_m\hat{\mc{P}}_\pi}{\tr\hat{\mc{P}}_\pi}\right)\llangle{M}|\hat{\Omega}_{m+1}\hat{\mc{P}}_\pi\hat{\Omega}_0|\rho_\syst\rrangle + \mc{O}(\tau_k^2),
    \label{eq: asf Markov plus nM corr}
\end{align}
where $\Omega_k:=\mc{I}+\tau_k\,\mc{D}^{(\mathsf{dd})}_{k,\syst}$, and where contributions with terms $\mc{O}(\tau_k^2)$ for any $k=0,\ldots,m+1$, are generally non-Markovian.

For concreteness, take the multi-qubit Clifford group, which is such that~\textsuperscript{\footnote{ Equivalently a unitary 2-design; the projectors are $\hat{\mc{P}}_1=\Psi$ and $\hat{\mc{P}}_2=\mbb1-\Psi$ where $\Psi:=\sum|ii\rangle\!\langle{jj}|/\dimS$. See e.g.~\cite{PhysRevA.73.062314}. For details on the particular case of the \gls{asf} in the Clifford case, see~\cite{figueroaromero2022general}.}} $\mc{F}_m = A p_1p_2\cdots{p}_m + B + \mc{O}(\tau_k^2)$, with
\begin{align}
    p_k &= \f{\tr(\hat{\Omega}_k)-1}{\dimS^2-1} = 1 + \tau_k\f{\tr\left(\hat{\mc{D}}_{k,\syst}^{(\mathsf{dd})}\right)}{\dimS^2-1} 
    = 1 - \tau_k \f{\dimS|\mbb{V}|\sum\gamma_k\tr\left(L^{(\syst)}_kL_k^{(\syst)\dg}\right)}{\dimS^2-1} = 1 - \tau_k\f{\dimS|\mbb{V}|\sum_{k_\syst}\gamma_{k_\syst}}{\dimS^2-1},
    \label{eq: noise strength Lind}\\[0.1in]
    A &= \llangle{M}|\hat{\Omega}_{m+1}\hat{\Omega}_{0}|\rho_\syst-\f{\mbb1}{\dimS}\rrangle,\qquad B = \llangle{M}|\hat{\Omega}_{m+1}\hat{\Omega}_0|\f{\mbb1}{\dimS}\rrangle\,
\end{align}
where $A$, $B$ are standard Markovian \gls{spam} noise contributions.

Finally, in the time-stationary case, $\Omega_0=\cdots=\Omega_{m+1}=\Omega$, the \gls{asf} would further reduce to an exponential plus higher order terms, $\mc{F}_m = A p^m + B + \mc{O}(\tau^2)$, with the single factor $p$ corresponding to the quality factor of the map $\Omega=\mc{I}+\tau\,\mc{D}_\syst^{(\mathsf{dd})}$. This is presented in the main text as $\mc{F}_m = A p_{\tau_\mathsf{dd}}^m + B + \mc{O}(\tau_\mathsf{dd}^2)$, where
\begin{align}
    p_{\tau_\mathsf{dd}} &= 1 - \tau_k\f{\dimS|\mbb{V}|\sum_{k_\syst}\gamma_{k_\syst}}{\dimS^2-1}, \label{eq: appendix p_tau}\\
    A &= \tr\left[M \Omega^2(\rho_\syst-\mbb1/\dimS)\right] \approx \tr\left[M(\rho_\syst-\mbb1/\dimS)\right] + 2\tau_\mathsf{dd}\tr\left[M\mc{D}_\syst^{(\tau_\mathsf{dd})}(\rho_\syst-\mbb1/\dimS)\right], \label{eq: appendix A}\\
    B &= \tr\left[M \Omega^2(\rho_\syst)\right] \approx \tr\left[M(\rho_\syst)\right] + 2\tau_\mathsf{dd}\tr\left[M\mc{D}_\syst^{(\tau_\mathsf{dd})}(\rho_\syst)\right] \label{eq: appendix B},
\end{align}
with first-order $\tau_\mathsf{dd}$ approximations in $A$, $B$. This means that when such approximation holds and the underlying noise has the form described above, the data of a \gls{rb} experiment, originally displaying non-Markovian deviations, will fall to an exponential plus higher-order time corrections if the ideal \gls{udd} pulses are applied between short time-windows $\tau$ (or $\tau_\mathsf{dd}$); the average gate-fidelity extracted in such a protocol would then contain that of $\Omega$ with respect to the identity $\mc{I}$ to the first time-order.

\subsection{Remarks}\label{Sec: Appendix Remarks}
\subsubsection{Noise model}
The main reason to model the noise maps $\Lambda_k$ as a dynamical map is twofold: one is to set a time scale between the action of the control gates, parametrized by the time-intervals $\tau_k$, and the other is to express both the generators of non-Markovianity (the unitary part) and Markovianity (the dissipating part) within the global $\syst\env$ noise dynamics. This can encompass a broad class of physically motivated noise models.

The only assumption made in order to obtain Eq.~\eqref{eq: asf Markov plus nM corr} was that either there is no global $\syst\env$ dissipation contribution in the noise dynamics, or that it is negligible to the first order in $\tau$, as discussed below Eq.~\eqref{eq: lindblad dd}. Notice that if there is no dissipation at all, local or otherwise, to the first order in $\tau$, the \gls{asf} would be trivial (i.e., look noiseless) and all higher-order contributions to the decay are generally non-Markovian or from the local Hamiltonian. Numerically, in such case while it may still be seen that the \gls{asf} approximates a decaying exponential to a large extent --depending mainly on how fast the \gls{dd} sequences are applied--, only higher-orders would describe this decay.

Finally, the presence or absence of dissipator terms has to be explicitly stated, as it is really the full $\Lambda^{(\tau)}$ map that generates noise. For example, one could think of phrasing Result~\ref{result: markovianized rb} in the main text for a dilated Hamiltonian $H^\prime$ acting in a larger Hilbert space $\syst\env\env^\prime$ so that the Lindbladian $\mc{L}^\prime(\cdot)=-i[H^\prime,\cdot]$ contains no explicit dissipators whatsoever; the point is that the actual original $\Lambda^{(\tau)}$ will have to be written in terms of a partial tracing of $\env^\prime$ and some fiducial initial state $\rho_{\env^\prime}$, i.e., $\Lambda^{(\tau)}(\cdot)=\tr_{\env^\prime}\left[\ex^{\tau\mc{L}^\prime}(\rho_{\env^\prime}\otimes\cdot)\right]$, so that the dynamics could in general still contain global $\syst\env$ dissipators, albeit hidden implicitly through $H^\prime$. In other words, the presence or absence of dissipation in the dynamics as stated for Result~\ref{result: markovianized rb} is really a statement of non-unitarity (incoherence) or unitarity (full coherence) on $\syst\env$, respectively, of the noise map $\Lambda^{(\tau)}$.

\subsubsection{Non-ideal dynamical decoupling}
The assumption of instantaneous, infinitely strong \gls{dd} pulses has been studied extensively~\cite{LidarCDD, ezzell_2022}, with the main consequence being that decoupling is not achieved even at the first small-time-order; however, as long as the width of the pulses is sufficiently small (relative to the inverse magnitude of the largest singular value of the evolution Hamiltonian), decoupling is approximately achieved. So in a sense, finite-width \gls{dd} would still be effective in Markovianizing \gls{rb}, as long as pulse widths are sufficiently small.

Similarly, one can relax the assumption of noiseless \gls{dd} operators in a straightforward way: as long as the noise is approximately local, time-stationary and gate-independent, at each time-step of the \gls{rb} protocol, this would amount to a relabeling of the noise maps $\Lambda_i\to\Lambda_i^\prime$, and Eq.~\eqref{eq: noise strength Lind} would hold at least up to the second equality with a different map $\mc{D}_\syst^{\prime(\mathsf{dd})}$ accounting for noise in the \gls{dd}.

\subsubsection{Single-qubit subsystem}
For a single-qubit system $\syst$, the Pauli group $\mbb{V}=\{\mbb1,X,Y,Z\}$ can be employed with either the so-called $XY4$ or the $XZ4$ sequences. Letting $\mc{X}(\cdot):=X(\cdot)X$, and similarly for the remaining Pauli operators, the ideal $XY(Z)4$ sequence takes the form
\begin{equation}
\mc{Y}(\mc{Z})\circ\Lambda^{(\tau)}\circ\mc{X}\circ\Lambda^{(\tau)}\circ\mc{Y}(\mc{Z})\circ\Lambda^{(\tau)}\circ\mc{X}\circ\Lambda^{(\tau)},
\end{equation}
or equivalently, $\Lambda^{(\tau/2)}\mc{Y}(\mc{Z})\Lambda^{(\tau)}\mc{X}\Lambda^{(\tau)}\mc{Y}(\mc{Z})\Lambda^{(\tau)}\mc{X}\Lambda^{(\tau/2)}$, where we removed the composition symbols for simplicity. These are arguably the simplest \gls{udd} sequences; the reason why they implement the property in Eq.~\eqref{eq: decoupling def} is because, up to a phase, they are both equivalent to interleaving all of the single-qubit Pauli operators among the noise, which amounts to averaging its generators over the subsystem qubit to a first-time order, as 
shown in Appendix~\ref{appendix: rbdd ideal}. Other variants of the $XY(Z)4$ sequences achieving higher-order decoupling exist, such as Concatenated \gls{dd}~\cite{LidarCDD}, but we will not employ them here.

\subsubsection{\texorpdfstring{Numerical example and the decoupling time $\tau_\mathsf{dd}$}{Numerical example and the decoupling time "tau\_dd"}}\label{appendix: rbdd decoupling times}
The numerical example we consider in the main text is a 1$\syst$+1$\env$ qubit model with noise dynamics of the form  $\Lambda^{(t)}=\ex^{t\mc{L}}$, where $\mc{L}(\cdot):=-i[H,\cdot]+\mc{D}_\syst(\cdot)$ for $H=J XX + h_x (XI + IX) + h_y (YI + IY)$, and we choose the parameters $\tau_\mathsf{rb}=0.03$, $\rho_\syst=M=|0\rangle\!\langle0|$, $J=1.7$, $h_x=1.47$, $h_y=-1.05$ and Lindblad terms on the $\syst$-dissipator $L_0=X$, $\gamma_0=0.002$, $L_1=Z$ and $\gamma_1=0.007$. We approximated the action of the exponential in the Lindblad evolution up to $\mc{O}(t^{10})$.

\begin{figure}[ht!]
\centering
\includegraphics[width=\textwidth]{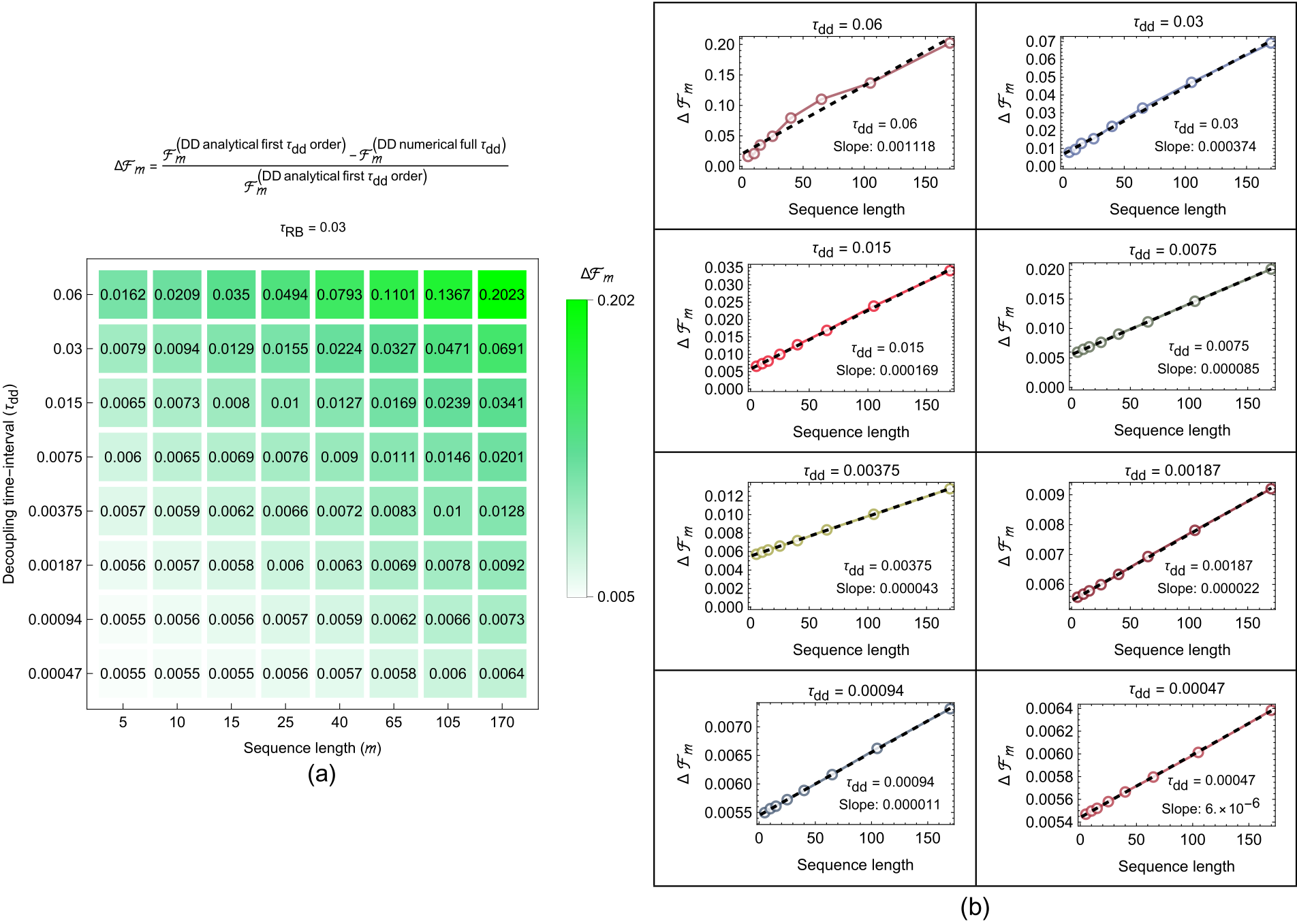}
\caption{\label{fig: tau_dd}\textbf{Behavior of the Markovianized full \gls{asf} with respect to the first time-order with $XY4$ in $\tau_\mathsf{dd}$}: We consider a relative difference, $\Delta\mc{F}_m$, of the full numerical \gls{asf} (i.e., to all time-orders) with single interleaved $XY4$ sequences with pulses on time-intervals $\tau_\mathsf{dd}$, without uncertainties, with respect to the analytical exponential first time-order, as given by Eq.~\eqref{eq: dd asf analytical result}; in (a) we plot the values of $\Delta\mc{F}_m$ with respect to $\tau_\mathsf{dd}$ (with $\tau_\mathsf{dd}\approx\tau_\mathsf{rb}$ and $\tau_\mathsf{dd}\ll\tau_\mathsf{rb}$) and sequence length $m$, showing it generally decreases together with $\tau_\mathsf{dd}$ and increases in $m$, while in (b) it is shown that $\Delta\mc{F}_m$ increases linearly in most cases by fitting a linear function with a rate proportionally small in $\tau_\mathsf{dd}$.}
\end{figure}

In particular, here we discuss the statement that \emph{how well \gls{dd} Markovianizes --whether it outputs decays closed to the purely exponential $\mc{O}(\tau_\mathsf{dd})$ decay--, depends mainly on whether the time-scale $\tau_\mathsf{dd}$ is small relative to that of the noise, $\tau_\mathsf{rb}$}. This can be assessed by the separation between the full-time-order \gls{asf} incorporating \gls{dd} and the corresponding first-time-order. This is easily tractable in our numerical example by inspecting this separation of the outputs in Fig.~\ref{fig: rb dd numerical} and/or other samples. This is done in Fig.~\ref{fig: tau_dd}, where the convergence of the \gls{asf} to the first time-order is clear, with the separation in sequence length being linear and suppressed in smaller $\tau_\mathsf{dd}$.

The behavior in larger, $\tau_\mathsf{dd}>2\tau_\mathsf{rb}$ can quickly become unreliable as uncertainties increase and the \gls{asf} drops quickly as the sequence length increases, so generally, in practice the timescale of the \gls{rb} sequences, $\tau_\mathsf{rb}$, can serve as the reference timescale with which $\tau_\mathsf{dd}$ should be minimized.

\section{Subsystem Pauli-twirled noise}\label{appendix: rc and rb}
Other Markovianizing strategies different from \gls{dd} could also be conceivable. In particular, techniques tailoring noise into a so-called \emph{Pauli channel} have been shown to reduce the impact of non-Markovian noise~\cite{PauliFR_2021, winick2022RC, Hashim_2023} in a statistically significant way.

Let us consider for now a closed $n$-qubit system. Let $\mbb{P}_n$ be the $n$-qubit Pauli group, i.e., that generated by $n$-fold Kronecker products of single-qubit Pauli operators. The so-called twirl, $\mc{T}_{\mbb{P}_n}$, over $\mbb{P}_n$, of a \gls{cp} map between $n$-qubits, $\Phi(\cdot):=\sum_\mu\phi_\mu(\cdot)\phi_\mu^\dg$, can be written as
\begin{align}   \mc{T}_{\mbb{P}_n}\left(\hat{\Phi}\right) &:= \f{1}{4^n}\sum_{P\in\mbb{P}_n}{\sum_{\mu}} P^\dg\phi_\mu{P}\otimes{P}^\mathrm{T}\phi_\mu^*{P}^*,
\end{align}
where $\hat{\Phi}:=\sum_\mu\phi_\mu\otimes\phi_\mu^*$ the vectorized form of $\Phi$, as defined in Appendix~\ref{appendix: notation}. Hence, expanding the Kraus operators in the Pauli basis, $\phi_\mu=\sum_i\alpha_i^\mu{P}_i$, we can rewrite the Pauli-twirled channel as
\begin{align}   \hat{\Phi}^{\mathsf{P}}:=\mc{T}_{\mbb{P}_n}\left(\hat{\Phi}\right) &= \f{1}{4^n}\sum\alpha_i^\mu\alpha_j^{*\mu}P_k^\dg{P}_i{P}_k\otimes{P}_k^\mathrm{T}P_j^*{P}_k^*=\sum{p}_iP_i\otimes{P}_i^*,
\label{eq: Pauli channel}
\end{align}
with $p_i:=\sum_\mu|\alpha_i^\mu|^2$. In general, any quantum channel of the form of Eq.~\eqref{eq: Pauli channel} is called an \emph{Pauli channel}. Such Pauli channel $\Phi^{\mathsf{P}}$ has Kraus operators $\sqrt{p_i}P_i$, where $p_i=\sum_\mu|\alpha_i^\mu|^2$ define a probability distribution satisfying $p_i\geq0$ and $\sum_ip_i\leq1$ (saturated if the channel $\hat{\Phi}$ is also \gls{tp}). Colloquially, then, we can say that Pauli-twirling a quantum channel, has the effect of tailoring it into a Pauli channel. In the specific case where we think of quantum channels as modeling quantum noise, we may refer to $p_i$ as the \emph{Pauli error-rates}, i.e., the probability distribution of any Pauli error from happening.

Throughout, we will reserve the zero\textsuperscript{th} Pauli to denote the identity, $P_0:=\mbb{1}$.

A particularly useful representation when dealing with multi-qubit systems is the so-called \gls{ptm} representation; in such representation, Pauli-twirled channels take a manifestly diagonal form. Explicitly, the \gls{ptm} representation of an $n$-qubit quantum channel $\Phi$, is also a vectorized form (i.e., matrix-form) of the channel, but expanded in the Pauli basis,
\begin{equation}
    R_\Phi := 2^{-n}\sum_{Q,P\in\mbb{P}_n}\llangle{Q}|\Phi(P)\rrangle\,|Q\rrangle\!\llangle{P}|,
\end{equation}
that is, the \gls{ptm} of $\Phi$ is defined by the elements 
\begin{equation}
    (R_\Phi)_{ij} = 2^{-n}\llangle{P_i}|\Phi(P_j)\rrangle = 2^{-n}\tr\left[P_i\,\Phi(P_j)\right],\qquad{P}_i,P_j\in\mbb{P}_n.
\end{equation}
In the case of a Pauli channel, $\Phi^{\mathsf{P}}$, the corresponding \gls{ptm} matrix $R_{\Phi^{\mathsf{P}}}$ is diagonal, with its (real) elements being related to the probabilities $p_i$ in Eq.~\eqref{eq: Pauli channel} by a so-called Walsh-Hadamard transform (see e.g.~\cite{flammia_aces}).

Remarkably, a Pauli-twirled noise channel has the same average gate-fidelity as the bare noise channel; this can be seen directly in Eq.~\eqref{eq: Markov asf}, i.e., $\mathfrak{F}_\Phi=\mathfrak{F}_{\Phi^\mathsf{P}}$; as the average gate-fidelity only takes into account the diagonal error rates. Naturally the same property holds if Pauli-twirling is done only on a subsystem. This, in turn, implies that any technique estimating average gate-fidelities will remain reliable under Pauli noise, and not greatly overestimate gate-quality by missing out on all other non-diagonal terms: indeed, the diamond norm for quantum channels with a given fidelity is minimized by Pauli channels~\cite{Sacchi_2005, rbconfidence_2014, Sanders_2015} and can be explicitly computed in such case through the trace-norm of the vector of Pauli error rates~\cite{PhysRevA.85.042311}. Pauli-twirling can be operationally implemented through protocols such as Pauli Frame Randomization~\cite{PFR_2005, PauliFR_2021} or Randomized Compiling~\cite{RC_2016, winick2022RC}.

\begin{figure}[t!]
    \centering
\includegraphics[width=0.9\textwidth]{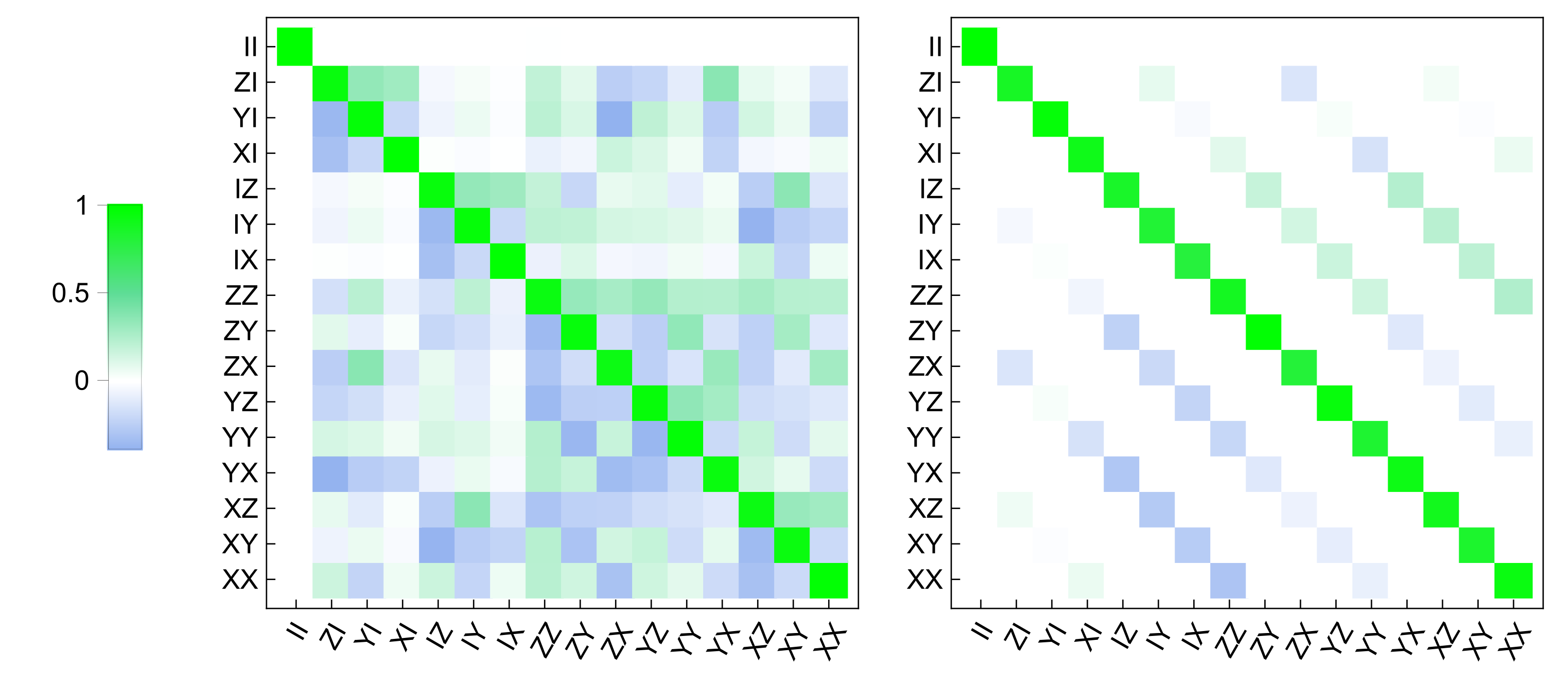}
    \caption{\textbf{Effect of twirling a subsystem with the Pauli group}: (Left) \gls{ptm} of the model used in the main text, $\Lambda=\exp(\tau\mc{L})$ where $\mc{L}(\cdot):=-i[H,\cdot]+\mc{D}_\syst(\cdot)$ for fixed $\tau=0.03$, $H=J XX + h_x (XI + IX) + h_y (YI + IY)$, and chosen parameters $\rho_\syst=M=|0\rangle\!\langle0|$, $J=1.7$, $h_x=1.47$, $h_y=-1.05$ and Lindblad terms $L_0=X$, $\gamma_0=0.002$, $L_1=Z$, $\gamma_1=0.007$; (Right) \gls{ptm} of the corresponding $\syst$-Pauli-twirled channel, $\Lambda^{\mathsf{P}_\syst}$.}
    \label{fig: subsystem twirl PTM}
\end{figure}

We can thus now consider our more general open-system scenario where our subsystem $\syst$ is a $n_s$-qubit system, with the limiting case of perfect twirling of some noise channel $\Lambda$ with the Pauli group solely on subsystem $\syst$. We refer to this as the $\syst$-Pauli-twirled channel, $\Lambda^\mathsf{P}$, associated to $\Lambda$. For simplicity, let us assume $\env$ is also multi-qubit; otherwise Pauli operators should be replaced with a corresponding $\env$ operator basis. Let $\hat{\Lambda}=\sum_\nu\lambda_\nu\otimes\lambda_\nu^*$, and expand the Kraus operators $\lambda_\nu$ in the Pauli basis as $\lambda_\nu:=\sum_{\mu,i}\alpha_{\mu,i}^\nu{P}_{\mu}\otimes{P}_i$. Define $\sum_\nu\alpha_{\mu,i}^\nu\alpha_{\sigma,j}^{*\nu}:=\chi_{\mu\sigma,ij}$. Then we can write~\textsuperscript{\footnote{ Strictly, the ordering in the vectorized (hat $\hat{\mc{X}}$) representation, as we defined it, is $\env\syst\otimes\env\syst$; we are slightly abusing notation for clarity of presentation.}}
\begin{align}
\hat{\Lambda}^{\mathsf{P}_\syst} := \left(\mc{I}_\env\otimes\mc{T}_{\mbb{P}_{n_\syst}}\right)\hat{\Lambda} &= \sum\chi_{\mu\sigma,ij}\left(P_\mu\otimes{P}_{\sigma}^*\right)\otimes\mc{T}_{\mbb{P}_{n_\syst}}\left(P_i\otimes{P}^*_j\right) \nonumber\\
&= \sum\chi_{\mu\sigma,ii}\left(P_\mu\otimes{P}_{\sigma}^*\right)\otimes\left(P_i\otimes{P}^*_i\right),
\label{eq: subsystem twirl}
\end{align}
which means that twirling solely over subsystem $\syst$ generates a Pauli-like channel on system $\syst$ that retains the dependence on the environment $\env$ only on the local Pauli error rates. The corresponding \gls{ptm} is not diagonal, despite nullifying many entries on the $\syst$ subspace, as exemplified in Fig.~\ref{fig: subsystem twirl PTM}. If the environment is traced out, then the reduced channel is a Pauli channel with $\env$-dependent Pauli error rates, $\tr_\env[\Lambda^{\mathsf{P}_\syst}(\cdot)]\propto\sum{p}_i(\rho^\prime_\env)P_i\otimes{P}_i$ for some reduced state of the environment $\rho^\prime_\env$; in multi-time processes, however, non-Markovianity is still carried through this environmental dependence. 

\subsection{\texorpdfstring{\gls{asf} under $\syst$-Pauli-twirled noise}{ASF under S-Pauli-twirled noise}}\label{appendix: asf Pauli-twirled}

Consider the limiting case where Pauli-twirling can be implemented effectively perfectly, via Randomized Compiling or otherwise, alongside \gls{rb} for non-Markovian noise, so that quality maps $\mc{Q}_{\pi,m}$ are defined in terms of maps $\hat{\Lambda}_i^{\mathsf{P}_\syst}$ of the form of Eq.~\eqref{eq: subsystem twirl}.

We now denote time-step indices as superscripts in $\chi$ and as a labeling for each sum index, i.e.,
\begin{equation}
\hat{\Lambda}^{\mathsf{P}_\syst}_{i}=\sum\chi_{\mu_i\sigma_i,k_ik_i}^{(i)}\left(P_{\mu_i}\otimes{P}_{\sigma_i}^*\right)\otimes\left(P_{k_i}\otimes{P}_{k_i}^{*}\right),
\end{equation}
then, plugging a set of $\{\Lambda_1^{\mathsf{P}_\syst},\ldots,\Lambda_m^{\mathsf{P}_\syst}\}$ noise maps into a quality map of length $m$ (defined in Eq.~\eqref{eq: quality factor}), denoting by $\{|\epsilon_i\rangle,|\epsilon_i^\prime\}_{i=0}^m$ basis vectors of $\env$, we have 
\begin{align}
    \hat{\mc{Q}}_{m,\pi} &= \sum_{\{\epsilon_j,\epsilon_j^\prime\}_{j=0}^{m}}^{\dimE} \f{\prod_{i=m-1}^{0}\tr\left(\langle\epsilon_{i+1}\epsilon_{i+1}^\prime|\hat{\Lambda}^{\mathsf{P}_\syst}_{i+1}|\epsilon_{i}\epsilon_{i}^\prime\rangle\hat{\mc{P}}_\pi\right)}{\tr\left(\hat{\mc{P}}_\pi\right)^m}|\epsilon_m\epsilon_m^\prime\rangle\!\langle\epsilon_0\epsilon_0^\prime| \nonumber\\
    &= \sum \f{\prod_{i=m-1}^{0}\chi^{(i+1)}_{\mu_{i+1}\sigma_{i+1},{k}_{i+1}{k}_{i+1}}\langle\epsilon_{i+1}|P_{\mu_{i+1}}|\epsilon_{i}\rangle\!\langle\epsilon_{i+1}^\prime|P^{*}_{\sigma_{i+1}}|\epsilon_{i}^\prime\rangle\tr\left[\left(P_{k_{i+1}}\otimes{P}^{*}_{k_{i+1}}\right)\hat{\mc{P}}_\pi\right]}{\tr\left(\hat{\mc{P}}_\pi\right)^m}|\epsilon_m\epsilon_m^\prime\rangle\!\langle\epsilon_0\epsilon_0^\prime| \nonumber\\
    &= \sum \f{\prod_{i=m-1}^{0}\chi^{(i+1)}_{\mu_{i+1}\sigma_{i+1},{k}_{i+1}{k}_{i+1}}\tr\left[\left(P_{k_{i+1}}\otimes{P}^{*}_{k_{i+1}}\right)\hat{\mc{P}}_\pi\right]}{\tr\left(\hat{\mc{P}}_\pi\right)^m}\left(P_{\mu_m}\cdots{P}_{k_1}\otimes{P}^{*}_{\sigma_m}\cdots{P}^{*}_{\sigma_1}\right),
\end{align}
which we can now plug-in into the \gls{asf} of Eq.~\eqref{eq: non-Markov asf}. Now, assuming that the \gls{spam} noise maps $\Lambda_{m+1}$ and $\Lambda_0$ have also been Pauli-twirled to $\Lambda_{m+1}^{\mathsf{P}_\syst}$ and $\Lambda_0^{\mathsf{P}_\syst}$, respectively, this gives us the $\env$-trace term
\begin{align}
\hat{\tr}\left(P_{\mu_{m+1}}\cdots{P}_{\mu_0}\otimes{P}^{*}_{\sigma_{m+1}}\cdots{P}^{*}_{\sigma_0}\right)|\rho_\env\rrangle = \tr\left[P_{\mu_{m+1}}\cdots{P}_{\mu_0}(\rho_\env){P}^{\dg}_{\sigma_0}\cdots{P}^{\dg}_{\sigma_{m+1}}\right] := \xi_m(\vec{\mu},\vec{\sigma}),
\end{align}
which is generally different from 1, and effectively will equal (up to a phase) the expectation of a Pauli observable on $\rho_\env$, which we call $\xi_m(\vec{\mu},\vec{\sigma})$, with the arguments $\vec{\mu},\vec{\sigma}$ containing all $\mu$ and $\sigma$ indices. The value of this expectation just depends on how the respective Paulis (anti)commute. Were $\env$ not a multi-qubit system, we would have the equivalent statement in terms of the respective $\env$ basis operators. Thus, the \gls{asf} has the form
\begin{align}
    \mc{F}_m 
    &= \sum_{\vec{\mu},\vec{\sigma},\vec{k}}\xi_m(\vec{\mu},\vec{\sigma})\sum_{\pi\in{R}_{\mbb{G}}\prod_{i=1}^{m}}\left(\f{\tr\left[\hat{\Phi}_{m}^{(\mu\sigma,kk)}\hat{\mc{P}}_\pi\right]}{\tr\hat{\mc{P}}_\pi}\right)\cdots\left(\f{\tr\left[\hat{\Phi}_1^{(\mu\sigma,kk)}\hat{\mc{P}}_\pi\right]}{\tr\hat{\mc{P}}_\pi}\right)\llangle{M}|\hat{\Phi}_{m+1}^{(\mu\sigma,kk)}\hat{\mc{P}}_\pi\hat{\Phi}_0^{(\mu\sigma,kk)}|\rho_\syst\rrangle,
    \label{eq: asf subsyst Pauli-twirl}
\end{align}
where we defined
\begin{equation}
\hat{\Phi}_i^{(\mu\sigma,jk)}:=\chi^{(i)}_{\mu_i\sigma_i,k_ij_i}P_{j_i}\otimes{P}^{*}_{k_i}.
\end{equation}

Notice that Eq.~\eqref{eq: asf subsyst Pauli-twirl}, i.e., the \gls{asf} with $\syst$-Pauli-twirled noise, is just the particular case with $\hat{\Phi}_i^{(\mu\sigma,kk)}$ for all $i$. That is, we could have done the same procedure with $\Lambda_i$ instead of $\Lambda_i^{\mathsf{P}_\syst}$, simply considering all remaining $\Phi$ terms. Furthermore, memory contributing to deviations in a non-Markovian \gls{asf}, and hence the non-trivial behavior of average gate-fidelities, remains the same; however, the impact these have on the \gls{asf} can be different for either case.

For concreteness, consider the case for \gls{rb} with multi-qubit Clifford gates, so that
\begin{align}
    \mc{F}_m = \sum_{\vec{\mu},\vec{\sigma},\vec{k}}\xi_m(\vec{\mu},\vec{\sigma}) \left(A_{0,m+1}^{(\mu\sigma,kk)} p_1^{(\mu\sigma)}p_2^{(\mu\sigma)}\cdots{p}_m^{(\mu\sigma)} + B_{0,m+1}^{(\mu\sigma,kk)}\right),
    \label{eq: nM asf time-dep form (appendix)}
\end{align}
where
\begin{align}
p_i^{(\mu\sigma)} = \f{\tr\left(\hat{\Phi}_i^{(\mu\sigma,kk)}\right)-1}{d_\syst^2-1} = \f{\chi^{(i)}_{\mu\sigma,00}-1}{\dimS^2-1},
\label{eq: nM p time-dep form (appendix)}
\end{align}
as all Paulis are traceless except the identity~\textsuperscript{\footnote{ Notice that the time-stationary case, $\Lambda_i=\Lambda$ for all $i$, would leave the \gls{asf} in a \emph{time-non-stationary} noise form due to the environment dependence, as we would drop all single $i$ indices, but the remaining $\env$ indices, $\mu,\sigma$, and thus all quality factors $p^{(\mu\sigma)}$ too, would remain distinct for each time-step.}}, while the \gls{spam} terms are
\begin{align}
A^{(\mu\sigma,kk)}_{0,m+1} &= \llangle{M}|\hat{\Phi}_{m+1}^{(\mu\sigma,kk)}\hat{\Phi}_{0}^{(\mu\sigma,kk)}|\rho_\syst-\f{\mbb1}{\dimS}\rrangle\nonumber\\
&= \chi^{(0)}_{\mu_0\sigma_0,k_0k_0}\chi^{(m+1)}_{\mu_{m+1}\sigma_{m+1},k_{m+1}k_{m+1}}\llangle{M}|(P_{k_{m+1}}P_{k_0}\otimes{P}_{k_{m+1}}^*P_{k_0}^*)|\,\rho_\syst-\f{\mbb1}{\dimS}\rrangle\nonumber\\
&= \chi^{(0)}_{\mu_0\sigma_0,k_0k_0}\chi^{(m+1)}_{\mu_{m+1}\sigma_{m+1},k_{m+1}k_{m+1}}\left(\llangle{M}|(P_{k_{m+1}}P_{k_0}\otimes{P}_{k_{m+1}}^*P_{k_0}^*)|\,\rho_\syst\rrangle-\f{\tr(M)}{\dimS}\right),\\
B^{(\mu\sigma,kk)}_{0,m+1} &= \llangle{M}|\hat{\Phi}^{(\mu\sigma,kk)}\hat{\Phi}^{(\mu\sigma,kk)}|\f{\mbb1}{\dimS}\rrangle = \chi^{(0)}_{\mu_0\sigma_0,k_0k_0}\chi^{(m+1)}_{\mu_{m+1}\sigma_{m+1},k_{m+1}k_{m+1}}\f{\tr(M)}{\dimS}.
\end{align}

In the Clifford case, notice that the quality factors $p_i^{(\mu\sigma)}$ are manifestly independent of the Pauli indices: this is just a consequence of average gate-fidelity only taking into account the Pauli error rate corresponding to the identity on system $\syst$, i.e., the probability of no error happening on $\syst$. This means these terms remain the same for an \gls{asf} with noise $\Lambda$ that has not been $\syst$-Pauli-twirled. The \gls{spam} terms, however, remain generally distinct and fully dependent on all Pauli indices.

In this sense, we can say that non-Markovian Clifford \gls{asf} decays under bare noise $\Lambda$ and under $\syst$-Pauli-twirled noise $\Lambda^\mathsf{P}$ only differ by non-Markovian \gls{spam} terms, as we could simply replace the $A$ and $B$ terms in Eq.~\eqref{eq: nM asf time-dep form (appendix)} with $A^{(\mu\sigma,jk)}_{0,m+1}$ and $B^{(\mu\sigma,jk)}_{0,m+1}$, summed over $j$, to obtain the \gls{asf} under $\Lambda$. This result is expressed as Eq. (\ref{eq: pauli asf (main)}) in Result~\ref{result: rc and rb} in the main text.

\subsection{\texorpdfstring{Variance Sequence Fidelity under $\syst$-Pauli-twirled noise}{Variance Sequence Fidelity under S-Pauli-twirled noise}}

While Pauli-twirling on a subspace does not necessarily Markovianize the \gls{asf} in the sense of removing non-exponential deviations, it must certainly reduce its statistical impact on other figures of merit sensible to all other error rates, similar to the case of full Pauli-twirling. As previously mentioned, this was pointed out in~\cite{PauliFR_2021, winick2022RC}. Indeed, within \gls{rb}, it is still unclear whether Pauli-twirling generally can increase the \gls{asf} or in which cases it is sufficient to remove non-exponential deviations.

However, a noticeable effect of Pauli-twirled noise is to have a reduced variance sequence fidelity. This can be understood intuitively because the variance over gates of \gls{rb} sequences, $\mathrm{f}_m=\llangle{M}|\hat{\tr}_\env\hat{\Lambda}_{m+1}\hat{\mc{G}}_{m+1}\cdots\hat{\Lambda}_1\hat{\mc{G}}_1|\tilde{\rho}\rrangle$ is by definition $\mathbf{V}\left(\mathrm{f}_m\right):=\mathbf{E}\left(\mathrm{f}_m^2\right)-\mc{F}_m^2$, and the first term, $\mathbf{E}\left(\mathrm{f}_m^2\right)$, is precisely proportional to the so-called \emph{unitarity} of the noise~\cite{Wallman_2015}, which is generally very small for Pauli-noise (albeit not necessarily zero).

We can make this statement mathematically concrete by decomposing $\hat{\Lambda}_i:=\hat{\Lambda}_i^{\mathsf{nd}}+\hat{\Lambda}_i^{\mathsf{P}_\syst}$, where
\begin{equation}
\hat{\Lambda}^{\mathsf{nd}}_{i}=\sum_{j\neq{k}}\chi_{\mu_i\sigma_i,j_ik_i}^{(i)}\left(P_{\mu_i}\otimes{P}_{\sigma_i}^*\right)\otimes\left(P_{j_i}\otimes{P}_{k_i}^{*}\right),
\end{equation}
is the operator corresponding to the map with all the off-diagonal terms in $\syst$. Denoting explicitly the dependence on the noise for the sequence fidelity $\mathrm{f}_m$, we can then write
\begin{align}
\mathrm{f}_m(\Lambda_0,\Lambda_1,\ldots,\Lambda_{m+1}) &= \mathrm{f}_m(\Lambda_0^{\mathsf{nd}},\Lambda_1,\ldots,\Lambda_{m+1}) + \mathrm{f}_m(\Lambda_0^{\mathsf{P}_\syst},\Lambda_1,\ldots,\Lambda_{m+1}),\\
\mathbf{V}[\mathrm{f}_m(\Lambda_0,\Lambda_1,\ldots,\Lambda_{m+1})] &= \mathbf{V}[\mathrm{f}_m(\Lambda_0^{\mathsf{nd}},\Lambda_1,\ldots,\Lambda_{m+1})] + \mathbf{V}[\mathrm{f}_m(\Lambda_0^{\mathsf{P}_\syst},\Lambda_1,\ldots,\Lambda_{m+1})],
\end{align}
and we can keep expanding all $\Lambda_i$ terms to get, informally,
\begin{equation}
    \mathbf{V}[\mathrm{f}_m(\Lambda_0,\Lambda_1,\ldots,\Lambda_{m+1})] = \mathbf{V}[\mathrm{f}_m(\Lambda_0^{\mathsf{P}_\syst},\Lambda_1^{\mathsf{P}_\syst},\ldots,\Lambda_{m+1}^{\mathsf{P}_\syst})] + \sum\mathbf{V}[\mathrm{f}_m(\text{at least one }\Lambda_i^{\mathsf{nd}}\text{, rest }\Lambda_j^{\mathsf{P}_\syst})].
\end{equation}

All fidelities ``$\mathrm{f}_m(\text{at least one }\Lambda_i^{\mathsf{nd}}\text{, rest }\Lambda_j^{\mathsf{P}_\syst})$'' are real numbers because $\Lambda_i^{\mathsf{nd}}$, the Pauli non-diagonal contribution to $\Lambda$, remains a Hermitian map. This implies that $\mathbf{V}[\mathrm{f}_m(\text{at least one }\Lambda_i^{\mathsf{nd}}\text{, rest }\Lambda_j^{\mathsf{P}_\syst})]\geq0$, and so it follows that
\begin{equation}
    \mathbf{V}[\mathrm{f}_m(\Lambda_0,\Lambda_1,\ldots,\Lambda_{m+1})] \geq \mathbf{V}[\mathrm{f}_m(\Lambda_0^{\mathsf{P}_\syst},\Lambda_1^{\mathsf{P}_\syst},\ldots,\Lambda_{m+1}^{\mathsf{P}_\syst})].
    \label{eq: variance pauli ineq (appendix)}
\end{equation}

This statement remains true for more general $\mathrm{f}_m$, in particular for any expectation value, not necessarily a \gls{rb} sequence fidelity. Notice the same cannot be said, however, about the \gls{asf} because $\Lambda_i^{\mathsf{nd}}$ does not necessarily remain completely-positive or even simply positive~\textsuperscript{\footnote{ (Complete) Positivity of a quantum channel $\Lambda$ does not correspond to positive semi-definiteness of $\hat{\Lambda}$; however it does correspond to positive semi-definiteness of the so-called Choi state of $\Lambda$, which can be defined as $\Upsilon_\Lambda:=(\Lambda\otimes\mc{I})\Psi$, where $\Psi=\sum|ii\rangle\!\langle{jj}|$ is a maximally entangled operator on $\syst^{\otimes2}$.}}, so the averages of these sequence fidelities can generally be negative too.

We exemplify the effect of Eq.~\eqref{eq: variance pauli ineq (appendix)} in the main text in Fig.~\ref{fig: rb rc numerical}, where for Pauli-twirled noise, in contrast to the bare noise data, and despite of still displaying non-exponential deviations, the sampled averages from the \gls{rb} protocol fall almost exactly at the true analytical average. While this example was done with a single-qubit Clifford group, the statement for the variance being reduced under Pauli-twirl is independent of the gate set.

This effect in the variance due to smaller unitarity was already pointed out in~\cite{Wallman_2015}, albeit not directly in terms of Pauli noise or a non-Markovian scenario. However, as pointed out there too, deriving a concrete bound is less straightforward.

Given the relatively low overhead of techniques enforcing Pauli-twirled noise, such as Randomized Compiling~\cite{RC_superc_2021}, and that of the various versions of \gls{dd}~\cite{ezzell_2022}, it is conceivable that both techniques or an optimal version thereof could be implemented efficiently to simultaneously exploit the benefits of both.


\begin{thebibliography}{94}%
\makeatletter
\providecommand \@ifxundefined [1]{%
 \@ifx{#1\undefined}
}%
\providecommand \@ifnum [1]{%
 \ifnum #1\expandafter \@firstoftwo
 \else \expandafter \@secondoftwo
 \fi
}%
\providecommand \@ifx [1]{%
 \ifx #1\expandafter \@firstoftwo
 \else \expandafter \@secondoftwo
 \fi
}%
\providecommand \natexlab [1]{#1}%
\providecommand \enquote  [1]{``#1''}%
\providecommand \bibnamefont  [1]{#1}%
\providecommand \bibfnamefont [1]{#1}%
\providecommand \citenamefont [1]{#1}%
\providecommand \href@noop [0]{\@secondoftwo}%
\providecommand \href [0]{\begingroup \@sanitize@url \@href}%
\providecommand \@href[1]{\@@startlink{#1}\@@href}%
\providecommand \@@href[1]{\endgroup#1\@@endlink}%
\providecommand \@sanitize@url [0]{\catcode `\\12\catcode `\$12\catcode
  `\&12\catcode `\#12\catcode `\^12\catcode `\_12\catcode `\%12\relax}%
\providecommand \@@startlink[1]{}%
\providecommand \@@endlink[0]{}%
\providecommand \url  [0]{\begingroup\@sanitize@url \@url }%
\providecommand \@url [1]{\endgroup\@href {#1}{\urlprefix }}%
\providecommand \urlprefix  [0]{URL }%
\providecommand \Eprint [0]{\href }%
\providecommand \doibase [0]{http://dx.doi.org/}%
\providecommand \selectlanguage [0]{\@gobble}%
\providecommand \bibinfo  [0]{\@secondoftwo}%
\providecommand \bibfield  [0]{\@secondoftwo}%
\providecommand \translation [1]{[#1]}%
\providecommand \BibitemOpen [0]{}%
\providecommand \bibitemStop [0]{}%
\providecommand \bibitemNoStop [0]{.\EOS\space}%
\providecommand \EOS [0]{\spacefactor3000\relax}%
\providecommand \BibitemShut  [1]{\csname bibitem#1\endcsname}%
\let\auto@bib@innerbib\@empty
\bibitem [{\citenamefont {Eisert}\ \emph {et~al.}(2020)\citenamefont {Eisert},
  \citenamefont {Hangleiter}, \citenamefont {Walk}, \citenamefont {Roth},
  \citenamefont {Markham}, \citenamefont {Parekh}, \citenamefont {Chabaud},\
  and\ \citenamefont {Kashefi}}]{kashefi_2020}%
  \BibitemOpen
  \bibfield  {author} {\bibinfo {author} {\bibfnamefont {J.}~\bibnamefont
  {Eisert}}, \bibinfo {author} {\bibfnamefont {D.}~\bibnamefont {Hangleiter}},
  \bibinfo {author} {\bibfnamefont {N.}~\bibnamefont {Walk}}, \bibinfo {author}
  {\bibfnamefont {I.}~\bibnamefont {Roth}}, \bibinfo {author} {\bibfnamefont
  {D.}~\bibnamefont {Markham}}, \bibinfo {author} {\bibfnamefont
  {R.}~\bibnamefont {Parekh}}, \bibinfo {author} {\bibfnamefont
  {U.}~\bibnamefont {Chabaud}}, \ and\ \bibinfo {author} {\bibfnamefont
  {E.}~\bibnamefont {Kashefi}},\ }\bibfield  {title} {\enquote {\bibinfo
  {title} {Quantum certification and benchmarking},}\ }\href {\doibase
  10.1038/s42254-020-0186-4} {\bibfield  {journal} {\bibinfo  {journal} {Nat.
  Rev. Phys.}\ }\textbf {\bibinfo {volume} {2}},\ \bibinfo {pages} {382–390}
  (\bibinfo {year} {2020})}\BibitemShut {NoStop}%
\bibitem [{\citenamefont {Kliesch}\ and\ \citenamefont
  {Roth}(2021)}]{QCVV_2021}%
  \BibitemOpen
  \bibfield  {author} {\bibinfo {author} {\bibfnamefont {M.}~\bibnamefont
  {Kliesch}}\ and\ \bibinfo {author} {\bibfnamefont {I.}~\bibnamefont {Roth}},\
  }\bibfield  {title} {\enquote {\bibinfo {title} {Theory of quantum system
  certification},}\ }\href {\doibase 10.1103/PRXQuantum.2.010201} {\bibfield
  {journal} {\bibinfo  {journal} {PRX Quantum}\ }\textbf {\bibinfo {volume}
  {2}},\ \bibinfo {pages} {010201} (\bibinfo {year} {2021})}\BibitemShut
  {NoStop}%
\bibitem [{\citenamefont {Emerson}\ \emph {et~al.}(2005)\citenamefont
  {Emerson}, \citenamefont {Alicki},\ and\ \citenamefont
  {{\.{Z}}yczkowski}}]{Emerson_2005}%
  \BibitemOpen
  \bibfield  {author} {\bibinfo {author} {\bibfnamefont {J.}~\bibnamefont
  {Emerson}}, \bibinfo {author} {\bibfnamefont {R.}~\bibnamefont {Alicki}}, \
  and\ \bibinfo {author} {\bibfnamefont {K.}~\bibnamefont {{\.{Z}}yczkowski}},\
  }\bibfield  {title} {\enquote {\bibinfo {title} {Scalable noise estimation
  with random unitary operators},}\ }\href {\doibase
  10.1088/1464-4266/7/10/021} {\bibfield  {journal} {\bibinfo  {journal} {J.
  Optics B: Quantum Semiclass. Opt.}\ }\textbf {\bibinfo {volume} {7}},\
  \bibinfo {pages} {S347} (\bibinfo {year} {2005})}\BibitemShut {NoStop}%
\bibitem [{\citenamefont {Magesan}\ \emph {et~al.}(2012)\citenamefont
  {Magesan}, \citenamefont {Gambetta},\ and\ \citenamefont
  {Emerson}}]{PhysRevA.85.042311}%
  \BibitemOpen
  \bibfield  {author} {\bibinfo {author} {\bibfnamefont {E.}~\bibnamefont
  {Magesan}}, \bibinfo {author} {\bibfnamefont {J.~M.}\ \bibnamefont
  {Gambetta}}, \ and\ \bibinfo {author} {\bibfnamefont {J.}~\bibnamefont
  {Emerson}},\ }\bibfield  {title} {\enquote {\bibinfo {title} {Characterizing
  quantum gates via randomized benchmarking},}\ }\href {\doibase
  10.1103/PhysRevA.85.042311} {\bibfield  {journal} {\bibinfo  {journal} {Phys.
  Rev. A}\ }\textbf {\bibinfo {volume} {85}},\ \bibinfo {pages} {042311}
  (\bibinfo {year} {2012})}\BibitemShut {NoStop}%
\bibitem [{\citenamefont {Knill}\ \emph {et~al.}(2008)\citenamefont {Knill},
  \citenamefont {Leibfried}, \citenamefont {Reichle}, \citenamefont {Britton},
  \citenamefont {Blakestad}, \citenamefont {Jost}, \citenamefont {Langer},
  \citenamefont {Ozeri}, \citenamefont {Seidelin},\ and\ \citenamefont
  {Wineland}}]{Knill_2008}%
  \BibitemOpen
  \bibfield  {author} {\bibinfo {author} {\bibfnamefont {E.}~\bibnamefont
  {Knill}}, \bibinfo {author} {\bibfnamefont {D.}~\bibnamefont {Leibfried}},
  \bibinfo {author} {\bibfnamefont {R.}~\bibnamefont {Reichle}}, \bibinfo
  {author} {\bibfnamefont {J.}~\bibnamefont {Britton}}, \bibinfo {author}
  {\bibfnamefont {R.~B.}\ \bibnamefont {Blakestad}}, \bibinfo {author}
  {\bibfnamefont {J.~D.}\ \bibnamefont {Jost}}, \bibinfo {author}
  {\bibfnamefont {C.}~\bibnamefont {Langer}}, \bibinfo {author} {\bibfnamefont
  {R.}~\bibnamefont {Ozeri}}, \bibinfo {author} {\bibfnamefont
  {S.}~\bibnamefont {Seidelin}}, \ and\ \bibinfo {author} {\bibfnamefont
  {D.~J.}\ \bibnamefont {Wineland}},\ }\bibfield  {title} {\enquote {\bibinfo
  {title} {Randomized benchmarking of quantum gates},}\ }\href {\doibase
  10.1103/PhysRevA.77.012307} {\bibfield  {journal} {\bibinfo  {journal} {Phys.
  Rev. A}\ }\textbf {\bibinfo {volume} {77}},\ \bibinfo {pages} {012307}
  (\bibinfo {year} {2008})}\BibitemShut {NoStop}%
\bibitem [{\citenamefont {Gambetta}\ \emph {et~al.}(2012)\citenamefont
  {Gambetta}, \citenamefont {C\'orcoles}, \citenamefont {Merkel}, \citenamefont
  {Johnson}, \citenamefont {Smolin}, \citenamefont {Chow}, \citenamefont
  {Ryan}, \citenamefont {Rigetti}, \citenamefont {Poletto}, \citenamefont
  {Ohki}, \citenamefont {Ketchen},\ and\ \citenamefont
  {Steffen}}]{PhysRevLett.109.240504}%
  \BibitemOpen
  \bibfield  {author} {\bibinfo {author} {\bibfnamefont {J.~M.}\ \bibnamefont
  {Gambetta}}, \bibinfo {author} {\bibfnamefont {A.~D.}\ \bibnamefont
  {C\'orcoles}}, \bibinfo {author} {\bibfnamefont {S.~T.}\ \bibnamefont
  {Merkel}}, \bibinfo {author} {\bibfnamefont {B.~R.}\ \bibnamefont {Johnson}},
  \bibinfo {author} {\bibfnamefont {J.~A.}\ \bibnamefont {Smolin}}, \bibinfo
  {author} {\bibfnamefont {J.~M.}\ \bibnamefont {Chow}}, \bibinfo {author}
  {\bibfnamefont {C.~A.}\ \bibnamefont {Ryan}}, \bibinfo {author}
  {\bibfnamefont {C.}~\bibnamefont {Rigetti}}, \bibinfo {author} {\bibfnamefont
  {S.}~\bibnamefont {Poletto}}, \bibinfo {author} {\bibfnamefont {T.~A.}\
  \bibnamefont {Ohki}}, \bibinfo {author} {\bibfnamefont {M.~B.}\ \bibnamefont
  {Ketchen}}, \ and\ \bibinfo {author} {\bibfnamefont {M.}~\bibnamefont
  {Steffen}},\ }\bibfield  {title} {\enquote {\bibinfo {title}
  {Characterization of addressability by simultaneous randomized
  benchmarking},}\ }\href {\doibase 10.1103/PhysRevLett.109.240504} {\bibfield
  {journal} {\bibinfo  {journal} {Phys. Rev. Lett.}\ }\textbf {\bibinfo
  {volume} {109}},\ \bibinfo {pages} {240504} (\bibinfo {year}
  {2012})}\BibitemShut {NoStop}%
\bibitem [{\citenamefont {Wallman}\ \emph {et~al.}(2015)\citenamefont
  {Wallman}, \citenamefont {Granade}, \citenamefont {Harper},\ and\
  \citenamefont {Flammia}}]{Wallman_2015}%
  \BibitemOpen
  \bibfield  {author} {\bibinfo {author} {\bibfnamefont {J.}~\bibnamefont
  {Wallman}}, \bibinfo {author} {\bibfnamefont {C.}~\bibnamefont {Granade}},
  \bibinfo {author} {\bibfnamefont {R.}~\bibnamefont {Harper}}, \ and\ \bibinfo
  {author} {\bibfnamefont {S.~T.}\ \bibnamefont {Flammia}},\ }\bibfield
  {title} {\enquote {\bibinfo {title} {Estimating the coherence of noise},}\
  }\href {\doibase 10.1088/1367-2630/17/11/113020} {\bibfield  {journal}
  {\bibinfo  {journal} {New J. Phys.}\ }\textbf {\bibinfo {volume} {17}},\
  \bibinfo {pages} {113020} (\bibinfo {year} {2015})}\BibitemShut {NoStop}%
\bibitem [{\citenamefont {Wood}\ and\ \citenamefont
  {Gambetta}(2018)}]{PhysRevA.97.032306}%
  \BibitemOpen
  \bibfield  {author} {\bibinfo {author} {\bibfnamefont {C.~J.}\ \bibnamefont
  {Wood}}\ and\ \bibinfo {author} {\bibfnamefont {J.~M.}\ \bibnamefont
  {Gambetta}},\ }\bibfield  {title} {\enquote {\bibinfo {title} {Quantification
  and characterization of leakage errors},}\ }\href {\doibase
  10.1103/PhysRevA.97.032306} {\bibfield  {journal} {\bibinfo  {journal} {Phys.
  Rev. A}\ }\textbf {\bibinfo {volume} {97}},\ \bibinfo {pages} {032306}
  (\bibinfo {year} {2018})}\BibitemShut {NoStop}%
\bibitem [{\citenamefont {Proctor}\ \emph {et~al.}(2022)\citenamefont
  {Proctor}, \citenamefont {Seritan}, \citenamefont {Rudinger}, \citenamefont
  {Nielsen}, \citenamefont {Blume-Kohout},\ and\ \citenamefont
  {Young}}]{proctor_2021}%
  \BibitemOpen
  \bibfield  {author} {\bibinfo {author} {\bibfnamefont {T.}~\bibnamefont
  {Proctor}}, \bibinfo {author} {\bibfnamefont {S.}~\bibnamefont {Seritan}},
  \bibinfo {author} {\bibfnamefont {K.}~\bibnamefont {Rudinger}}, \bibinfo
  {author} {\bibfnamefont {E.}~\bibnamefont {Nielsen}}, \bibinfo {author}
  {\bibfnamefont {R.}~\bibnamefont {Blume-Kohout}}, \ and\ \bibinfo {author}
  {\bibfnamefont {K.}~\bibnamefont {Young}},\ }\bibfield  {title} {\enquote
  {\bibinfo {title} {Scalable randomized benchmarking of quantum computers
  using mirror circuits},}\ }\href {\doibase 10.1103/PhysRevLett.129.150502}
  {\bibfield  {journal} {\bibinfo  {journal} {Phys. Rev. Lett.}\ }\textbf
  {\bibinfo {volume} {129}},\ \bibinfo {pages} {150502} (\bibinfo {year}
  {2022})}\BibitemShut {NoStop}%
\bibitem [{\citenamefont {Erhard}\ \emph {et~al.}(2019)\citenamefont {Erhard},
  \citenamefont {Wallman}, \citenamefont {Postler}, \citenamefont {Meth},
  \citenamefont {Stricker}, \citenamefont {Martinez}, \citenamefont
  {Schindler}, \citenamefont {Monz}, \citenamefont {Emerson},\ and\
  \citenamefont {Blatt}}]{cycle_2019}%
  \BibitemOpen
  \bibfield  {author} {\bibinfo {author} {\bibfnamefont {A.}~\bibnamefont
  {Erhard}}, \bibinfo {author} {\bibfnamefont {J.~J.}\ \bibnamefont {Wallman}},
  \bibinfo {author} {\bibfnamefont {L.}~\bibnamefont {Postler}}, \bibinfo
  {author} {\bibfnamefont {M.}~\bibnamefont {Meth}}, \bibinfo {author}
  {\bibfnamefont {R.}~\bibnamefont {Stricker}}, \bibinfo {author}
  {\bibfnamefont {E.~A.}\ \bibnamefont {Martinez}}, \bibinfo {author}
  {\bibfnamefont {P.}~\bibnamefont {Schindler}}, \bibinfo {author}
  {\bibfnamefont {T.}~\bibnamefont {Monz}}, \bibinfo {author} {\bibfnamefont
  {J.}~\bibnamefont {Emerson}}, \ and\ \bibinfo {author} {\bibfnamefont
  {R.}~\bibnamefont {Blatt}},\ }\bibfield  {title} {\enquote {\bibinfo {title}
  {Characterizing large-scale quantum computers via cycle benchmarking},}\
  }\href {\doibase 10.1038/s41467-019-13068-7} {\bibfield  {journal} {\bibinfo
  {journal} {Nat. Commun.}\ }\textbf {\bibinfo {volume} {10}} (\bibinfo {year}
  {2019}),\ 10.1038/s41467-019-13068-7}\BibitemShut {NoStop}%
\bibitem [{\citenamefont {Helsen}\ \emph {et~al.}(2023)\citenamefont {Helsen},
  \citenamefont {Ioannou}, \citenamefont {Kitzinger}, \citenamefont {Onorati},
  \citenamefont {Werner}, \citenamefont {Eisert},\ and\ \citenamefont
  {Roth}}]{helsen_seq_2021}%
  \BibitemOpen
  \bibfield  {author} {\bibinfo {author} {\bibfnamefont {J.}~\bibnamefont
  {Helsen}}, \bibinfo {author} {\bibfnamefont {M.}~\bibnamefont {Ioannou}},
  \bibinfo {author} {\bibfnamefont {J.}~\bibnamefont {Kitzinger}}, \bibinfo
  {author} {\bibfnamefont {E.}~\bibnamefont {Onorati}}, \bibinfo {author}
  {\bibfnamefont {A.~H.}\ \bibnamefont {Werner}}, \bibinfo {author}
  {\bibfnamefont {J.}~\bibnamefont {Eisert}}, \ and\ \bibinfo {author}
  {\bibfnamefont {I.}~\bibnamefont {Roth}},\ }\bibfield  {title} {\enquote
  {\bibinfo {title} {Shadow estimation of gate-set properties from random
  sequences},}\ }\href {\doibase 10.1038/s41467-023-39382-9} {\bibfield
  {journal} {\bibinfo  {journal} {Nature Communications}\ }\textbf {\bibinfo
  {volume} {14}},\ \bibinfo {pages} {5039} (\bibinfo {year}
  {2023})}\BibitemShut {NoStop}%
\bibitem [{\citenamefont {Flammia}(2021)}]{flammia_aces}%
  \BibitemOpen
  \bibfield  {author} {\bibinfo {author} {\bibfnamefont {S.~T.}\ \bibnamefont
  {Flammia}},\ }\href@noop {} {\enquote {\bibinfo {title} {Averaged circuit
  eigenvalue sampling},}\ } (\bibinfo {year} {2021}),\ \Eprint
  {http://arxiv.org/abs/2108.05803} {arXiv:2108.05803 [quant-ph]} \BibitemShut
  {NoStop}%
\bibitem [{\citenamefont {Harper}\ and\ \citenamefont
  {Flammia}(2023)}]{harper2023learning}%
  \BibitemOpen
  \bibfield  {author} {\bibinfo {author} {\bibfnamefont {R.}~\bibnamefont
  {Harper}}\ and\ \bibinfo {author} {\bibfnamefont {S.~T.}\ \bibnamefont
  {Flammia}},\ }\bibfield  {title} {\enquote {\bibinfo {title} {Learning
  correlated noise in a 39-qubit quantum processor},}\ }\href {\doibase
  10.1103/PRXQuantum.4.040311} {\bibfield  {journal} {\bibinfo  {journal} {PRX
  Quantum}\ }\textbf {\bibinfo {volume} {4}},\ \bibinfo {pages} {040311}
  (\bibinfo {year} {2023})}\BibitemShut {NoStop}%
\bibitem [{\citenamefont {Chen}\ \emph {et~al.}(2022)\citenamefont {Chen},
  \citenamefont {Ding},\ and\ \citenamefont {Huang}}]{PRXQuantum.3.030320}%
  \BibitemOpen
  \bibfield  {author} {\bibinfo {author} {\bibfnamefont {J.}~\bibnamefont
  {Chen}}, \bibinfo {author} {\bibfnamefont {D.}~\bibnamefont {Ding}}, \ and\
  \bibinfo {author} {\bibfnamefont {C.}~\bibnamefont {Huang}},\ }\bibfield
  {title} {\enquote {\bibinfo {title} {Randomized benchmarking beyond
  groups},}\ }\href {\doibase 10.1103/PRXQuantum.3.030320} {\bibfield
  {journal} {\bibinfo  {journal} {PRX Quantum}\ }\textbf {\bibinfo {volume}
  {3}},\ \bibinfo {pages} {030320} (\bibinfo {year} {2022})}\BibitemShut
  {NoStop}%
\bibitem [{\citenamefont {Helsen}\ \emph {et~al.}(2022)\citenamefont {Helsen},
  \citenamefont {Roth}, \citenamefont {Onorati}, \citenamefont {Werner},\ and\
  \citenamefont {Eisert}}]{helsen_general_2022}%
  \BibitemOpen
  \bibfield  {author} {\bibinfo {author} {\bibfnamefont {J.}~\bibnamefont
  {Helsen}}, \bibinfo {author} {\bibfnamefont {I.}~\bibnamefont {Roth}},
  \bibinfo {author} {\bibfnamefont {E.}~\bibnamefont {Onorati}}, \bibinfo
  {author} {\bibfnamefont {A.}~\bibnamefont {Werner}}, \ and\ \bibinfo {author}
  {\bibfnamefont {J.}~\bibnamefont {Eisert}},\ }\bibfield  {title} {\enquote
  {\bibinfo {title} {General framework for randomized benchmarking},}\ }\href
  {\doibase 10.1103/PRXQuantum.3.020357} {\bibfield  {journal} {\bibinfo
  {journal} {PRX Quantum}\ }\textbf {\bibinfo {volume} {3}},\ \bibinfo {pages}
  {020357} (\bibinfo {year} {2022})}\BibitemShut {NoStop}%
\bibitem [{\citenamefont {Rivas}\ \emph {et~al.}(2014)\citenamefont {Rivas},
  \citenamefont {Huelga},\ and\ \citenamefont {Plenio}}]{rivas_2014}%
  \BibitemOpen
  \bibfield  {author} {\bibinfo {author} {\bibfnamefont {A.}~\bibnamefont
  {Rivas}}, \bibinfo {author} {\bibfnamefont {S.~F.}\ \bibnamefont {Huelga}}, \
  and\ \bibinfo {author} {\bibfnamefont {M.~B.}\ \bibnamefont {Plenio}},\
  }\bibfield  {title} {\enquote {\bibinfo {title} {Quantum non-{M}arkovianity:
  characterization, quantification and detection},}\ }\href {\doibase
  10.1088/0034-4885/77/9/094001} {\bibfield  {journal} {\bibinfo  {journal}
  {Rep. Prog. Phys.}\ }\textbf {\bibinfo {volume} {77}},\ \bibinfo {pages}
  {094001} (\bibinfo {year} {2014})}\BibitemShut {NoStop}%
\bibitem [{\citenamefont {Breuer}\ \emph {et~al.}(2016)\citenamefont {Breuer},
  \citenamefont {Laine}, \citenamefont {Piilo},\ and\ \citenamefont
  {Vacchini}}]{RevModPhys.88.021002}%
  \BibitemOpen
  \bibfield  {author} {\bibinfo {author} {\bibfnamefont {H.-P.}\ \bibnamefont
  {Breuer}}, \bibinfo {author} {\bibfnamefont {E.-M.}\ \bibnamefont {Laine}},
  \bibinfo {author} {\bibfnamefont {J.}~\bibnamefont {Piilo}}, \ and\ \bibinfo
  {author} {\bibfnamefont {B.}~\bibnamefont {Vacchini}},\ }\bibfield  {title}
  {\enquote {\bibinfo {title} {Colloquium: Non-{M}arkovian dynamics in open
  quantum systems},}\ }\href {\doibase 10.1103/RevModPhys.88.021002} {\bibfield
   {journal} {\bibinfo  {journal} {Rev. Mod. Phys.}\ }\textbf {\bibinfo
  {volume} {88}},\ \bibinfo {pages} {021002} (\bibinfo {year}
  {2016})}\BibitemShut {NoStop}%
\bibitem [{\citenamefont {de~Vega}\ and\ \citenamefont
  {Alonso}(2017)}]{RevModPhys.89.015001}%
  \BibitemOpen
  \bibfield  {author} {\bibinfo {author} {\bibfnamefont {I.}~\bibnamefont
  {de~Vega}}\ and\ \bibinfo {author} {\bibfnamefont {D.}~\bibnamefont
  {Alonso}},\ }\bibfield  {title} {\enquote {\bibinfo {title} {Dynamics of
  non-{M}arkovian open quantum systems},}\ }\href {\doibase
  10.1103/RevModPhys.89.015001} {\bibfield  {journal} {\bibinfo  {journal}
  {Rev. Mod. Phys.}\ }\textbf {\bibinfo {volume} {89}},\ \bibinfo {pages}
  {015001} (\bibinfo {year} {2017})}\BibitemShut {NoStop}%
\bibitem [{\citenamefont {Pollock}\ \emph
  {et~al.}(2018{\natexlab{a}})\citenamefont {Pollock}, \citenamefont
  {Rodr\'{\i}guez-Rosario}, \citenamefont {Frauenheim}, \citenamefont
  {Paternostro},\ and\ \citenamefont {Modi}}]{PhysRevA.97.012127}%
  \BibitemOpen
  \bibfield  {author} {\bibinfo {author} {\bibfnamefont {F.~A.}\ \bibnamefont
  {Pollock}}, \bibinfo {author} {\bibfnamefont {C.}~\bibnamefont
  {Rodr\'{\i}guez-Rosario}}, \bibinfo {author} {\bibfnamefont {T.}~\bibnamefont
  {Frauenheim}}, \bibinfo {author} {\bibfnamefont {M.}~\bibnamefont
  {Paternostro}}, \ and\ \bibinfo {author} {\bibfnamefont {K.}~\bibnamefont
  {Modi}},\ }\bibfield  {title} {\enquote {\bibinfo {title} {Non-{M}arkovian
  quantum processes: Complete framework and efficient characterization},}\
  }\href {\doibase 10.1103/PhysRevA.97.012127} {\bibfield  {journal} {\bibinfo
  {journal} {Phys. Rev. A}\ }\textbf {\bibinfo {volume} {97}},\ \bibinfo
  {pages} {012127} (\bibinfo {year} {2018}{\natexlab{a}})}\BibitemShut
  {NoStop}%
\bibitem [{\citenamefont {Pollock}\ \emph
  {et~al.}(2018{\natexlab{b}})\citenamefont {Pollock}, \citenamefont
  {Rodr\'{\i}guez-Rosario}, \citenamefont {Frauenheim}, \citenamefont
  {Paternostro},\ and\ \citenamefont {Modi}}]{PhysRevLett.120.040405}%
  \BibitemOpen
  \bibfield  {author} {\bibinfo {author} {\bibfnamefont {F.~A.}\ \bibnamefont
  {Pollock}}, \bibinfo {author} {\bibfnamefont {C.}~\bibnamefont
  {Rodr\'{\i}guez-Rosario}}, \bibinfo {author} {\bibfnamefont {T.}~\bibnamefont
  {Frauenheim}}, \bibinfo {author} {\bibfnamefont {M.}~\bibnamefont
  {Paternostro}}, \ and\ \bibinfo {author} {\bibfnamefont {K.}~\bibnamefont
  {Modi}},\ }\bibfield  {title} {\enquote {\bibinfo {title} {Operational markov
  condition for quantum processes},}\ }\href {\doibase
  10.1103/PhysRevLett.120.040405} {\bibfield  {journal} {\bibinfo  {journal}
  {Phys. Rev. Lett.}\ }\textbf {\bibinfo {volume} {120}},\ \bibinfo {pages}
  {040405} (\bibinfo {year} {2018}{\natexlab{b}})}\BibitemShut {NoStop}%
\bibitem [{\citenamefont {Milz}\ and\ \citenamefont
  {Modi}(2021)}]{PRXQuantum.2.030201}%
  \BibitemOpen
  \bibfield  {author} {\bibinfo {author} {\bibfnamefont {S.}~\bibnamefont
  {Milz}}\ and\ \bibinfo {author} {\bibfnamefont {K.}~\bibnamefont {Modi}},\
  }\bibfield  {title} {\enquote {\bibinfo {title} {Quantum stochastic processes
  and quantum non-{M}arkovian phenomena},}\ }\href {\doibase
  10.1103/PRXQuantum.2.030201} {\bibfield  {journal} {\bibinfo  {journal} {PRX
  Quantum}\ }\textbf {\bibinfo {volume} {2}},\ \bibinfo {pages} {030201}
  (\bibinfo {year} {2021})}\BibitemShut {NoStop}%
\bibitem [{\citenamefont {Epstein}\ \emph {et~al.}(2014)\citenamefont
  {Epstein}, \citenamefont {Cross}, \citenamefont {Magesan},\ and\
  \citenamefont {Gambetta}}]{limits_RB_2014}%
  \BibitemOpen
  \bibfield  {author} {\bibinfo {author} {\bibfnamefont {J.~M.}\ \bibnamefont
  {Epstein}}, \bibinfo {author} {\bibfnamefont {A.~W.}\ \bibnamefont {Cross}},
  \bibinfo {author} {\bibfnamefont {E.}~\bibnamefont {Magesan}}, \ and\
  \bibinfo {author} {\bibfnamefont {J.~M.}\ \bibnamefont {Gambetta}},\
  }\bibfield  {title} {\enquote {\bibinfo {title} {Investigating the limits of
  randomized benchmarking protocols},}\ }\href {\doibase
  10.1103/PhysRevA.89.062321} {\bibfield  {journal} {\bibinfo  {journal} {Phys.
  Rev. A}\ }\textbf {\bibinfo {volume} {89}},\ \bibinfo {pages} {062321}
  (\bibinfo {year} {2014})}\BibitemShut {NoStop}%
\bibitem [{\citenamefont {Ball}\ \emph {et~al.}(2016)\citenamefont {Ball},
  \citenamefont {Stace}, \citenamefont {Flammia},\ and\ \citenamefont
  {Biercuk}}]{noise_corr_RB_2016}%
  \BibitemOpen
  \bibfield  {author} {\bibinfo {author} {\bibfnamefont {H.}~\bibnamefont
  {Ball}}, \bibinfo {author} {\bibfnamefont {T.~M.}\ \bibnamefont {Stace}},
  \bibinfo {author} {\bibfnamefont {S.~T.}\ \bibnamefont {Flammia}}, \ and\
  \bibinfo {author} {\bibfnamefont {M.~J.}\ \bibnamefont {Biercuk}},\
  }\bibfield  {title} {\enquote {\bibinfo {title} {Effect of noise correlations
  on randomized benchmarking},}\ }\href {\doibase 10.1103/PhysRevA.93.022303}
  {\bibfield  {journal} {\bibinfo  {journal} {Phys. Rev. A}\ }\textbf {\bibinfo
  {volume} {93}},\ \bibinfo {pages} {022303} (\bibinfo {year}
  {2016})}\BibitemShut {NoStop}%
\bibitem [{\citenamefont {Figueroa-Romero}\ \emph {et~al.}(2021)\citenamefont
  {Figueroa-Romero}, \citenamefont {Modi}, \citenamefont {Harris},
  \citenamefont {Stace},\ and\ \citenamefont {Hsieh}}]{PRXQuantum.2.040351}%
  \BibitemOpen
  \bibfield  {author} {\bibinfo {author} {\bibfnamefont {P.}~\bibnamefont
  {Figueroa-Romero}}, \bibinfo {author} {\bibfnamefont {K.}~\bibnamefont
  {Modi}}, \bibinfo {author} {\bibfnamefont {R.~J.}\ \bibnamefont {Harris}},
  \bibinfo {author} {\bibfnamefont {T.~M.}\ \bibnamefont {Stace}}, \ and\
  \bibinfo {author} {\bibfnamefont {M.-H.}\ \bibnamefont {Hsieh}},\ }\bibfield
  {title} {\enquote {\bibinfo {title} {Randomized benchmarking for
  non-markovian noise},}\ }\href {\doibase 10.1103/PRXQuantum.2.040351}
  {\bibfield  {journal} {\bibinfo  {journal} {PRX Quantum}\ }\textbf {\bibinfo
  {volume} {2}},\ \bibinfo {pages} {040351} (\bibinfo {year}
  {2021})}\BibitemShut {NoStop}%
\bibitem [{\citenamefont {Figueroa-Romero}\ \emph {et~al.}(2022)\citenamefont
  {Figueroa-Romero}, \citenamefont {Modi},\ and\ \citenamefont
  {Hsieh}}]{figueroaromero2022general}%
  \BibitemOpen
  \bibfield  {author} {\bibinfo {author} {\bibfnamefont {P.}~\bibnamefont
  {Figueroa-Romero}}, \bibinfo {author} {\bibfnamefont {K.}~\bibnamefont
  {Modi}}, \ and\ \bibinfo {author} {\bibfnamefont {M.-H.}\ \bibnamefont
  {Hsieh}},\ }\bibfield  {title} {\enquote {\bibinfo {title} {Towards a general
  framework of {R}andomized {B}enchmarking incorporating non-{M}arkovian
  {N}oise},}\ }\href {\doibase 10.22331/q-2022-12-01-868} {\bibfield  {journal}
  {\bibinfo  {journal} {{Quantum}}\ }\textbf {\bibinfo {volume} {6}},\ \bibinfo
  {pages} {868} (\bibinfo {year} {2022})}\BibitemShut {NoStop}%
\bibitem [{\citenamefont {Ceasura}\ \emph {et~al.}(2022)\citenamefont
  {Ceasura}, \citenamefont {Iyer}, \citenamefont {Wallman},\ and\ \citenamefont
  {Pashayan}}]{ceasura2022nonexponential}%
  \BibitemOpen
  \bibfield  {author} {\bibinfo {author} {\bibfnamefont {A.}~\bibnamefont
  {Ceasura}}, \bibinfo {author} {\bibfnamefont {P.}~\bibnamefont {Iyer}},
  \bibinfo {author} {\bibfnamefont {J.~J.}\ \bibnamefont {Wallman}}, \ and\
  \bibinfo {author} {\bibfnamefont {H.}~\bibnamefont {Pashayan}},\ }\href@noop
  {} {\enquote {\bibinfo {title} {Non-exponential behaviour in logical
  randomized benchmarking},}\ } (\bibinfo {year} {2022}),\ \Eprint
  {http://arxiv.org/abs/2212.05488} {arXiv:2212.05488 [quant-ph]} \BibitemShut
  {NoStop}%
\bibitem [{\citenamefont {Chen}\ \emph {et~al.}(2020)\citenamefont {Chen},
  \citenamefont {Ma}, \citenamefont {Zheng}, \citenamefont {Allcock},
  \citenamefont {Zhang},\ and\ \citenamefont
  {Hsieh}}]{PhysRevApplied.13.034045}%
  \BibitemOpen
  \bibfield  {author} {\bibinfo {author} {\bibfnamefont {Y.-Q.}\ \bibnamefont
  {Chen}}, \bibinfo {author} {\bibfnamefont {K.-L.}\ \bibnamefont {Ma}},
  \bibinfo {author} {\bibfnamefont {Y.-C.}\ \bibnamefont {Zheng}}, \bibinfo
  {author} {\bibfnamefont {J.}~\bibnamefont {Allcock}}, \bibinfo {author}
  {\bibfnamefont {S.}~\bibnamefont {Zhang}}, \ and\ \bibinfo {author}
  {\bibfnamefont {C.-Y.}\ \bibnamefont {Hsieh}},\ }\bibfield  {title} {\enquote
  {\bibinfo {title} {Non-{M}arkovian noise characterization with the transfer
  tensor method},}\ }\href {\doibase 10.1103/PhysRevApplied.13.034045}
  {\bibfield  {journal} {\bibinfo  {journal} {Phys. Rev. Appl.}\ }\textbf
  {\bibinfo {volume} {13}},\ \bibinfo {pages} {034045} (\bibinfo {year}
  {2020})}\BibitemShut {NoStop}%
\bibitem [{\citenamefont {White}\ \emph {et~al.}(2020)\citenamefont {White},
  \citenamefont {Hill}, \citenamefont {Pollock}, \citenamefont {Hollenberg},\
  and\ \citenamefont {Modi}}]{White2020demonstration}%
  \BibitemOpen
  \bibfield  {author} {\bibinfo {author} {\bibfnamefont {G.~A.~L.}\
  \bibnamefont {White}}, \bibinfo {author} {\bibfnamefont {C.~D.}\ \bibnamefont
  {Hill}}, \bibinfo {author} {\bibfnamefont {F.~A.}\ \bibnamefont {Pollock}},
  \bibinfo {author} {\bibfnamefont {L.~C.~L.}\ \bibnamefont {Hollenberg}}, \
  and\ \bibinfo {author} {\bibfnamefont {K.}~\bibnamefont {Modi}},\ }\bibfield
  {title} {\enquote {\bibinfo {title} {Demonstration of non-{M}arkovian process
  characterisation and control on a quantum processor},}\ }\href {\doibase
  10.1038/s41467-020-20113-3} {\bibfield  {journal} {\bibinfo  {journal} {Nat.
  Commun.}\ }\textbf {\bibinfo {volume} {11}},\ \bibinfo {pages} {6301}
  (\bibinfo {year} {2020})}\BibitemShut {NoStop}%
\bibitem [{\citenamefont {Goswami}\ \emph {et~al.}(2021)\citenamefont
  {Goswami}, \citenamefont {Giarmatzi}, \citenamefont {Monterola},
  \citenamefont {Shrapnel}, \citenamefont {Romero},\ and\ \citenamefont
  {Costa}}]{PhysRevA.104.022432}%
  \BibitemOpen
  \bibfield  {author} {\bibinfo {author} {\bibfnamefont {K.}~\bibnamefont
  {Goswami}}, \bibinfo {author} {\bibfnamefont {C.}~\bibnamefont {Giarmatzi}},
  \bibinfo {author} {\bibfnamefont {C.}~\bibnamefont {Monterola}}, \bibinfo
  {author} {\bibfnamefont {S.}~\bibnamefont {Shrapnel}}, \bibinfo {author}
  {\bibfnamefont {J.}~\bibnamefont {Romero}}, \ and\ \bibinfo {author}
  {\bibfnamefont {F.}~\bibnamefont {Costa}},\ }\bibfield  {title} {\enquote
  {\bibinfo {title} {Experimental characterization of a non-markovian quantum
  process},}\ }\href {\doibase 10.1103/PhysRevA.104.022432} {\bibfield
  {journal} {\bibinfo  {journal} {Phys. Rev. A}\ }\textbf {\bibinfo {volume}
  {104}},\ \bibinfo {pages} {022432} (\bibinfo {year} {2021})}\BibitemShut
  {NoStop}%
\bibitem [{\citenamefont {White}\ \emph {et~al.}(2022)\citenamefont {White},
  \citenamefont {Pollock}, \citenamefont {Hollenberg}, \citenamefont {Modi},\
  and\ \citenamefont {Hill}}]{white2021nonmarkovian}%
  \BibitemOpen
  \bibfield  {author} {\bibinfo {author} {\bibfnamefont {G.}~\bibnamefont
  {White}}, \bibinfo {author} {\bibfnamefont {F.}~\bibnamefont {Pollock}},
  \bibinfo {author} {\bibfnamefont {L.}~\bibnamefont {Hollenberg}}, \bibinfo
  {author} {\bibfnamefont {K.}~\bibnamefont {Modi}}, \ and\ \bibinfo {author}
  {\bibfnamefont {C.}~\bibnamefont {Hill}},\ }\bibfield  {title} {\enquote
  {\bibinfo {title} {Non-markovian quantum process tomography},}\ }\href
  {\doibase 10.1103/PRXQuantum.3.020344} {\bibfield  {journal} {\bibinfo
  {journal} {PRX Quantum}\ }\textbf {\bibinfo {volume} {3}},\ \bibinfo {pages}
  {020344} (\bibinfo {year} {2022})}\BibitemShut {NoStop}%
\bibitem [{\citenamefont {Papi\ifmmode~\check{c}\else \v{c}\fi{}}\ and\
  \citenamefont {de~Vega}(2022)}]{PhysRevA.105.022605}%
  \BibitemOpen
  \bibfield  {author} {\bibinfo {author} {\bibfnamefont {M.}~\bibnamefont
  {Papi\ifmmode~\check{c}\else \v{c}\fi{}}}\ and\ \bibinfo {author}
  {\bibfnamefont {I.}~\bibnamefont {de~Vega}},\ }\bibfield  {title} {\enquote
  {\bibinfo {title} {Neural-network-based qubit-environment
  characterization},}\ }\href {\doibase 10.1103/PhysRevA.105.022605} {\bibfield
   {journal} {\bibinfo  {journal} {Phys. Rev. A}\ }\textbf {\bibinfo {volume}
  {105}},\ \bibinfo {pages} {022605} (\bibinfo {year} {2022})}\BibitemShut
  {NoStop}%
\bibitem [{\citenamefont {Guo}(2022)}]{Guo_2022}%
  \BibitemOpen
  \bibfield  {author} {\bibinfo {author} {\bibfnamefont {C.}~\bibnamefont
  {Guo}},\ }\bibfield  {title} {\enquote {\bibinfo {title} {Quantifying
  non-{M}arkovianity in open quantum dynamics},}\ }\href {\doibase
  10.21468/SciPostPhys.13.2.028} {\bibfield  {journal} {\bibinfo  {journal}
  {SciPost Phys.}\ }\textbf {\bibinfo {volume} {13}},\ \bibinfo {pages} {028}
  (\bibinfo {year} {2022})}\BibitemShut {NoStop}%
\bibitem [{\citenamefont {White}(2022)}]{White2022}%
  \BibitemOpen
  \bibfield  {author} {\bibinfo {author} {\bibfnamefont {G.}~\bibnamefont
  {White}},\ }\bibfield  {title} {\enquote {\bibinfo {title} {Characterization
  and control of non-{M}arkovian quantum noise},}\ }\href {\doibase
  10.1038/s42254-022-00446-2} {\bibfield  {journal} {\bibinfo  {journal} {Nat.
  Rev. Phys.}\ }\textbf {\bibinfo {volume} {4}},\ \bibinfo {pages} {287}
  (\bibinfo {year} {2022})}\BibitemShut {NoStop}%
\bibitem [{\citenamefont {Ezzell}\ \emph {et~al.}(2023)\citenamefont {Ezzell},
  \citenamefont {Pokharel}, \citenamefont {Tewala}, \citenamefont {Quiroz},\
  and\ \citenamefont {Lidar}}]{ezzell_2022}%
  \BibitemOpen
  \bibfield  {author} {\bibinfo {author} {\bibfnamefont {N.}~\bibnamefont
  {Ezzell}}, \bibinfo {author} {\bibfnamefont {B.}~\bibnamefont {Pokharel}},
  \bibinfo {author} {\bibfnamefont {L.}~\bibnamefont {Tewala}}, \bibinfo
  {author} {\bibfnamefont {G.}~\bibnamefont {Quiroz}}, \ and\ \bibinfo {author}
  {\bibfnamefont {D.~A.}\ \bibnamefont {Lidar}},\ }\bibfield  {title} {\enquote
  {\bibinfo {title} {Dynamical decoupling for superconducting qubits: A
  performance survey},}\ }\href {\doibase 10.1103/PhysRevApplied.20.064027}
  {\bibfield  {journal} {\bibinfo  {journal} {Phys. Rev. Appl.}\ }\textbf
  {\bibinfo {volume} {20}},\ \bibinfo {pages} {064027} (\bibinfo {year}
  {2023})}\BibitemShut {NoStop}%
\bibitem [{\citenamefont {Winick}\ \emph {et~al.}(2022)\citenamefont {Winick},
  \citenamefont {Wallman}, \citenamefont {Dahlen}, \citenamefont {Hincks},
  \citenamefont {Ospadov},\ and\ \citenamefont {Emerson}}]{winick2022RC}%
  \BibitemOpen
  \bibfield  {author} {\bibinfo {author} {\bibfnamefont {A.}~\bibnamefont
  {Winick}}, \bibinfo {author} {\bibfnamefont {J.~J.}\ \bibnamefont {Wallman}},
  \bibinfo {author} {\bibfnamefont {D.}~\bibnamefont {Dahlen}}, \bibinfo
  {author} {\bibfnamefont {I.}~\bibnamefont {Hincks}}, \bibinfo {author}
  {\bibfnamefont {E.}~\bibnamefont {Ospadov}}, \ and\ \bibinfo {author}
  {\bibfnamefont {J.}~\bibnamefont {Emerson}},\ }\href@noop {} {\enquote
  {\bibinfo {title} {Concepts and conditions for error suppression through
  randomized compiling},}\ } (\bibinfo {year} {2022}),\ \Eprint
  {http://arxiv.org/abs/2212.07500} {arXiv:2212.07500 [quant-ph]} \BibitemShut
  {NoStop}%
\bibitem [{\citenamefont {Berk}\ \emph {et~al.}(2023)\citenamefont {Berk},
  \citenamefont {Milz}, \citenamefont {Pollock},\ and\ \citenamefont
  {Modi}}]{berk2021extracting}%
  \BibitemOpen
  \bibfield  {author} {\bibinfo {author} {\bibfnamefont {G.~D.}\ \bibnamefont
  {Berk}}, \bibinfo {author} {\bibfnamefont {S.}~\bibnamefont {Milz}}, \bibinfo
  {author} {\bibfnamefont {F.~A.}\ \bibnamefont {Pollock}}, \ and\ \bibinfo
  {author} {\bibfnamefont {K.}~\bibnamefont {Modi}},\ }\bibfield  {title}
  {\enquote {\bibinfo {title} {Extracting quantum dynamical resources:
  consumption of non-markovianity for noise reduction},}\ }\href {\doibase
  10.1038/s41534-023-00774-w} {\bibfield  {journal} {\bibinfo  {journal} {npj
  Quantum Information}\ }\textbf {\bibinfo {volume} {9}},\ \bibinfo {pages}
  {104} (\bibinfo {year} {2023})}\BibitemShut {NoStop}%
\bibitem [{\citenamefont {Ng}\ \emph {et~al.}(2011)\citenamefont {Ng},
  \citenamefont {Lidar},\ and\ \citenamefont {Preskill}}]{DD_FT_Preskill}%
  \BibitemOpen
  \bibfield  {author} {\bibinfo {author} {\bibfnamefont {H.~K.}\ \bibnamefont
  {Ng}}, \bibinfo {author} {\bibfnamefont {D.~A.}\ \bibnamefont {Lidar}}, \
  and\ \bibinfo {author} {\bibfnamefont {J.}~\bibnamefont {Preskill}},\
  }\bibfield  {title} {\enquote {\bibinfo {title} {Combining dynamical
  decoupling with fault-tolerant quantum computation},}\ }\href {\doibase
  10.1103/PhysRevA.84.012305} {\bibfield  {journal} {\bibinfo  {journal} {Phys.
  Rev. A}\ }\textbf {\bibinfo {volume} {84}},\ \bibinfo {pages} {012305}
  (\bibinfo {year} {2011})}\BibitemShut {NoStop}%
\bibitem [{\citenamefont {Pokharel}\ \emph {et~al.}(2018)\citenamefont
  {Pokharel}, \citenamefont {Anand}, \citenamefont {Fortman},\ and\
  \citenamefont {Lidar}}]{PhysRevLett.121.220502}%
  \BibitemOpen
  \bibfield  {author} {\bibinfo {author} {\bibfnamefont {B.}~\bibnamefont
  {Pokharel}}, \bibinfo {author} {\bibfnamefont {N.}~\bibnamefont {Anand}},
  \bibinfo {author} {\bibfnamefont {B.}~\bibnamefont {Fortman}}, \ and\
  \bibinfo {author} {\bibfnamefont {D.~A.}\ \bibnamefont {Lidar}},\ }\bibfield
  {title} {\enquote {\bibinfo {title} {Demonstration of fidelity improvement
  using dynamical decoupling with superconducting qubits},}\ }\href {\doibase
  10.1103/PhysRevLett.121.220502} {\bibfield  {journal} {\bibinfo  {journal}
  {Phys. Rev. Lett.}\ }\textbf {\bibinfo {volume} {121}},\ \bibinfo {pages}
  {220502} (\bibinfo {year} {2018})}\BibitemShut {NoStop}%
\bibitem [{\citenamefont {Ware}\ \emph {et~al.}(2021)\citenamefont {Ware},
  \citenamefont {Ribeill}, \citenamefont {Rist\`e}, \citenamefont {Ryan},
  \citenamefont {Johnson},\ and\ \citenamefont {da~Silva}}]{PauliFR_2021}%
  \BibitemOpen
  \bibfield  {author} {\bibinfo {author} {\bibfnamefont {M.}~\bibnamefont
  {Ware}}, \bibinfo {author} {\bibfnamefont {G.}~\bibnamefont {Ribeill}},
  \bibinfo {author} {\bibfnamefont {D.}~\bibnamefont {Rist\`e}}, \bibinfo
  {author} {\bibfnamefont {C.~A.}\ \bibnamefont {Ryan}}, \bibinfo {author}
  {\bibfnamefont {B.}~\bibnamefont {Johnson}}, \ and\ \bibinfo {author}
  {\bibfnamefont {M.~P.}\ \bibnamefont {da~Silva}},\ }\bibfield  {title}
  {\enquote {\bibinfo {title} {Experimental {P}auli-frame randomization on a
  superconducting qubit},}\ }\href {\doibase 10.1103/PhysRevA.103.042604}
  {\bibfield  {journal} {\bibinfo  {journal} {Phys. Rev. A}\ }\textbf {\bibinfo
  {volume} {103}},\ \bibinfo {pages} {042604} (\bibinfo {year}
  {2021})}\BibitemShut {NoStop}%
\bibitem [{\citenamefont {Hashim}\ \emph {et~al.}(2021)\citenamefont {Hashim},
  \citenamefont {Naik}, \citenamefont {Morvan}, \citenamefont {Ville},
  \citenamefont {Mitchell}, \citenamefont {Kreikebaum}, \citenamefont {Davis},
  \citenamefont {Smith}, \citenamefont {Iancu}, \citenamefont {O'Brien},
  \citenamefont {Hincks}, \citenamefont {Wallman}, \citenamefont {Emerson},\
  and\ \citenamefont {Siddiqi}}]{RC_superc_2021}%
  \BibitemOpen
  \bibfield  {author} {\bibinfo {author} {\bibfnamefont {A.}~\bibnamefont
  {Hashim}}, \bibinfo {author} {\bibfnamefont {R.~K.}\ \bibnamefont {Naik}},
  \bibinfo {author} {\bibfnamefont {A.}~\bibnamefont {Morvan}}, \bibinfo
  {author} {\bibfnamefont {J.-L.}\ \bibnamefont {Ville}}, \bibinfo {author}
  {\bibfnamefont {B.}~\bibnamefont {Mitchell}}, \bibinfo {author}
  {\bibfnamefont {J.~M.}\ \bibnamefont {Kreikebaum}}, \bibinfo {author}
  {\bibfnamefont {M.}~\bibnamefont {Davis}}, \bibinfo {author} {\bibfnamefont
  {E.}~\bibnamefont {Smith}}, \bibinfo {author} {\bibfnamefont
  {C.}~\bibnamefont {Iancu}}, \bibinfo {author} {\bibfnamefont {K.~P.}\
  \bibnamefont {O'Brien}}, \bibinfo {author} {\bibfnamefont {I.}~\bibnamefont
  {Hincks}}, \bibinfo {author} {\bibfnamefont {J.~J.}\ \bibnamefont {Wallman}},
  \bibinfo {author} {\bibfnamefont {J.}~\bibnamefont {Emerson}}, \ and\
  \bibinfo {author} {\bibfnamefont {I.}~\bibnamefont {Siddiqi}},\ }\bibfield
  {title} {\enquote {\bibinfo {title} {Randomized compiling for scalable
  quantum computing on a noisy superconducting quantum processor},}\ }\href
  {\doibase 10.1103/PhysRevX.11.041039} {\bibfield  {journal} {\bibinfo
  {journal} {Phys. Rev. X}\ }\textbf {\bibinfo {volume} {11}},\ \bibinfo
  {pages} {041039} (\bibinfo {year} {2021})}\BibitemShut {NoStop}%
\bibitem [{\citenamefont {Kelly}\ \emph {et~al.}(2014)\citenamefont {Kelly},
  \citenamefont {Barends}, \citenamefont {Campbell}, \citenamefont {Chen},
  \citenamefont {Chen}, \citenamefont {Chiaro}, \citenamefont {Dunsworth},
  \citenamefont {Fowler}, \citenamefont {Hoi}, \citenamefont {Jeffrey},
  \citenamefont {Megrant}, \citenamefont {Mutus}, \citenamefont {Neill},
  \citenamefont {O'Malley}, \citenamefont {Quintana}, \citenamefont {Roushan},
  \citenamefont {Sank}, \citenamefont {Vainsencher}, \citenamefont {Wenner},
  \citenamefont {White}, \citenamefont {Cleland},\ and\ \citenamefont
  {Martinis}}]{orbit_2014}%
  \BibitemOpen
  \bibfield  {author} {\bibinfo {author} {\bibfnamefont {J.}~\bibnamefont
  {Kelly}}, \bibinfo {author} {\bibfnamefont {R.}~\bibnamefont {Barends}},
  \bibinfo {author} {\bibfnamefont {B.}~\bibnamefont {Campbell}}, \bibinfo
  {author} {\bibfnamefont {Y.}~\bibnamefont {Chen}}, \bibinfo {author}
  {\bibfnamefont {Z.}~\bibnamefont {Chen}}, \bibinfo {author} {\bibfnamefont
  {B.}~\bibnamefont {Chiaro}}, \bibinfo {author} {\bibfnamefont
  {A.}~\bibnamefont {Dunsworth}}, \bibinfo {author} {\bibfnamefont {A.~G.}\
  \bibnamefont {Fowler}}, \bibinfo {author} {\bibfnamefont {I.-C.}\
  \bibnamefont {Hoi}}, \bibinfo {author} {\bibfnamefont {E.}~\bibnamefont
  {Jeffrey}}, \bibinfo {author} {\bibfnamefont {A.}~\bibnamefont {Megrant}},
  \bibinfo {author} {\bibfnamefont {J.}~\bibnamefont {Mutus}}, \bibinfo
  {author} {\bibfnamefont {C.}~\bibnamefont {Neill}}, \bibinfo {author}
  {\bibfnamefont {P.~J.~J.}\ \bibnamefont {O'Malley}}, \bibinfo {author}
  {\bibfnamefont {C.}~\bibnamefont {Quintana}}, \bibinfo {author}
  {\bibfnamefont {P.}~\bibnamefont {Roushan}}, \bibinfo {author} {\bibfnamefont
  {D.}~\bibnamefont {Sank}}, \bibinfo {author} {\bibfnamefont {A.}~\bibnamefont
  {Vainsencher}}, \bibinfo {author} {\bibfnamefont {J.}~\bibnamefont {Wenner}},
  \bibinfo {author} {\bibfnamefont {T.~C.}\ \bibnamefont {White}}, \bibinfo
  {author} {\bibfnamefont {A.~N.}\ \bibnamefont {Cleland}}, \ and\ \bibinfo
  {author} {\bibfnamefont {J.~M.}\ \bibnamefont {Martinis}},\ }\bibfield
  {title} {\enquote {\bibinfo {title} {Optimal {Q}uantum {C}ontrol using
  {R}andomized {B}enchmarking},}\ }\href {\doibase
  10.1103/PhysRevLett.112.240504} {\bibfield  {journal} {\bibinfo  {journal}
  {Phys. Rev. Lett.}\ }\textbf {\bibinfo {volume} {112}},\ \bibinfo {pages}
  {240504} (\bibinfo {year} {2014})}\BibitemShut {NoStop}%
\bibitem [{\citenamefont {Souza}\ \emph {et~al.}(2015)\citenamefont {Souza},
  \citenamefont {Sarthour}, \citenamefont {Oliveira},\ and\ \citenamefont
  {Suter}}]{PhysRevA.92.062332}%
  \BibitemOpen
  \bibfield  {author} {\bibinfo {author} {\bibfnamefont {A.~M.}\ \bibnamefont
  {Souza}}, \bibinfo {author} {\bibfnamefont {R.~S.}\ \bibnamefont {Sarthour}},
  \bibinfo {author} {\bibfnamefont {I.~S.}\ \bibnamefont {Oliveira}}, \ and\
  \bibinfo {author} {\bibfnamefont {D.}~\bibnamefont {Suter}},\ }\bibfield
  {title} {\enquote {\bibinfo {title} {High-fidelity gate operations for
  quantum computing beyond dephasing time limits},}\ }\href {\doibase
  10.1103/PhysRevA.92.062332} {\bibfield  {journal} {\bibinfo  {journal} {Phys.
  Rev. A}\ }\textbf {\bibinfo {volume} {92}},\ \bibinfo {pages} {062332}
  (\bibinfo {year} {2015})}\BibitemShut {NoStop}%
\bibitem [{Note1()}]{Note1}%
  \BibitemOpen
  \bibinfo {note} {We point out that the average gate-fidelity of a quantum
  channel $\protect \tilde {\protect \mathcal {G}}$ with respect to a gate
  $\protect \mathcal {G}$, is a measure of their ``average orthogonality'',
  rather than their distinguishability (which albeit related, are not quite the
  same thing): $\protect \mathrm {F}_\protect \mathsf {avg}:=\DOTSI \intop
  \ilimits@ {d}\psi \protect \tr [\protect \tilde {\protect \mathcal {G}}(\psi
  )\protect \mathcal {G}(\psi )]$, where $\psi $ are (uniformly distributed)
  pure states.}\BibitemShut {Stop}%
\bibitem [{\citenamefont {White}\ \emph {et~al.}(2023)\citenamefont {White},
  \citenamefont {Modi},\ and\ \citenamefont {Hill}}]{crosstalk_nM_White}%
  \BibitemOpen
  \bibfield  {author} {\bibinfo {author} {\bibfnamefont {G.~A.~L.}\
  \bibnamefont {White}}, \bibinfo {author} {\bibfnamefont {K.}~\bibnamefont
  {Modi}}, \ and\ \bibinfo {author} {\bibfnamefont {C.~D.}\ \bibnamefont
  {Hill}},\ }\bibfield  {title} {\enquote {\bibinfo {title} {Filtering
  crosstalk from bath non-markovianity via spacetime classical shadows},}\
  }\href {\doibase 10.1103/PhysRevLett.130.160401} {\bibfield  {journal}
  {\bibinfo  {journal} {Phys. Rev. Lett.}\ }\textbf {\bibinfo {volume} {130}},\
  \bibinfo {pages} {160401} (\bibinfo {year} {2023})}\BibitemShut {NoStop}%
\bibitem [{\citenamefont {Milz}\ \emph {et~al.}(2020)\citenamefont {Milz},
  \citenamefont {Sakuldee}, \citenamefont {Pollock},\ and\ \citenamefont
  {Modi}}]{Milz2020kolmogorov}%
  \BibitemOpen
  \bibfield  {author} {\bibinfo {author} {\bibfnamefont {S.}~\bibnamefont
  {Milz}}, \bibinfo {author} {\bibfnamefont {F.}~\bibnamefont {Sakuldee}},
  \bibinfo {author} {\bibfnamefont {F.~A.}\ \bibnamefont {Pollock}}, \ and\
  \bibinfo {author} {\bibfnamefont {K.}~\bibnamefont {Modi}},\ }\bibfield
  {title} {\enquote {\bibinfo {title} {Kolmogorov extension theorem for
  (quantum) causal modelling and general probabilistic theories},}\ }\href
  {\doibase 10.22331/q-2020-04-20-255} {\bibfield  {journal} {\bibinfo
  {journal} {{Quantum}}\ }\textbf {\bibinfo {volume} {4}},\ \bibinfo {pages}
  {255} (\bibinfo {year} {2020})}\BibitemShut {NoStop}%
\bibitem [{Note2()}]{Note2}%
  \BibitemOpen
  \bibinfo {note} {Non-Markovianity \protect \emph {can} be said to introduce a
  type of gate-dependence; nevertheless, here by explicit gate-independence we
  mean noise associated to $\protect \mathcal {I}_\protect \mathsf {E}\otimes
  \protect \mathcal {G}$, where $\protect \mathcal {I}_\protect \mathsf {E}$ is
  an identity map on $\protect \mathsf {E}$ and $\protect \mathcal {G}$ is an
  ideal gate, is not explicitly dependent on $\protect \mathcal
  {G}$.}\BibitemShut {Stop}%
\bibitem [{Note3()}]{Note3}%
  \BibitemOpen
  \bibinfo {note} {Here with noise being time-stationary, albeit non-Markovian,
  we mean that noise associated to $\protect \mathcal {I}_\protect \mathsf
  {E}\otimes \protect \mathcal {G}_i$, where $\protect \mathcal {I}_\protect
  \mathsf {E}$ is an identity channel on $\protect \mathsf {E}$ and $\protect
  \mathcal {G}_i$ is the ideal gate in $\protect \mathsf {S}$ at timestep $i$,
  is independent of the timestep $i$.}\BibitemShut {Stop}%
\bibitem [{\citenamefont {Khodjasteh}\ and\ \citenamefont
  {Lidar}(2007)}]{LidarCDD}%
  \BibitemOpen
  \bibfield  {author} {\bibinfo {author} {\bibfnamefont {K.}~\bibnamefont
  {Khodjasteh}}\ and\ \bibinfo {author} {\bibfnamefont {D.~A.}\ \bibnamefont
  {Lidar}},\ }\bibfield  {title} {\enquote {\bibinfo {title} {Performance of
  deterministic dynamical decoupling schemes: Concatenated and periodic pulse
  sequences},}\ }\href {\doibase 10.1103/PhysRevA.75.062310} {\bibfield
  {journal} {\bibinfo  {journal} {Phys. Rev. A}\ }\textbf {\bibinfo {volume}
  {75}},\ \bibinfo {pages} {062310} (\bibinfo {year} {2007})}\BibitemShut
  {NoStop}%
\bibitem [{\citenamefont {Arenz}\ \emph {et~al.}(2017)\citenamefont {Arenz},
  \citenamefont {Burgarth},\ and\ \citenamefont {Hillier}}]{Arenz_2017}%
  \BibitemOpen
  \bibfield  {author} {\bibinfo {author} {\bibfnamefont {C.}~\bibnamefont
  {Arenz}}, \bibinfo {author} {\bibfnamefont {D.}~\bibnamefont {Burgarth}}, \
  and\ \bibinfo {author} {\bibfnamefont {R.}~\bibnamefont {Hillier}},\
  }\bibfield  {title} {\enquote {\bibinfo {title} {Dynamical decoupling and
  homogenization of continuous variable systems},}\ }\href {\doibase
  10.1088/1751-8121/aa6017} {\bibfield  {journal} {\bibinfo  {journal} {J.
  Phys. A: Math. Theor.}\ }\textbf {\bibinfo {volume} {50}},\ \bibinfo {pages}
  {135303} (\bibinfo {year} {2017})}\BibitemShut {NoStop}%
\bibitem [{\citenamefont {Arenz}\ \emph {et~al.}(2018)\citenamefont {Arenz},
  \citenamefont {Burgarth}, \citenamefont {Facchi},\ and\ \citenamefont
  {Hillier}}]{DD_unbounded_2018}%
  \BibitemOpen
  \bibfield  {author} {\bibinfo {author} {\bibfnamefont {C.}~\bibnamefont
  {Arenz}}, \bibinfo {author} {\bibfnamefont {D.}~\bibnamefont {Burgarth}},
  \bibinfo {author} {\bibfnamefont {P.}~\bibnamefont {Facchi}}, \ and\ \bibinfo
  {author} {\bibfnamefont {R.}~\bibnamefont {Hillier}},\ }\bibfield  {title}
  {\enquote {\bibinfo {title} {Dynamical decoupling of unbounded
  hamiltonians},}\ }\href {\doibase 10.1063/1.5016495} {\bibfield  {journal}
  {\bibinfo  {journal} {J. Math. Phys.}\ }\textbf {\bibinfo {volume} {59}}
  (\bibinfo {year} {2018}),\ 10.1063/1.5016495}\BibitemShut {NoStop}%
\bibitem [{\citenamefont {Luchnikov}\ \emph {et~al.}(2019)\citenamefont
  {Luchnikov}, \citenamefont {Vintskevich}, \citenamefont {Ouerdane},\ and\
  \citenamefont {Filippov}}]{PhysRevLett.122.160401}%
  \BibitemOpen
  \bibfield  {author} {\bibinfo {author} {\bibfnamefont {I.~A.}\ \bibnamefont
  {Luchnikov}}, \bibinfo {author} {\bibfnamefont {S.~V.}\ \bibnamefont
  {Vintskevich}}, \bibinfo {author} {\bibfnamefont {H.}~\bibnamefont
  {Ouerdane}}, \ and\ \bibinfo {author} {\bibfnamefont {S.~N.}\ \bibnamefont
  {Filippov}},\ }\bibfield  {title} {\enquote {\bibinfo {title} {Simulation
  complexity of open quantum dynamics: Connection with tensor networks},}\
  }\href {\doibase 10.1103/PhysRevLett.122.160401} {\bibfield  {journal}
  {\bibinfo  {journal} {Phys. Rev. Lett.}\ }\textbf {\bibinfo {volume} {122}},\
  \bibinfo {pages} {160401} (\bibinfo {year} {2019})}\BibitemShut {NoStop}%
\bibitem [{\citenamefont {Burgarth}\ \emph
  {et~al.}(2021{\natexlab{a}})\citenamefont {Burgarth}, \citenamefont {Facchi},
  \citenamefont {Fraas},\ and\ \citenamefont
  {Hillier}}]{nM_cannot_decouple_2021}%
  \BibitemOpen
  \bibfield  {author} {\bibinfo {author} {\bibfnamefont {D.}~\bibnamefont
  {Burgarth}}, \bibinfo {author} {\bibfnamefont {P.}~\bibnamefont {Facchi}},
  \bibinfo {author} {\bibfnamefont {M.}~\bibnamefont {Fraas}}, \ and\ \bibinfo
  {author} {\bibfnamefont {R.}~\bibnamefont {Hillier}},\ }\bibfield  {title}
  {\enquote {\bibinfo {title} {{Non-Markovian noise that cannot be dynamically
  decoupled by periodic spin echo pulses}},}\ }\href {\doibase
  10.21468/SciPostPhys.11.2.027} {\bibfield  {journal} {\bibinfo  {journal}
  {SciPost Phys.}\ }\textbf {\bibinfo {volume} {11}},\ \bibinfo {pages} {027}
  (\bibinfo {year} {2021}{\natexlab{a}})}\BibitemShut {NoStop}%
\bibitem [{\citenamefont {Qi}\ \emph {et~al.}(2023)\citenamefont {Qi},
  \citenamefont {Xu}, \citenamefont {Poletti},\ and\ \citenamefont
  {Ng}}]{qi2022efficacy}%
  \BibitemOpen
  \bibfield  {author} {\bibinfo {author} {\bibfnamefont {J.}~\bibnamefont
  {Qi}}, \bibinfo {author} {\bibfnamefont {X.}~\bibnamefont {Xu}}, \bibinfo
  {author} {\bibfnamefont {D.}~\bibnamefont {Poletti}}, \ and\ \bibinfo
  {author} {\bibfnamefont {H.~K.}\ \bibnamefont {Ng}},\ }\bibfield  {title}
  {\enquote {\bibinfo {title} {Efficacy of noisy dynamical decoupling},}\
  }\href {\doibase 10.1103/PhysRevA.107.032615} {\bibfield  {journal} {\bibinfo
   {journal} {Phys. Rev. A}\ }\textbf {\bibinfo {volume} {107}},\ \bibinfo
  {pages} {032615} (\bibinfo {year} {2023})}\BibitemShut {NoStop}%
\bibitem [{\citenamefont {Hashim}\ \emph {et~al.}(2023)\citenamefont {Hashim},
  \citenamefont {Seritan}, \citenamefont {Proctor}, \citenamefont {Rudinger},
  \citenamefont {Goss}, \citenamefont {Naik}, \citenamefont {Kreikebaum},
  \citenamefont {Santiago},\ and\ \citenamefont {Siddiqi}}]{Hashim_2023}%
  \BibitemOpen
  \bibfield  {author} {\bibinfo {author} {\bibfnamefont {A.}~\bibnamefont
  {Hashim}}, \bibinfo {author} {\bibfnamefont {S.}~\bibnamefont {Seritan}},
  \bibinfo {author} {\bibfnamefont {T.}~\bibnamefont {Proctor}}, \bibinfo
  {author} {\bibfnamefont {K.}~\bibnamefont {Rudinger}}, \bibinfo {author}
  {\bibfnamefont {N.}~\bibnamefont {Goss}}, \bibinfo {author} {\bibfnamefont
  {R.~K.}\ \bibnamefont {Naik}}, \bibinfo {author} {\bibfnamefont {J.~M.}\
  \bibnamefont {Kreikebaum}}, \bibinfo {author} {\bibfnamefont {D.~I.}\
  \bibnamefont {Santiago}}, \ and\ \bibinfo {author} {\bibfnamefont
  {I.}~\bibnamefont {Siddiqi}},\ }\bibfield  {title} {\enquote {\bibinfo
  {title} {Benchmarking quantum logic operations relative to thresholds for
  fault tolerance},}\ }\href {\doibase 10.1038/s41534-023-00764-y} {\bibfield
  {journal} {\bibinfo  {journal} {npj Quantum Information}\ }\textbf {\bibinfo
  {volume} {9}} (\bibinfo {year} {2023}),\
  10.1038/s41534-023-00764-y}\BibitemShut {NoStop}%
\bibitem [{\citenamefont {Wallman}\ and\ \citenamefont
  {Emerson}(2016)}]{RC_2016}%
  \BibitemOpen
  \bibfield  {author} {\bibinfo {author} {\bibfnamefont {J.~J.}\ \bibnamefont
  {Wallman}}\ and\ \bibinfo {author} {\bibfnamefont {J.}~\bibnamefont
  {Emerson}},\ }\bibfield  {title} {\enquote {\bibinfo {title} {Noise tailoring
  for scalable quantum computation via randomized compiling},}\ }\href
  {\doibase 10.1103/PhysRevA.94.052325} {\bibfield  {journal} {\bibinfo
  {journal} {Phys. Rev. A}\ }\textbf {\bibinfo {volume} {94}},\ \bibinfo
  {pages} {052325} (\bibinfo {year} {2016})}\BibitemShut {NoStop}%
\bibitem [{Note4()}]{Note4}%
  \BibitemOpen
  \bibinfo {note} {In the case of of \gls {rc} and \gls {rb}, the number of
  samples compound as their product, and both techniques have at most a linear
  sampling complexity in system size; this contrasts with the case of \gls {rc}
  and GST employed in~\cite {Hashim_2023}, where they reported requiring 40
  hours to collect data for 100 \gls {rc} randomizations to characterize a
  2-qubit gate in a superconducting system.}\BibitemShut {Stop}%
\bibitem [{\citenamefont {Sanders}\ \emph {et~al.}(2015)\citenamefont
  {Sanders}, \citenamefont {Wallman},\ and\ \citenamefont
  {Sanders}}]{Sanders_2015}%
  \BibitemOpen
  \bibfield  {author} {\bibinfo {author} {\bibfnamefont {Y.~R.}\ \bibnamefont
  {Sanders}}, \bibinfo {author} {\bibfnamefont {J.~J.}\ \bibnamefont
  {Wallman}}, \ and\ \bibinfo {author} {\bibfnamefont {B.~C.}\ \bibnamefont
  {Sanders}},\ }\bibfield  {title} {\enquote {\bibinfo {title} {Bounding
  quantum gate error rate based on reported average fidelity},}\ }\href
  {\doibase 10.1088/1367-2630/18/1/012002} {\bibfield  {journal} {\bibinfo
  {journal} {New J. Phys.}\ }\textbf {\bibinfo {volume} {18}},\ \bibinfo
  {pages} {012002} (\bibinfo {year} {2015})}\BibitemShut {NoStop}%
\bibitem [{Note5()}]{Note5}%
  \BibitemOpen
  \bibinfo {note} {The statement in Eq.~\protect \textup {\hbox {\mathsurround
  \z@ \protect \normalfont (\ignorespaces \ref {eq: variance pauli
  (main)}\unskip \@@italiccorr )}} holds more generally for $\protect \mathrm
  {f}_m$ being noisy expectation values.}\BibitemShut {Stop}%
\bibitem [{\citenamefont {Dirkse}\ \emph {et~al.}(2019)\citenamefont {Dirkse},
  \citenamefont {Helsen},\ and\ \citenamefont {Wehner}}]{PhysRevA.99.012315}%
  \BibitemOpen
  \bibfield  {author} {\bibinfo {author} {\bibfnamefont {B.}~\bibnamefont
  {Dirkse}}, \bibinfo {author} {\bibfnamefont {J.}~\bibnamefont {Helsen}}, \
  and\ \bibinfo {author} {\bibfnamefont {S.}~\bibnamefont {Wehner}},\
  }\bibfield  {title} {\enquote {\bibinfo {title} {Efficient unitarity
  randomized benchmarking of few-qubit clifford gates},}\ }\href {\doibase
  10.1103/PhysRevA.99.012315} {\bibfield  {journal} {\bibinfo  {journal} {Phys.
  Rev. A}\ }\textbf {\bibinfo {volume} {99}},\ \bibinfo {pages} {012315}
  (\bibinfo {year} {2019})}\BibitemShut {NoStop}%
\bibitem [{Note6()}]{Note6}%
  \BibitemOpen
  \bibinfo {note} {In the $\protect \mathsf {S}\protect \mathsf {E}$ case, it
  is unclear whether $\protect \mathsf {S}$-Pauli-twirling \protect \emph
  {indeed will always} either only decrease or leave the total unitarity
  unchanged, due precisely to $\protect \mathsf {E}$; here the role of the
  average sequence fidelity would also come into play to always decrease the
  variance.}\BibitemShut {Stop}%
\bibitem [{\citenamefont {Abdurakhimov}\ \emph {et~al.}(2022)\citenamefont
  {Abdurakhimov}, \citenamefont {Mahboob}, \citenamefont {Toida}, \citenamefont
  {Kakuyanagi}, \citenamefont {Matsuzaki},\ and\ \citenamefont
  {Saito}}]{PRXQuantum.3.040332}%
  \BibitemOpen
  \bibfield  {author} {\bibinfo {author} {\bibfnamefont {L.~V.}\ \bibnamefont
  {Abdurakhimov}}, \bibinfo {author} {\bibfnamefont {I.}~\bibnamefont
  {Mahboob}}, \bibinfo {author} {\bibfnamefont {H.}~\bibnamefont {Toida}},
  \bibinfo {author} {\bibfnamefont {K.}~\bibnamefont {Kakuyanagi}}, \bibinfo
  {author} {\bibfnamefont {Y.}~\bibnamefont {Matsuzaki}}, \ and\ \bibinfo
  {author} {\bibfnamefont {S.}~\bibnamefont {Saito}},\ }\bibfield  {title}
  {\enquote {\bibinfo {title} {Identification of different types of
  high-frequency defects in superconducting qubits},}\ }\href {\doibase
  10.1103/PRXQuantum.3.040332} {\bibfield  {journal} {\bibinfo  {journal} {PRX
  Quantum}\ }\textbf {\bibinfo {volume} {3}},\ \bibinfo {pages} {040332}
  (\bibinfo {year} {2022})}\BibitemShut {NoStop}%
\bibitem [{\citenamefont {Addis}\ \emph {et~al.}(2015)\citenamefont {Addis},
  \citenamefont {Ciccarello}, \citenamefont {Cascio}, \citenamefont {Palma},\
  and\ \citenamefont {Maniscalco}}]{Addis_2015}%
  \BibitemOpen
  \bibfield  {author} {\bibinfo {author} {\bibfnamefont {C.}~\bibnamefont
  {Addis}}, \bibinfo {author} {\bibfnamefont {F.}~\bibnamefont {Ciccarello}},
  \bibinfo {author} {\bibfnamefont {M.}~\bibnamefont {Cascio}}, \bibinfo
  {author} {\bibfnamefont {G.~M.}\ \bibnamefont {Palma}}, \ and\ \bibinfo
  {author} {\bibfnamefont {S.}~\bibnamefont {Maniscalco}},\ }\bibfield  {title}
  {\enquote {\bibinfo {title} {Dynamical decoupling efficiency versus quantum
  non-markovianity},}\ }\href {\doibase 10.1088/1367-2630/17/12/123004}
  {\bibfield  {journal} {\bibinfo  {journal} {New Journal of Physics}\ }\textbf
  {\bibinfo {volume} {17}},\ \bibinfo {pages} {123004} (\bibinfo {year}
  {2015})}\BibitemShut {NoStop}%
\bibitem [{\citenamefont {D'Arrigo}\ \emph {et~al.}(2019)\citenamefont
  {D'Arrigo}, \citenamefont {Falci},\ and\ \citenamefont
  {Paladino}}]{DArrigo2019}%
  \BibitemOpen
  \bibfield  {author} {\bibinfo {author} {\bibfnamefont {A.}~\bibnamefont
  {D'Arrigo}}, \bibinfo {author} {\bibfnamefont {G.}~\bibnamefont {Falci}}, \
  and\ \bibinfo {author} {\bibfnamefont {E.}~\bibnamefont {Paladino}},\
  }\bibfield  {title} {\enquote {\bibinfo {title} {Quantum zeno and anti-zeno
  effect on a two-qubit gate by dynamical decoupling},}\ }\href {\doibase
  10.1140/epjst/e2018-800081-0} {\bibfield  {journal} {\bibinfo  {journal} {The
  European Physical Journal Special Topics}\ }\textbf {\bibinfo {volume}
  {227}},\ \bibinfo {pages} {2189} (\bibinfo {year} {2019})}\BibitemShut
  {NoStop}%
\bibitem [{\citenamefont {Burgarth}\ \emph
  {et~al.}(2021{\natexlab{b}})\citenamefont {Burgarth}, \citenamefont {Facchi},
  \citenamefont {Ligab\`o},\ and\ \citenamefont
  {Lonigro}}]{hidden_nM_PhysRevA.103.012203}%
  \BibitemOpen
  \bibfield  {author} {\bibinfo {author} {\bibfnamefont {D.}~\bibnamefont
  {Burgarth}}, \bibinfo {author} {\bibfnamefont {P.}~\bibnamefont {Facchi}},
  \bibinfo {author} {\bibfnamefont {M.}~\bibnamefont {Ligab\`o}}, \ and\
  \bibinfo {author} {\bibfnamefont {D.}~\bibnamefont {Lonigro}},\ }\bibfield
  {title} {\enquote {\bibinfo {title} {Hidden non-markovianity in open quantum
  systems},}\ }\href {\doibase 10.1103/PhysRevA.103.012203} {\bibfield
  {journal} {\bibinfo  {journal} {Phys. Rev. A}\ }\textbf {\bibinfo {volume}
  {103}},\ \bibinfo {pages} {012203} (\bibinfo {year}
  {2021}{\natexlab{b}})}\BibitemShut {NoStop}%
\bibitem [{\citenamefont {Burgarth}\ \emph
  {et~al.}(2021{\natexlab{c}})\citenamefont {Burgarth}, \citenamefont {Facchi},
  \citenamefont {Lonigro},\ and\ \citenamefont
  {Modi}}]{elusive_nM_PhysRevA.104.L050404}%
  \BibitemOpen
  \bibfield  {author} {\bibinfo {author} {\bibfnamefont {D.}~\bibnamefont
  {Burgarth}}, \bibinfo {author} {\bibfnamefont {P.}~\bibnamefont {Facchi}},
  \bibinfo {author} {\bibfnamefont {D.}~\bibnamefont {Lonigro}}, \ and\
  \bibinfo {author} {\bibfnamefont {K.}~\bibnamefont {Modi}},\ }\bibfield
  {title} {\enquote {\bibinfo {title} {Quantum non-markovianity elusive to
  interventions},}\ }\href {\doibase 10.1103/PhysRevA.104.L050404} {\bibfield
  {journal} {\bibinfo  {journal} {Phys. Rev. A}\ }\textbf {\bibinfo {volume}
  {104}},\ \bibinfo {pages} {L050404} (\bibinfo {year}
  {2021}{\natexlab{c}})}\BibitemShut {NoStop}%
\bibitem [{\citenamefont {Claes}\ \emph {et~al.}(2021)\citenamefont {Claes},
  \citenamefont {Rieffel},\ and\ \citenamefont {Wang}}]{PRXQuantum.2.010351}%
  \BibitemOpen
  \bibfield  {author} {\bibinfo {author} {\bibfnamefont {J.}~\bibnamefont
  {Claes}}, \bibinfo {author} {\bibfnamefont {E.}~\bibnamefont {Rieffel}}, \
  and\ \bibinfo {author} {\bibfnamefont {Z.}~\bibnamefont {Wang}},\ }\bibfield
  {title} {\enquote {\bibinfo {title} {Character randomized benchmarking for
  non-multiplicity-free groups with applications to subspace, leakage, and
  matchgate randomized benchmarking},}\ }\href {\doibase
  10.1103/PRXQuantum.2.010351} {\bibfield  {journal} {\bibinfo  {journal} {PRX
  Quantum}\ }\textbf {\bibinfo {volume} {2}},\ \bibinfo {pages} {010351}
  (\bibinfo {year} {2021})}\BibitemShut {NoStop}%
\bibitem [{\citenamefont {Chru\ifmmode \acute{s}\else
  \'{s}\fi{}ci\ifmmode~\acute{n}\else \'{n}\fi{}ski}\ and\ \citenamefont
  {Kossakowski}(2006)}]{PhysRevA.73.062314}%
  \BibitemOpen
  \bibfield  {author} {\bibinfo {author} {\bibfnamefont {D.}~\bibnamefont
  {Chru\ifmmode \acute{s}\else \'{s}\fi{}ci\ifmmode~\acute{n}\else
  \'{n}\fi{}ski}}\ and\ \bibinfo {author} {\bibfnamefont {A.}~\bibnamefont
  {Kossakowski}},\ }\bibfield  {title} {\enquote {\bibinfo {title}
  {Multipartite invariant states. i. unitary symmetry},}\ }\href {\doibase
  10.1103/PhysRevA.73.062314} {\bibfield  {journal} {\bibinfo  {journal} {Phys.
  Rev. A}\ }\textbf {\bibinfo {volume} {73}},\ \bibinfo {pages} {062314}
  (\bibinfo {year} {2006})}\BibitemShut {NoStop}%
\bibitem [{\citenamefont {Chiribella}\ \emph {et~al.}(2008)\citenamefont
  {Chiribella}, \citenamefont {D'Ariano},\ and\ \citenamefont
  {Perinotti}}]{PhysRevLett.101.060401}%
  \BibitemOpen
  \bibfield  {author} {\bibinfo {author} {\bibfnamefont {G.}~\bibnamefont
  {Chiribella}}, \bibinfo {author} {\bibfnamefont {G.~M.}\ \bibnamefont
  {D'Ariano}}, \ and\ \bibinfo {author} {\bibfnamefont {P.}~\bibnamefont
  {Perinotti}},\ }\bibfield  {title} {\enquote {\bibinfo {title} {Quantum
  circuit architecture},}\ }\href {\doibase 10.1103/PhysRevLett.101.060401}
  {\bibfield  {journal} {\bibinfo  {journal} {Phys. Rev. Lett.}\ }\textbf
  {\bibinfo {volume} {101}},\ \bibinfo {pages} {060401} (\bibinfo {year}
  {2008})}\BibitemShut {NoStop}%
\bibitem [{\citenamefont {Chiribella}\ \emph {et~al.}(2009)\citenamefont
  {Chiribella}, \citenamefont {D'Ariano},\ and\ \citenamefont
  {Perinotti}}]{PhysRevA.80.022339}%
  \BibitemOpen
  \bibfield  {author} {\bibinfo {author} {\bibfnamefont {G.}~\bibnamefont
  {Chiribella}}, \bibinfo {author} {\bibfnamefont {G.~M.}\ \bibnamefont
  {D'Ariano}}, \ and\ \bibinfo {author} {\bibfnamefont {P.}~\bibnamefont
  {Perinotti}},\ }\bibfield  {title} {\enquote {\bibinfo {title} {Theoretical
  framework for quantum networks},}\ }\href {\doibase
  10.1103/PhysRevA.80.022339} {\bibfield  {journal} {\bibinfo  {journal} {Phys.
  Rev. A}\ }\textbf {\bibinfo {volume} {80}},\ \bibinfo {pages} {022339}
  (\bibinfo {year} {2009})}\BibitemShut {NoStop}%
\bibitem [{\citenamefont {Portmann}\ \emph {et~al.}(2017)\citenamefont
  {Portmann}, \citenamefont {Matt}, \citenamefont {Maurer}, \citenamefont
  {Renner},\ and\ \citenamefont {Tackmann}}]{Portmann_causal}%
  \BibitemOpen
  \bibfield  {author} {\bibinfo {author} {\bibfnamefont {C.}~\bibnamefont
  {Portmann}}, \bibinfo {author} {\bibfnamefont {C.}~\bibnamefont {Matt}},
  \bibinfo {author} {\bibfnamefont {U.}~\bibnamefont {Maurer}}, \bibinfo
  {author} {\bibfnamefont {R.}~\bibnamefont {Renner}}, \ and\ \bibinfo {author}
  {\bibfnamefont {B.}~\bibnamefont {Tackmann}},\ }\bibfield  {title} {\enquote
  {\bibinfo {title} {Causal boxes: Quantum information-processing systems
  closed under composition},}\ }\href {\doibase 10.1109/tit.2017.2676805}
  {\bibfield  {journal} {\bibinfo  {journal} {IEEE Trans. Inf. Theory}\ ,\
  \bibinfo {pages} {1–1}} (\bibinfo {year} {2017})}\BibitemShut {NoStop}%
\bibitem [{\citenamefont {Nurdin}\ and\ \citenamefont
  {Gough}(2021)}]{nurdin2021heisenberg}%
  \BibitemOpen
  \bibfield  {author} {\bibinfo {author} {\bibfnamefont {H.~I.}\ \bibnamefont
  {Nurdin}}\ and\ \bibinfo {author} {\bibfnamefont {J.}~\bibnamefont {Gough}},\
  }\bibfield  {title} {\enquote {\bibinfo {title} {From the heisenberg to the
  schrödinger picture: Quantum stochastic processes and process tensors},}\
  }\href {\doibase 10.1109/cdc45484.2021.9683765} {\bibfield  {journal}
  {\bibinfo  {journal} {2021 60th IEEE Conference on Decision and Control
  (CDC)}\ } (\bibinfo {year} {2021}),\
  10.1109/cdc45484.2021.9683765}\BibitemShut {NoStop}%
\bibitem [{\citenamefont {Costa}\ and\ \citenamefont
  {Shrapnel}(2016)}]{Costa_2016}%
  \BibitemOpen
  \bibfield  {author} {\bibinfo {author} {\bibfnamefont {F.}~\bibnamefont
  {Costa}}\ and\ \bibinfo {author} {\bibfnamefont {S.}~\bibnamefont
  {Shrapnel}},\ }\bibfield  {title} {\enquote {\bibinfo {title} {Quantum causal
  modelling},}\ }\href {\doibase 10.1088/1367-2630/18/6/063032} {\bibfield
  {journal} {\bibinfo  {journal} {New J. Phys.}\ }\textbf {\bibinfo {volume}
  {18}},\ \bibinfo {pages} {063032} (\bibinfo {year} {2016})}\BibitemShut
  {NoStop}%
\bibitem [{\citenamefont {Kretschmann}\ and\ \citenamefont
  {Werner}(2005)}]{PhysRevA.72.062323}%
  \BibitemOpen
  \bibfield  {author} {\bibinfo {author} {\bibfnamefont {D.}~\bibnamefont
  {Kretschmann}}\ and\ \bibinfo {author} {\bibfnamefont {R.~F.}\ \bibnamefont
  {Werner}},\ }\bibfield  {title} {\enquote {\bibinfo {title} {Quantum channels
  with memory},}\ }\href {\doibase 10.1103/PhysRevA.72.062323} {\bibfield
  {journal} {\bibinfo  {journal} {Phys. Rev. A}\ }\textbf {\bibinfo {volume}
  {72}},\ \bibinfo {pages} {062323} (\bibinfo {year} {2005})}\BibitemShut
  {NoStop}%
\bibitem [{\citenamefont {Gutoski}\ and\ \citenamefont
  {Watrous}(2007)}]{Gutoski_games}%
  \BibitemOpen
  \bibfield  {author} {\bibinfo {author} {\bibfnamefont {G.}~\bibnamefont
  {Gutoski}}\ and\ \bibinfo {author} {\bibfnamefont {J.}~\bibnamefont
  {Watrous}},\ }\bibfield  {title} {\enquote {\bibinfo {title} {Toward a
  general theory of quantum games},}\ }\href {\doibase 10.1145/1250790.1250873}
  {\bibfield  {journal} {\bibinfo  {journal} {STOC '07}\ ,\ \bibinfo {pages}
  {565–574}} (\bibinfo {year} {2007})}\BibitemShut {NoStop}%
\bibitem [{\citenamefont {Milz}\ \emph {et~al.}(2017)\citenamefont {Milz},
  \citenamefont {Pollock},\ and\ \citenamefont {Modi}}]{Milz_2017}%
  \BibitemOpen
  \bibfield  {author} {\bibinfo {author} {\bibfnamefont {S.}~\bibnamefont
  {Milz}}, \bibinfo {author} {\bibfnamefont {F.~A.}\ \bibnamefont {Pollock}}, \
  and\ \bibinfo {author} {\bibfnamefont {K.}~\bibnamefont {Modi}},\ }\bibfield
  {title} {\enquote {\bibinfo {title} {An introduction to operational quantum
  dynamics},}\ }\href {\doibase 10.1142/s1230161217400169} {\bibfield
  {journal} {\bibinfo  {journal} {Open Syst. Inf. Dyn.}\ }\textbf {\bibinfo
  {volume} {24}},\ \bibinfo {pages} {1740016} (\bibinfo {year}
  {2017})}\BibitemShut {NoStop}%
\bibitem [{\citenamefont {Taranto}(2020)}]{Taranto_2020}%
  \BibitemOpen
  \bibfield  {author} {\bibinfo {author} {\bibfnamefont {P.}~\bibnamefont
  {Taranto}},\ }\bibfield  {title} {\enquote {\bibinfo {title} {Memory effects
  in quantum processes},}\ }\href {\doibase 10.1142/s0219749919410028}
  {\bibfield  {journal} {\bibinfo  {journal} {Int. J. Quantum Inf.}\ }\textbf
  {\bibinfo {volume} {18}},\ \bibinfo {pages} {1941002} (\bibinfo {year}
  {2020})}\BibitemShut {NoStop}%
\bibitem [{\citenamefont {Milz}\ \emph {et~al.}(2019)\citenamefont {Milz},
  \citenamefont {Kim}, \citenamefont {Pollock},\ and\ \citenamefont
  {Modi}}]{PhysRevLett.123.040401}%
  \BibitemOpen
  \bibfield  {author} {\bibinfo {author} {\bibfnamefont {S.}~\bibnamefont
  {Milz}}, \bibinfo {author} {\bibfnamefont {M.~S.}\ \bibnamefont {Kim}},
  \bibinfo {author} {\bibfnamefont {F.~A.}\ \bibnamefont {Pollock}}, \ and\
  \bibinfo {author} {\bibfnamefont {K.}~\bibnamefont {Modi}},\ }\bibfield
  {title} {\enquote {\bibinfo {title} {Completely positive divisibility does
  not mean {M}arkovianity},}\ }\href {\doibase 10.1103/PhysRevLett.123.040401}
  {\bibfield  {journal} {\bibinfo  {journal} {Phys. Rev. Lett.}\ }\textbf
  {\bibinfo {volume} {123}},\ \bibinfo {pages} {040401} (\bibinfo {year}
  {2019})}\BibitemShut {NoStop}%
\bibitem [{Note7()}]{Note7}%
  \BibitemOpen
  \bibinfo {note} {The Clifford group on $n$-qubits is defined as the set of
  unitaries normalizing the $n$-qubit Pauli group, $\protect \mathds {P}_n^*$
  modulo the identity, i.e., $\protect \mathds {C}_n:=\{G:\protect \mathds
  {U}(2^n)|P\in \pm \protect \mathds {P}_n^*\protect \,\Rightarrow \protect
  \,GPG^\dagger \in \pm \protect \mathds {P}_n^*\}$. The crucial property for
  \gls {rb} is that the $n$-qubit Clifford group forms a so-called unitary
  2-design, while the main limitation to scalability comes from the fact that
  $n$-qubit Clifford gates are composite gates, with both the number of
  elements and elementary components scaling non-favorably in $n$.}\BibitemShut
  {Stop}%
\bibitem [{\citenamefont {Blume-Kohout}\ \emph {et~al.}(2022)\citenamefont
  {Blume-Kohout}, \citenamefont {da~Silva}, \citenamefont {Nielsen},
  \citenamefont {Proctor}, \citenamefont {Rudinger}, \citenamefont {Sarovar},\
  and\ \citenamefont {Young}}]{taxonomy}%
  \BibitemOpen
  \bibfield  {author} {\bibinfo {author} {\bibfnamefont {R.}~\bibnamefont
  {Blume-Kohout}}, \bibinfo {author} {\bibfnamefont {M.~P.}\ \bibnamefont
  {da~Silva}}, \bibinfo {author} {\bibfnamefont {E.}~\bibnamefont {Nielsen}},
  \bibinfo {author} {\bibfnamefont {T.}~\bibnamefont {Proctor}}, \bibinfo
  {author} {\bibfnamefont {K.}~\bibnamefont {Rudinger}}, \bibinfo {author}
  {\bibfnamefont {M.}~\bibnamefont {Sarovar}}, \ and\ \bibinfo {author}
  {\bibfnamefont {K.}~\bibnamefont {Young}},\ }\bibfield  {title} {\enquote
  {\bibinfo {title} {A taxonomy of small markovian errors},}\ }\href {\doibase
  10.1103/PRXQuantum.3.020335} {\bibfield  {journal} {\bibinfo  {journal} {PRX
  Quantum}\ }\textbf {\bibinfo {volume} {3}},\ \bibinfo {pages} {020335}
  (\bibinfo {year} {2022})}\BibitemShut {NoStop}%
\bibitem [{Note8()}]{Note8}%
  \BibitemOpen
  \bibinfo {note} {A unitary $t$-design is an ensemble of unitary matrices that
  reproduce up to the $t$\protect \textsuperscript {th} statistical moment of
  the unitary group with the uniform, so-called Haar measure; see e.g.~\cite
  {PhysRevA.85.042311}\label {footnote: unitary designs}}\BibitemShut {NoStop}%
\bibitem [{\citenamefont {Milz}\ \emph {et~al.}(2021)\citenamefont {Milz},
  \citenamefont {Spee}, \citenamefont {Xu}, \citenamefont {Pollock},
  \citenamefont {Modi},\ and\ \citenamefont {Gühne}}]{time_entanglement}%
  \BibitemOpen
  \bibfield  {author} {\bibinfo {author} {\bibfnamefont {S.}~\bibnamefont
  {Milz}}, \bibinfo {author} {\bibfnamefont {C.}~\bibnamefont {Spee}}, \bibinfo
  {author} {\bibfnamefont {Z.-P.}\ \bibnamefont {Xu}}, \bibinfo {author}
  {\bibfnamefont {F.}~\bibnamefont {Pollock}}, \bibinfo {author} {\bibfnamefont
  {K.}~\bibnamefont {Modi}}, \ and\ \bibinfo {author} {\bibfnamefont
  {O.}~\bibnamefont {Gühne}},\ }\bibfield  {title} {\enquote {\bibinfo {title}
  {Genuine multipartite entanglement in time},}\ }\href {\doibase
  10.21468/scipostphys.10.6.141} {\bibfield  {journal} {\bibinfo  {journal}
  {SciPost Phys.}\ }\textbf {\bibinfo {volume} {10}} (\bibinfo {year} {2021}),\
  10.21468/scipostphys.10.6.141}\BibitemShut {NoStop}%
\bibitem [{Note9()}]{Note9}%
  \BibitemOpen
  \bibinfo {note} {The only assumption we don't explicitly relax is
  gate-dependence, i.e., the noise maps $\Lambda _i$ at any given time-step $i$
  do not depend explicitly on which gate $\protect \mathcal {G}_i$ was
  applied.}\BibitemShut {Stop}%
\bibitem [{\citenamefont {Kong}(2021)}]{kong2021framework}%
  \BibitemOpen
  \bibfield  {author} {\bibinfo {author} {\bibfnamefont {L.}~\bibnamefont
  {Kong}},\ }\href@noop {} {\enquote {\bibinfo {title} {A framework for
  randomized benchmarking over compact groups},}\ } (\bibinfo {year} {2021}),\
  \Eprint {http://arxiv.org/abs/2111.10357} {arXiv:2111.10357 [quant-ph]}
  \BibitemShut {NoStop}%
\bibitem [{Note10()}]{Note10}%
  \BibitemOpen
  \bibinfo {note} {A representation of a finite group $\protect \mathds {G}$
  can be defined as a map from $\protect \mathds {G}$ to a vector space of
  unitary linear operators, e.g., that of $d$-dimensional complex matrices
  $\phi :\protect \mathds {G}\to \protect \mathds {U}(\protect \mathds
  {C}_{d\times {d}})$. We call a representation reducible if it can be
  expressed as a direct sum of irreducible representations (irreps), and we say
  this representation is multiplicity-free if it contains no more than a single
  copy of such irreps: $\phi =\protect \Moplus _\mu \phi _\mu ^{\otimes {n_\mu
  }}$ with $n_\mu =1$ for all labels $\mu $. This is a technical constraint
  that can potentially be relaxed~\cite {PRXQuantum.2.010351,
  helsen_general_2022} with some added but potentially non-crucial
  complications.\label {fn: multiplicity-free}}\BibitemShut {NoStop}%
\bibitem [{Note11()}]{Note11}%
  \BibitemOpen
  \bibinfo {note} {We use the notation $\protect \hat {\protect \mathcal
  {P}}_\pi $ in order to not excessively conflate terms in Eq.~\protect \textup
  {\hbox {\mathsurround \z@ \protect \normalfont (\ignorespaces \ref {eq:
  non-Markov asf}\unskip \@@italiccorr )}}, however, all $\protect \hat
  {\protect \mathcal {P}}_\pi $ here are literally projection operators, not
  stemming \protect \emph {by definition} from a quantum channel. Of course one
  can still also associate a channel $\protect \mathcal {P}_\pi $ to them,
  which however, does not necessarily has the action of a projection
  superoperator.}\BibitemShut {Stop}%
\bibitem [{\citenamefont {Helsen}\ \emph {et~al.}(2019)\citenamefont {Helsen},
  \citenamefont {Xue}, \citenamefont {Vandersypen},\ and\ \citenamefont
  {Wehner}}]{helsen2019character}%
  \BibitemOpen
  \bibfield  {author} {\bibinfo {author} {\bibfnamefont {J.}~\bibnamefont
  {Helsen}}, \bibinfo {author} {\bibfnamefont {X.}~\bibnamefont {Xue}},
  \bibinfo {author} {\bibfnamefont {L.~M.~K.}\ \bibnamefont {Vandersypen}}, \
  and\ \bibinfo {author} {\bibfnamefont {S.}~\bibnamefont {Wehner}},\
  }\bibfield  {title} {\enquote {\bibinfo {title} {A new class of efficient
  randomized benchmarking protocols},}\ }\href {\doibase
  10.1038/s41534-019-0182-7} {\bibfield  {journal} {\bibinfo  {journal} {npj
  Quantum Inf.}\ }\textbf {\bibinfo {volume} {5}},\ \bibinfo {pages} {71}
  (\bibinfo {year} {2019})}\BibitemShut {NoStop}%
\bibitem [{\citenamefont {Gough}\ and\ \citenamefont
  {Nurdin}(2017)}]{Nurdin_2017}%
  \BibitemOpen
  \bibfield  {author} {\bibinfo {author} {\bibfnamefont {J.~E.}\ \bibnamefont
  {Gough}}\ and\ \bibinfo {author} {\bibfnamefont {H.~I.}\ \bibnamefont
  {Nurdin}},\ }\bibfield  {title} {\enquote {\bibinfo {title} {Can quantum
  {M}arkov evolutions ever be dynamically decoupled?}}\ }in\ \href {\doibase
  10.1109/CDC.2017.8264587} {\emph {\bibinfo {booktitle} {2017 IEEE 56th Annual
  Conference on Decision and Control (CDC)}}}\ (\bibinfo  {publisher} {IEEE},\
  \bibinfo {year} {2017})\ pp.\ \bibinfo {pages} {6155--6160}\BibitemShut
  {NoStop}%
\bibitem [{Note12()}]{Note12}%
  \BibitemOpen
  \bibinfo {note} {Equivalently a unitary 2-design; the projectors are
  $\protect \hat {\protect \mathcal {P}}_1=\Psi $ and $\protect \hat {\protect
  \mathcal {P}}_2=\protect \mathds 1-\Psi $ where $\Psi :=\DOTSB \sum@
  \slimits@ |ii\rangle \protect \!\langle {jj}|/d_\protect \mathsf {S}$. See
  e.g.~\cite {PhysRevA.73.062314}. For details on the particular case of the
  \gls {asf} in the Clifford case, see~\cite
  {figueroaromero2022general}.}\BibitemShut {Stop}%
\bibitem [{\citenamefont {Sacchi}(2005)}]{Sacchi_2005}%
  \BibitemOpen
  \bibfield  {author} {\bibinfo {author} {\bibfnamefont {M.~F.}\ \bibnamefont
  {Sacchi}},\ }\bibfield  {title} {\enquote {\bibinfo {title} {Optimal
  discrimination of quantum operations},}\ }\href {\doibase
  10.1103/PhysRevA.71.062340} {\bibfield  {journal} {\bibinfo  {journal} {Phys.
  Rev. A}\ }\textbf {\bibinfo {volume} {71}},\ \bibinfo {pages} {062340}
  (\bibinfo {year} {2005})}\BibitemShut {NoStop}%
\bibitem [{\citenamefont {Wallman}\ and\ \citenamefont
  {Flammia}(2014)}]{rbconfidence_2014}%
  \BibitemOpen
  \bibfield  {author} {\bibinfo {author} {\bibfnamefont {J.~J.}\ \bibnamefont
  {Wallman}}\ and\ \bibinfo {author} {\bibfnamefont {S.~T.}\ \bibnamefont
  {Flammia}},\ }\bibfield  {title} {\enquote {\bibinfo {title} {Randomized
  benchmarking with confidence},}\ }\href {\doibase
  10.1088/1367-2630/16/10/103032} {\bibfield  {journal} {\bibinfo  {journal}
  {New J. Phys.}\ }\textbf {\bibinfo {volume} {16}},\ \bibinfo {pages} {103032}
  (\bibinfo {year} {2014})}\BibitemShut {NoStop}%
\bibitem [{\citenamefont {Kern}\ \emph {et~al.}(2005)\citenamefont {Kern},
  \citenamefont {Alber},\ and\ \citenamefont {Shepelyansky}}]{PFR_2005}%
  \BibitemOpen
  \bibfield  {author} {\bibinfo {author} {\bibfnamefont {O.}~\bibnamefont
  {Kern}}, \bibinfo {author} {\bibfnamefont {G.}~\bibnamefont {Alber}}, \ and\
  \bibinfo {author} {\bibfnamefont {D.~L.}\ \bibnamefont {Shepelyansky}},\
  }\bibfield  {title} {\enquote {\bibinfo {title} {Quantum error correction of
  coherent errors by randomization},}\ }\href {\doibase
  10.1140/epjd/e2004-00196-9} {\bibfield  {journal} {\bibinfo  {journal} {Eur.
  Phys. J. D.}\ }\textbf {\bibinfo {volume} {32}},\ \bibinfo {pages}
  {153–156} (\bibinfo {year} {2005})}\BibitemShut {NoStop}%
\bibitem [{Note13()}]{Note13}%
  \BibitemOpen
  \bibinfo {note} {Strictly, the ordering in the vectorized (hat $\protect \hat
  {\protect \mathcal {X}}$) representation, as we defined it, is $\protect
  \mathsf {E}\protect \mathsf {S}\otimes \protect \mathsf {E}\protect \mathsf
  {S}$; we are slightly abusing notation for clarity of
  presentation.}\BibitemShut {Stop}%
\bibitem [{Note14()}]{Note14}%
  \BibitemOpen
  \bibinfo {note} {Notice that the time-stationary case, $\Lambda _i=\Lambda $
  for all $i$, would leave the \gls {asf} in a \protect \emph
  {time-non-stationary} noise form due to the environment dependence, as we
  would drop all single $i$ indices, but the remaining $\protect \mathsf {E}$
  indices, $\mu ,\sigma $, and thus all quality factors $p^{(\mu \sigma )}$
  too, would remain distinct for each time-step.}\BibitemShut {Stop}%
\bibitem [{Note15()}]{Note15}%
  \BibitemOpen
  \bibinfo {note} {(Complete) Positivity of a quantum channel $\Lambda $ does
  not correspond to positive semi-definiteness of $\protect \hat {\Lambda }$;
  however it does correspond to positive semi-definiteness of the so-called
  Choi state of $\Lambda $, which can be defined as $\Upsilon _\Lambda
  :=(\Lambda \otimes \protect \mathcal {I})\Psi $, where $\Psi =\DOTSB \sum@
  \slimits@ |ii\rangle \protect \!\langle {jj}|$ is a maximally entangled
  operator on $\protect \mathsf {S}^{\otimes 2}$.}\BibitemShut {Stop}%
\end{thebibliography}
\end{document}